\documentclass[12pt]{article}

\usepackage[margin={1 in}]{geometry}

\usepackage{floatrow} %
\usepackage[utf8]{inputenc}
\usepackage{setspace}
\usepackage{multirow}
\usepackage[table,xcdraw]{xcolor}

\usepackage{comment}

\usepackage[normalem]{ulem}

\usepackage{amsmath}
\usepackage{amssymb}
\usepackage{amsthm}
\usepackage{graphicx}
\usepackage{float}
\usepackage{adjustbox}

\usepackage[
    sorting=ynt,
    backend=biber,
    style=apa,
    natbib
  ]{biblatex}
  
\usepackage[labelfont=bf]{caption}
\usepackage{enumitem}
\usepackage{xcolor}
\usepackage{graphicx}
\usepackage{setspace}
\usepackage[title]{appendix}
\usepackage{hyperref}
\hypersetup{
    colorlinks=true,
    linkcolor=blue,
    filecolor=magenta,      
    urlcolor=blue,
    citecolor=blue,
}
\usepackage{booktabs,caption}   %
\usepackage[flushleft]{threeparttable}
\usepackage{subfiles}

\usepackage{xr}
\makeatletter
\newcommand*{\addFileDependency}[1]{
  \typeout{(#1)}
  \@addtofilelist{#1}
  \IfFileExists{#1}{}{\typeout{No file #1.}}
}
\makeatother

\usepackage[labelformat=simple]{subcaption}

\makeatletter
\patchcmd{\ALG@step}{\addtocounter{ALG@line}{1}}{\refstepcounter{ALG@line}}{}{}
\newcommand{\ALG@lineautorefname}{Line}
\makeatother

\usepackage{etoolbox}
\makeatletter
\patchcmd{\hyper@makecurrent}{%
    \ifx\Hy@param\Hy@chapterstring
        \let\Hy@param\Hy@chapapp
    \fi
}{%
    \iftoggle{inappendix}{%
        \@checkappendixparam{chapter}%
        \@checkappendixparam{section}%
        \@checkappendixparam{subsection}%
        \@checkappendixparam{subsubsection}%
        \@checkappendixparam{paragraph}%
        \@checkappendixparam{subparagraph}%
    }{}%
}{}{\errmessage{failed to patch}}

\newcommand*{\@checkappendixparam}[1]{%
    \def\@checkappendixparamtmp{#1}%
    \ifx\Hy@param\@checkappendixparamtmp
        \let\Hy@param\Hy@appendixstring
    \fi
}
\makeatletter

\newtoggle{inappendix}
\togglefalse{inappendix}

\apptocmd{\appendix}{\toggletrue{inappendix}}{}{\errmessage{failed to patch}}
\apptocmd{\subappendices}{\toggletrue{inappendix}}{}{\errmessage{failed to patch}}

\newcommand*{\subfilesbibliography}[1]{%
 \expandafter\ifx\csname ver@subfiles.cls\endcsname\relax
   \expandafter\@secondoftwo
 \else
   \expandafter\@firstoftwo
  \fi
  {\bibliography{#1}}
  {}
}
\makeatother

\usepackage{xcolor}
\usepackage{makecell}

\newcommand{\aref}[1]{\hyperref[#1]{Appendix~\ref*{#1}}}

\def\figscale{1}

\title{\vspace{-3em}{\LARGE Describing~Deferred~Acceptance~and~Strategyproofness to Participants: Experimental Analysis}\footnote{Gonczarowski: Department of Economics and Department of Computer Science, Harvard University. (e-mail: \url{yannai@gonch.name}); Heffetz: Johnson Graduate School of Management, Cornell University, Bogen Department of Economics and Federmann Center for Rationality, The Hebrew University of Jerusalem, and NBER (e-mail: \url{oh33@cornell.edu}); Ishai: Bogen Department of Economics and Federmann Center for Rationality, The Hebrew University of Jerusalem (e-mail: \url{guy.ishai@mail.huji.ac.il}); Thomas: Microsoft Research (e-mail: \url{thomas.clay95@gmail.com}). The authors thank Keren-Or Barashi Gortler, Itamar Bellaiche, Yehonatan Caspi, Yael Cohen, Gabriela Cohen-Hadid, Ayala Goldfarb, Michael Khalfin, Ido Leshkowitz, Josef Mccrum, Shenhav Or, Yonatan Rahimi, and Ohad Weschler for excellent research assistance; Eric Budish, Ben Enke, Nicole Immorlica, David Laibson, Markus Mobius, Assaf Romm, Shigehiro Serizawa, Ran Shorrer, Alex Teytelboym, and Leeat Yariv
for helpful discussions;
participants at the Stanford Institute for Theoretical Economics (SITE) 2023 Experimental Economics, SITE 2023 Market Design, WZB Berlin Matching Workshop, Crown Family Israel Center for Innovation (ICI) 2024 Academic Conference, Virtual Market Design Seminar, 2024 Marketplace Innovations Workshop, 8th Solomon Lew Conference on Behavioral Economics (Tel Aviv), 1st Annual Chicago Booth Market Design Conference, EC 2024, Communicating Clearly to Market Participants Workshop (Stony Brook), and seminar participants at Bar Ilan, Cornell, the Hebrew University, and Microsoft Research for comments that significantly improved the paper; 
and Adam Chafee and his team at the Cornell Business Simulation Lab for their help with running the experiment. Pre-registration of our experiment can be found at
\url{https://aspredicted.org/7eq7e.pdf}. A one-page abstract of this paper appeared in the  proceedings of EC 2024. The authors gratefully acknowledge research support by the following sources. Gonczarowski: National Science Foundation (NSF-BSF grant No.\ 2343922), Harvard FAS Inequality in America Initiative, and Harvard FAS Dean’s Competitive Fund for Promising Scholarship. Heffetz: Israel Science Foundation (grant No.\ 2968/21), US-Israel Binational Science Foundation (NSF-BSF grant No.\ 2023676), Cornell’s S.C. Johnson School, and Cornell's Center for Social Sciences. Ishai: Barbara and Morton Mandel Doctoral Program, Bogen Family, and Federmann Center for Rationality. Thomas: NSF CCF-1955205, Wallace Memorial Fellowship in Engineering, and Siebel Scholar award; part of his work was carried out while in Princeton's Department of Computer Science. }}

\author{Yannai A. Gonczarowski \and Ori Heffetz \and Guy Ishai \and Clayton Thomas}

\date{
September 26, 2024
}

\begin{document}

\maketitle

\vspace{-2em}
\begin{abstract}
We conduct an incentivized lab experiment to test participants' ability to understand the DA matching mechanism and the strategyproofness property, conveyed in different ways.
We find that while many participants can (using a novel GUI) learn DA's mechanics and calculate its outcomes, such understanding does not imply understanding of strategyproofness (as measured by specially designed tests).
However, a novel \emph{menu} description of strategyproofness conveys this property significantly better than other treatments.
While behavioral effects are small on average, participants with levels of strategyproofness understanding above a certain threshold play the classical dominant strategy at very high rates.

\end{abstract}

\section{Introduction}
\label{sec:intro}

To what extent do participants in the widely used Deferred Acceptance matching mechanism (\citealp{GaleS62}; henceforth, DA) understand how it works? To what extent do they understand one of the mechanism's most celebrated properties, namely, its strategyproofness? 
Are there principled ways to change the structure of how DA and strategyproofness are described to participants that could improve their understanding? 
In this paper we experimentally study these questions. 
We compare a traditional DA description, and a description of the definition of strategyproofness itself, with new \emph{menu} versions of both, theoretically developed in \citet{GonczarowskiHT23}.
We study effects of changing descriptions on participants' understanding of the mechanism, on their understanding of strategyproofness, and on the strategies they play in the mechanism.

DA is a mechanism used in many real-world matching systems, from residency matching to school choice. 
A crucial property of DA is that it is strategyproof:
Under classical assumptions on participants' preferences, straightforward (henceforth, SF) reporting---i.e., ranking options from highest to lowest value---is a dominant strategy.\footnote{
    Throughout this paper, we refer to the term (non-)straightforward, abbreviated (N)SF, to refer to the strategy of (not) ranking options from highest to lowest value.
    The SF strategy is sometimes described in past research as the  ``truthtelling strategy,'' but we adopt different terminology to separate participants' behavior from any notions of dishonesty.
}
In theory, strategyproofness obviates any need to strategize. It thus promotes ease of participation, reduces inequality across those with different levels of strategic sophistication, and increases robustness of theoretical predictions on play. 
However, strategyproofness achieves all this only if participants are \emph{aware} that the mechanism is strategyproof.

Unfortunately, growing evidence from both the lab and the field shows that DA participants frequently do not play straightforwardly. 
This behavior has been the focus of much recent work in market design \citep[for surveys, see][]{HakimovK19,Rees-JonesS23}. 
Various theoretical explanations of non-straightforward (henceforth, NSF) strategies have been suggested---from cognitive failures to nonclassical preferences---and based on these explanations, several papers have suggested that (in some contexts and with important caveats) NSF play might be lowered by using dynamic, interactive mechanisms.\footnote{
   For examples in various mechanisms, \citet{Li17} and \citet{PyciaT23} suggest that NSF behavior can be explained by failures of contingent reasoning, and \citet{DreyfussHR22} and \citet{MeisnerW23} suggest expectations-based loss aversion.
   Based on their respective explanations, these papers propose changing the dynamic implementation of the mechanism.
   Other papers investigating dynamic implementations include \citet{KagelL93, BreitmoserS19, BoH23}. See discussion in \autoref{sec:related}.
} 
Other pragmatic approaches to mitigating NSF behavior have also been suggested, such as providing participants with trusted, concrete advice on how to determine their ranking.\footnote{
    See, e.g., \citet{GuillenH14, MasudaMSSW20, GuillenV21, ReesJonesS18}; for details, see \autoref{sec:related}.
}

Despite ample progress towards a better understanding of NSF play, our knowledge remains limited about one potential channel inducing it: To what extend do participants play NSF due to general \emph{misunderstandings}, of either the mechanism itself or the property of strategyproofness? 
This question, and the question of how to mitigate such misunderstanding, are the focus of the current paper.

We conduct an incentivized lab experiment with five between-subjects treatments.
We first expose participants to one of five descriptions of DA or its strategyproofness property, accompanied with specifically tailored training modules. 
We then elicit ranking behavior in ten incentivized DA rounds. 
Finally, we test understanding of strategyproofness. 
We use a static DA setting in which four participants---one human and three computerized---are matched to four prizes based on participants' submitted rankings of the prizes and their exogenous priorities for getting the prizes.

Our design features two novel aspects.
First, inspired by prior real-world approaches to explaining DA \citep[see, e.g.,][]{MatchingExplained20},
we construct a new GUI (graphical user interface) in which participants can perform the sequence of DA proposals and tentative acceptances for themselves to calculate the outcome.
This GUI gives us effective tools towards not only \emph{teaching} the DA algorithm to participants, but also \emph{assessing} their understanding of its (taught) mechanics---providing us with new, DA-understanding  outcome variables.
Second, we directly test participants' understanding of \emph{strategyproofness} (separately from assessing their understanding of DA mechanics, and from their actual play), via eighteen quiz-style questions that ask participants either (i) which outcomes are logically possible in different abstract scenarios, or 
(ii) how participants should practically play to maximize their earnings---all solvable using the definition of strategyproofness alone.
These tests provide a rich, second set of new outcome variables that allow us to assess measures of strategyproofness understanding (henceforth, SP understanding) separate from both (our new) DA-understanding outcomes and (routinely studied) SF-play outcomes.
Together with our experimental flow, our framework provides a general and extensive method of testing participants' understanding of different facets of mechanisms based on descriptions, which we view as a methodological contribution of our paper (in addition to the substantive contribution of our main findings below).

In our first treatment---Traditional DA Mechanics (Trad-DA)---we examine participants' response to the traditional, complete description of (the mechanics of, i.e., how the outcome is calculated in) the participant-proposing DA algorithm.

Our second treatment is motivated by recent theoretical work by \citet{GonczarowskiHT23} (henceforth, GHT). 
GHT construct novel descriptions of (static, direct-revelation) mechanisms where strategyproofness holds via a simple argument.
In particular, GHT constructs
\emph{menu descriptions}, which, inspired by \citet{Hammond79}, 
present the mechanism to participant $i$ via two steps:
\begin{itemize}[]
  \label{itemize:menu-description}
  \itemsep0em
  \item \textbf{Step (1)} uses only the reports of other participants to describe, in complete and explicit detail, the set of outcomes participant $i$ might receive, called $i$'s \emph{menu}.
  \item \textbf{Step (2)} describes 
    how to award participant $i$ her favorite 
    outcome (according to her report) from her menu.
\end{itemize}
The main idea behind menu descriptions is that strategyproofness for participant $i$ follows from a menu description via a one-sentence argument: $i$'s menu in Step~(1) cannot be affected by her report, and SF reporting guarantees~$i$ her favorite outcome from the menu in Step~(2).
We %
compare participants' reactions to such a description with their reactions to the traditional description of DA, where the proof is more challenging.\footnote{
  \citet{KatuscakK2020} also experiment with a menu description (of the Top Trading Cycle mechanism) in a matching environment, though without our new design features and outcome variables. For details and other related work, see \autoref{sec:related}.
}

This second treatment---Menu DA Mechanics (Menu-DA)---relays to participants a menu description of DA that follows the main positive result of GHT, i.e., it relays the two-step outline above while providing full details on (the mechanics of) how the menu is calculated in Step (1).
This menu description constructed by GHT is more involved than the traditional description, 
and we convey it to participants using a modified and extended version of the GUI from the first treatment. 
Both of our DA Mechanics treatments provide in-depth coaching to participants, who learn to calculate their DA outcomes for themselves within the GUI.
These two treatments are designed to be as directly comparable to each other as reasonably possible, including visually and in the mechanics-understanding measures they elicit.

Our third treatment---Menu SP Property (Menu-SP)---departs from our two DA Mechanics treatments by focusing on relaying \emph{only} that the matching mechanism satisfies the strategyproofness property, without specifying how one could calculate the mechanism's outcome.
In particular, the Menu-SP treatment relays the two-step outline above, while providing no details on \emph{how} the menu is calculated in Step (1).
This treatment provides a potential real-world approach to conveying strategyproofness, allowing comparison between Menu-DA and its stripped-down version that contains only the information directly relevant for knowing that the mechanism is strategyproof. (And, since every strategyproof mechanism has a menu description, this treatment conveys no more information than that the mechanism is strategyproof.)

Our fourth treatment---Textbook SP Property (Textbook-SP)---also relays (only) the fact that the mechanism is strategyproof.
It does so using a definition inspired by classical textbook ones (but written in simpler, everyday language), providing a mathematically equivalent version for comparison with Menu-SP.

Importantly, none of our treatments gives explicit strategic advice on how participants should construct their reported rankings. 
In particular, %
we distinguish between \emph{advice}---such as recommending that participants play the SF strategy---which all our treatments avoid; and \emph{the strategyproofness property}---a particular property of the mechanism, i.e., of the mapping from reported rankings to allocations---that the SP Property treatments aim to directly teach. 
These treatments also provide coaching to participants, using questions similar in nature to those in the SP-understanding tests, accompanied with detailed feedback. 
As with our two DA-Mechanics treatments, the two SP-Property treatments are designed to be as directly comparable to each other as reasonably possible.

Finally, our fifth treatment is a control, close-to-zero-information treatment, termed Null. It provides a short, generic description explaining that the mechanism generates a matching while attempting to accommodate participants' submitted rankings, without any details explaining either the mechanics of how the matching mechanism works or that it is strategyproof.

In all five treatments, the training module---which differs by treatment---is followed by an incentivized DA-rounds module---which is \emph{identical} across treatments (in both interface and underlying mechanism). 
Thus, for example, while the DA Mechanics treatments (Trad-DA and Menu-DA) convey to participants (only) that the mechanism is DA, this mechanism is (also) in fact strategyproof; similarly, while the SP Property treatments (Menu-SP and Textbook-SP) convey to participants (only) that the mechanism is strategyproof, this mechanism is (also) implemented using DA.
More generally, our experimental design tests how participants respond when different facets of a \emph{fixed environment} are described in different ways.

We conduct our experiment (combined $N=542$; pre-registered) on two different samples: US participants recruited on Prolific and participating remotely via live video sessions ($N=255$), and Cornell students participating physically via in-person lab sessions ($N=287$). These different settings, as well as populations, are meant to induce variation in focus and attention, as well as cognitive resources and education---all key moderators of our hypothesized mechanisms---to investigate the extent to which they affect our main results. 
Our main results replicate across the two samples, hence we pool them in our main analysis.

We highlight four main results. 
First, participant training scores in the Traditional DA Mechanics treatment suggest that our training is indeed effective at teaching many participants the mechanics of DA. 
For example, the last task in our DA Mechanics training modules asks participants to calculate their DA outcome using our GUI completely on their own for a specific DA scenario.
Three quarters of participants in Trad-DA (76\% ; SE $=$ 4\%) succeed with no mistakes on first attempt. Of the remaining quarter (24\%), who are only informed “Incorrect allocation. Please try again,” another third (8 percent of the total) succeed with no mistakes on the next attempt.
Crucial to this result is our new GUI and extensive instruction and quizzing of participants, which enables a much more systematic reporting and analysis of participants' understanding levels compared with prior work.

Second, we find that understanding the full, explicit mechanics of DA \emph{does not} imply understanding its strategyproofness.
Our SP-understanding tests contain 
(i)~thirteen \emph{abstract} questions asking participants to apply the logical definition of strategyproofness (e.g., ``...imagine that you have submitted the ranking \mbox{B--A--C--D} [of four possible prizes A, B, C, D], and got Prize C... If you had instead submitted \mbox{A--B--C--D}...\ is it possible (or certain) that you would have gotten Prize A? Yes/No''); %
and (ii)~five \emph{practical} questions asking participants about the implications of strategyproofness (e.g., ``If I want to maximize my earnings in a given round, then... Sometimes I might have to rank the prize that earns me the most in second place or lower. True/False'').

In Trad-DA, mean overall SP-understanding score is 56\%,
and in Menu-DA it is 58\% (both SEs $=2\%$).
Despite strategyproofness holding via a one-sentence argument in Menu-DA, the two are statistically indistinguishable, and are very close to the baseline set by the Null treatment,  of 54\% (SE $=1\%$).
Thus, to the extent that our tests are effective at measuring understanding of strategyproofness, instructing participants on the full mechanics of how their outcome is calculated has (perhaps startlingly) little effect on their SP understanding.

Third, we find that an SP Property treatment, and in particular Menu-SP, \emph{can} to some extent teach participants strategyproofness. 
Participants' mean scores on the SP-understanding tests in Menu-SP is 71\% (SE $=2\%$)---the highest of any treatment.
Participants in Textbook-SP (which is based on the classical definition of strategyproofness) have a mean score of 62\% (SE $=2\%$)---the second highest, but significantly lower than in Menu-SP.
Investigating the two subsets of SP-understanding questions separately, we find that participants in Menu-SP fare better in both the abstract and the practical questions.
We also find that scores on the SP-understanding test, particularly the practical ones, 
exhibit a form of all-or-nothing bimodality, suggesting a degree to which participants either consistently ``get strategyproofness'' or consistently ``get it wrong.''

Interestingly, despite significant shifts in participants' SP-understanding scores, 
we do \emph{not} see a similarly significant shift in the overall distributions of SF play.
In particular, the rate of SF play has a spread-out, somewhat bimodal distribution
across participants in all treatments, %
and mean rates of SF play across treatments range from 48\% in Null to 59\% in Menu-SP (SE $=$ 3\% in all treatments).\footnote{
  Mean rates of SF play are: 56\% in Trad-DA, 50\% in Menu-DA, 59\% in Menu-SP, 53\% in Textbook-SP, and 48\% in Null. 
} 
Hence, although treatment differences in mean SF rate are largely directionally consistent with mean SP-understanding rates, the overall distributions do not generally suggest large statistical differences between rates of SF play.

Fourth, we find that, despite the relatively similar distributions of rates of SF play across treatments, there is a strong positive relationship, across participants, between SP-understanding and rates of SF play.
In particular, there is a step-function-like pattern in the relation between these two variables:
At low and mid-rate SP-understanding levels, higher understanding score is not associated with significantly higher SF behavior rates (which are around 50\% on average);
however, for SP-understanding scores above roughly 75--80\%, SF behavior rates are dramatically higher (mostly 80--100\%). 
Moreover, underlying the findings above, we find that Menu-SP shifts the largest group of participants from the region of low-to-medium SP-understanding and relatively low SF rates to the region of high rates in both measures. 
When coupled with the bimodality we see in the SP-understanding test, this suggests that participants not only \emph{perceive} strategyproofness in a binary way, but that they additionally \emph{act on} their perceptions to play SF at dramatically different rates.

Our results section and appendices include further analysis. %
We find that in addition to its relatively low SP-understanding scores, Menu-DA makes DA mechanics substantially more complicated for participants relative to Trad-DA---as reflected by training scores and success at DA-outcome calculations using our GUI.
We investigate the joint distribution of participants' performance on the abstract and practical sets of questions of SP understanding and find that in our setting, abstract understanding of strategyproofness is mostly a necessary condition (albeit an insufficient one) to understanding its practical implications.
We investigate how participants' behavior relates to SP-understanding in more detail, and find that %
success at the practical questions is a far better predictor of  SF behavior compared to the abstract questions. 
We also show that our results are robust to controlling for a rich set of demographic characteristics and session-date fixed effects. 
Our findings also replicate across samples---while, notably, baseline understanding levels, as well as treatment effects, are lower on Prolific than at Cornell.\footnote{
    We find similar results when splitting the sample into top vs.\ bottom performers in cognitive-ability exit questions, and into more vs.\ less attentive participants as measured by attention-check questions embedded in our experiment.
}

Overall, our findings highlight that understanding the mechanics of calculating the outcome of a mechanism---in our case, DA---is very different from understanding that it is strategyproof. While traditional DA descriptions can be effective at teaching the former, i.e., how outcomes are calculated from inputs, we find that they do not make an effective difference in teaching the latter (relative to our close-to-zero-information Null treatment).
In contrast, a stripped-down menu description does make an effective difference in teaching strategyproofness.

Our paper proceeds as follows.
\autoref{sec:design} outlines our experimental design.
\autoref{sec:results} presents our results.
In \autoref{sec:related}, we overview much of the long line of literature in behavioral mechanism design related to and inspiring our work.
We conclude in \autoref{sec:conclusion} by discussing some implications of our findings. 
In particular, we see promise in our new Menu SP Property description, for both future research and real-world applications.

\section{Experimental Design}
\label{sec:design}

Our experiment studies the effects of traditional vs.\ menu descriptions of DA and SP on three outcomes of interest: (i) participants' understanding of the description; (ii) their understanding of strategyproofness; and (iii) their behavior. 

Our experiment (a) introduces the matching environment; (b) describes the matching mechanism---``allocation process''---using one of our five descriptions; (c) attempts to improve, and tests, participants' understanding of the mechanism using novel training questions and/or GUI; (d) elicits behavior in standard, incentivized DA rounds; (e) elicits understanding of strategyproofness using novel tests; and finally (f) elicits reflections, perceptions, cognitive-ability measures, and a set of demographic characteristics. 
Only steps (b) and (c) differ across treatments.

Participants earn money mainly from incentivized rounds of DA in (d) (up to \pounds /\$9.90 on Prolific/at Cornell, respectively), but they are also incentivized with a monetary comprehension bonus for all questions they answer correctly on the first attempt in training questions in (a)--(c) and in the SP-understanding tests in (e) (up to \pounds /\$4.50). 
We add these (somewhat nonstandard) comprehension bonuses in order to encourage participants to increase their attention,
and to incentivize each of our main outcome variables.
Participants are shown their cumulative DA-round earnings after each round; their understanding bonus is only calculated at the end of the experiment, based on their fraction of correct answers.

\begin{figure}[htbp]
    \centering
    \includegraphics[width=\textwidth]{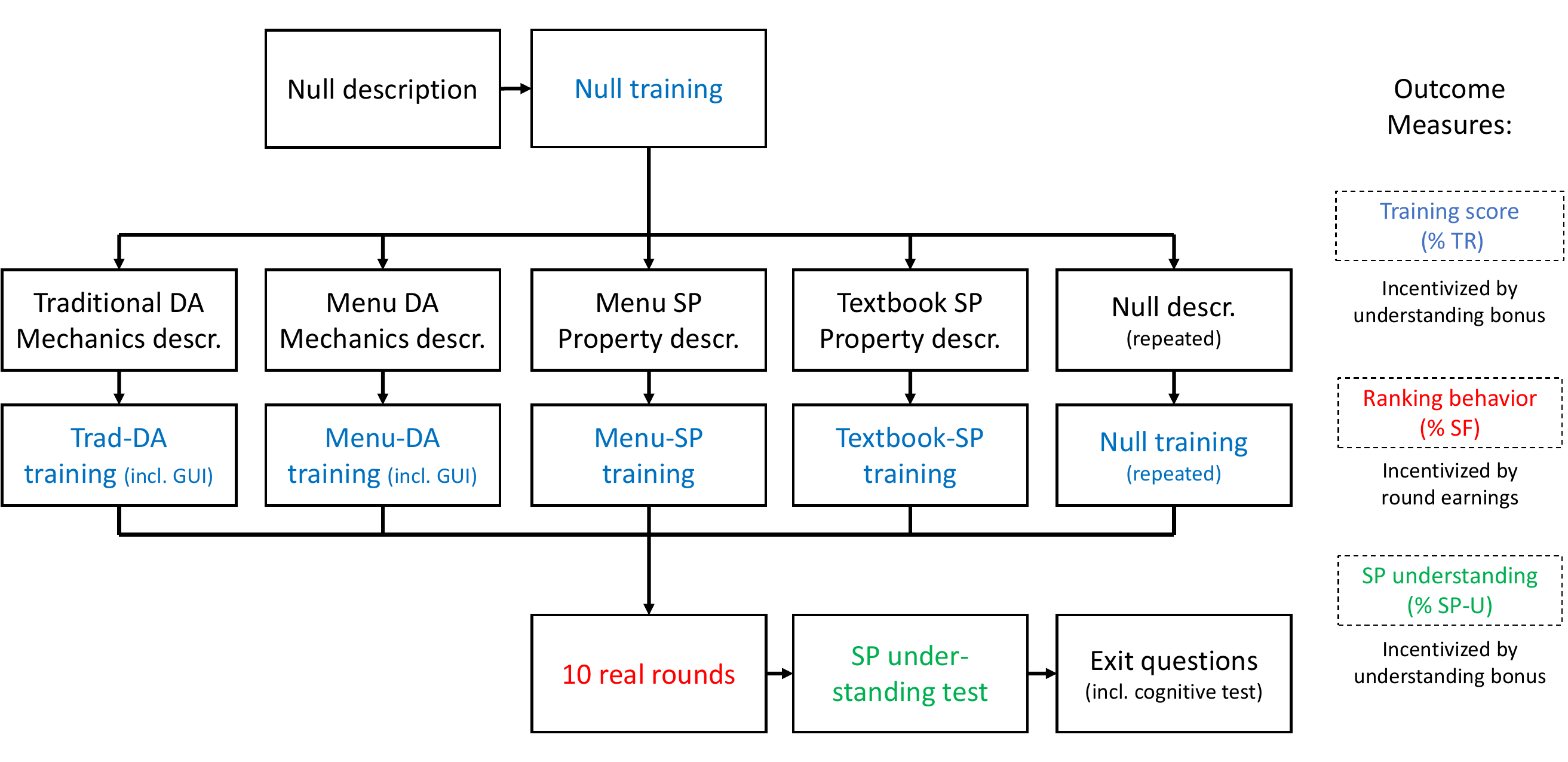}
    \caption{Overview of experimental flow}
    \label{fig:experiment-flow}
\end{figure}

\autoref{fig:experiment-flow} provides a summary of the overall experimental flow across the five treatments: two DA Mechanics treatments, two SP Property treatments, and the Null treatment.
DA Mechanics treatments describe the details of the allocation process using either a traditional algorithm (Trad-DA) or menu algorithm (Menu-DA). 
SP Property treatments convey only the fact that the mechanism is strategyproof---i.e., without specifying the details of the allocation process---using either a definition inspired by menu descriptions (Menu-SP) or a benchmark definition inspired by conventional textbook definitions of strategyproofness (Textbook-SP).
Our experiment's three main outcome variables are participants' performance on the training questions (\% TR), their performance on the strategyproofness understanding test (\% SP-U), and their fraction of straightforward ranking behavior (\% SF).

Our experiment is programmed in oTree \citep{CHEN201688}.
The rest of this section describes the experiment parts in chronological order.
For screenshots of all screens in all treatments, see %
\autoref{full-materials} in the Supplementary Materials.

\subsection{Setting (All Treatments)}
\label{sec:design-setting-and-environment}
    The experiment starts with consent and introductory screens.
    Then, participants see a description of the matching environment, i.e., the inputs and outputs of the allocation process and how the outcomes will earn them money, without any description of how the actual allocation will be determined.
    We call this the ``Null description.''

    The Null description presents the matching environment as a static setting with four participants: one human participant (labeled ``You,'' to whom we refer below simply as ``the participant'' when no confusion can arise) and three computerized participants (labeled ``Ruth,'' ``Shirley,'' and ``Theresa'').
    This setting is presented as a game in which the human and computerized participants each submit a ranking of the four prizes, and each participant wins one prize.
    The participant is informed that the prizes are worth different amounts of money to the different participants, and in each round the participant is shown how much each prize earns them.
   (See \autoref{fig:real-round-example} in \autoref{sec:design-rounds-of-play} for an example real round. The participant's prize values are presented immediately under ``Step 1: Round Information'' in the interface. The rest of the interface shown in the figure is now described.)

    The participant is told that the allocation process depends on their ranking, on the computerized participants' rankings, and on the ``prize priorities.''
    The prize priorities (which fill the role of the preferences of the prizes in the matching mechanism) are also shown to the participant before they submit their ranking.
    The computerized participants' rankings are not shown at any point.
    See \autoref{sec:design-rounds-of-play} for information on the distribution of the round parameters (the prize values for the human participant, prize priorities, and computerized participants’ rankings).
    To avoid potentially misleading language, we explain to the participant that computerized participants' rankings are ``determined beforehand''---rather than using language that refers to randomization, which may implicitly (and incorrectly) suggest uniform distributions. 
    We highlight to the participant that their ranking cannot affect the priorities or computerized participants' rankings, and provide an intuitive description of what the rankings and priorities mean.\footnote{
        These intuitive descriptions read as follows.
        For rankings: ``The allocation process attempts to give each participant a prize that they ranked higher rather than a prize that they ranked lower. However, this is not always possible, since the allocation process must take into account the rankings of all participants.''
        For priorities: ``These priorities can affect the allocation of prizes. The higher your priority is for getting some prize, the more likely you generally are to get that prize at the end of the process.''
    }

After learning about the environment (Null description), the participant plays two practice rounds (Null training) where they gain basic experience with the environment and answer a few basic training questions about the facts they have thus far learned (which count towards their understanding bonus).
These practice rounds also include an attention-check question.

\subsection{Descriptions (by Treatment)}
\label{sec:design-descriptions}

After completing the basic Null training, the participant is randomly assigned into one of five treatments.
Treatments differ in the way that they describe DA Mechanics or the SP Property; these descriptions are summarized in \autoref{tab:all-treatments-DA}.
For the full content of the main description text in all treatments, see \autoref{app:select-screenshots} in the Supplementary Materials.

\begin{table}[htb]
    \caption{Expository versions of main description texts, by treatment }
    \label{tab:all-treatments-DA}
    \centering
    \small
    \begingroup
\def\arraystretch{1}%
\newcommand{\tablewidthscale}{0.9}
\begin{tabular}{llllll}
\toprule
    \begin{minipage}{\tablewidthscale\textwidth}
    \textbf{Traditional DA Mechanics (Trad-DA):}
    You will receive a prize according to the following process:
    [The participant-proposing DA algorithm is then explained in detail.]
    \end{minipage}
    \\
    \\
    \begin{minipage}{\tablewidthscale\textwidth}
     \textbf{Menu DA Mechanics (Menu-DA): }
      A temporary allocation will be calculated using the submitted rankings of all the participants \emph{except for you}, according to the following process: [The prize-proposing DA algorithm excluding the participant is then explained in detail.] Your Obtainable Prizes include each prize which your priority of getting is higher than that of the participant paired to it in the temporary allocation. You will receive your highest-ranked Obtainable Prize.
    \end{minipage}
    \\
    \\
    \begin{minipage}{\tablewidthscale\textwidth}
    \textbf{Menu SP Property (Menu-SP): }
    Some set of Obtainable Prizes will be calculated using the submitted rankings of all the participants \emph{except for you}. You will receive your highest-ranked Obtainable Prize.
    \end{minipage}
    \\
    \\
    \begin{minipage}{\tablewidthscale\textwidth}
    \textbf{Textbook SP Property (Textbook-SP): }
      The prize you receive upon submitting a ranking $L$ is always \emph{at least as high}, according to ranking $L$, compared to the prize you would receive submitting another ranking.
    \end{minipage}
    \\
    \\
    \begin{minipage}{\tablewidthscale\textwidth}
    \textbf{Null: }
    [A repeat of the Null description initially shown to all participants in order to convey the basic environment (see \autoref{sec:design-setting-and-environment}).]
    \end{minipage}
    \\
    \bottomrule
\end{tabular}

\endgroup

\end{table}

In the Trad-DA description, the mechanism is described via the participant-proposing DA algorithm, similarly to how it is described in real-world contexts and in previous experiments (see, e.g., \citealp{MatchingExplained20, ChenS06}). 
In the Menu-DA description, the mechanism is described via the menu description of GHT (\citeyear{GonczarowskiHT23}).
This algorithic description proceeds in two steps: (1) a modified DA algorithm is run (with the proposing and receiving sides flipped, and with the human participant excluded), without using the human participant's ranking, in order to determine a menu of prizes; and (2) the human participant gets their highest-ranked prize out of the menu.

In the Menu-SP description, the participant is informed only that the mechanism is strategyproof, by explaining that \emph{some} menu of prizes will be determined without using the participant's ranking (with no details on how this is done), and the participant will get their highest-ranked prize out of the menu.
In the Textbook-SP description, participants are informed of the mechanism's strategyproofness via a description based on a standard game-theory definition. 
As in all other description treatments, we use plain English with no mathematical terms; here we explain that ranking prizes according to some ranking gets you a prize at least as high on that ranking compared to any different ranking.

\begin{figure}[htbp]
    \centering
    \caption{Charts illustrating to participants the overall allocation process}
    \label{fig:main-principle-graphics}

    \begin{minipage}[t]{0.49\textwidth}
    \centering
        \begin{minipage}{0.8\textwidth}
        \subcaption{Trad-DA and Textbook-SP}
        \end{minipage}
    \end{minipage}
    \begin{minipage}[t]{0.49\textwidth}
    \centering
        \begin{minipage}{0.8\textwidth}
        \subcaption{Menu-DA and Menu-SP}
        \end{minipage}
    \end{minipage}

    \begin{minipage}[t]{0.4\textwidth}
        \vspace{0em}
        \frame{\marginbox{0.1in}{\includegraphics[width=\textwidth]{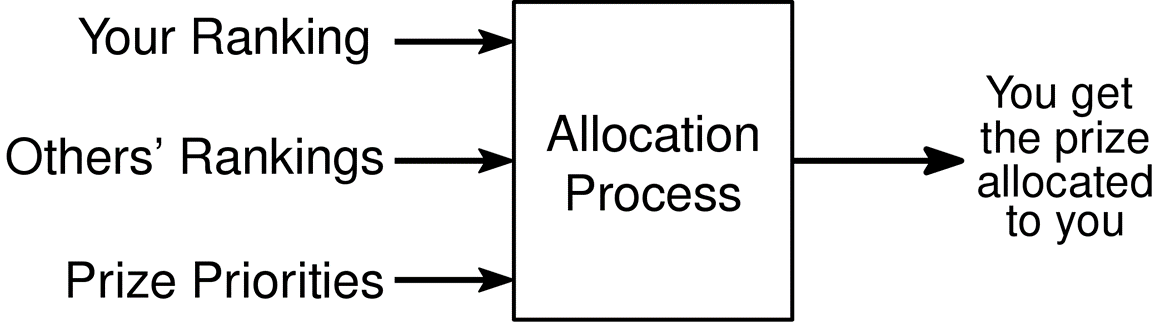}}}
    \end{minipage}
    \qquad
    \begin{minipage}[t]{0.47\textwidth}
        \vspace{0em}
        \frame{\marginbox{0.1in}{\includegraphics[width=\textwidth]{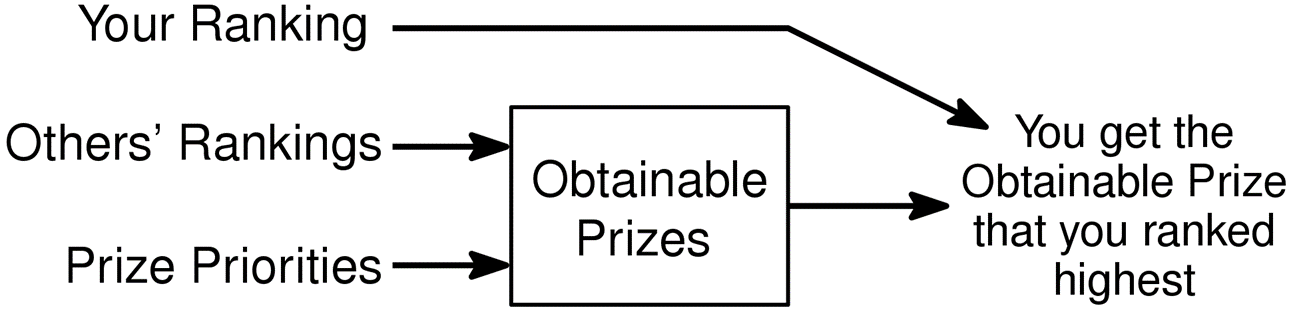}}}
    \end{minipage}
\end{figure}

Each non-Null description is also accompanied by a chart that conveys either: the fact that the participant's allocation is determined based on their rankings, the computerized participants' rankings, and the prize priorities (in Trad-DA and Textbook-SP); or the fact that the menu is determined based on the computerized participants' rankings and the prize priorities, and the human participant receives the prize they rank highest among these (in Menu-DA and Menu-SP).
See \autoref{fig:main-principle-graphics} for screenshots.

Our Null treatment gives no additional information on the allocation mechanism. Instead, it repeats the Null description discussed in \autoref{sec:design-setting-and-environment}.

\paragraph{SP Property vs.\ ranking advice.}

Our SP Property treatments investigate the participant's responses to learning (only) that the mechanism is strategyproof, but without providing any recommendation on how the participant should rank the prizes.
In other words, in both SP Property treatments, we focus on the \emph{definition} of strategyproofness, and avoid  providing participants with any \emph{advice} on how participants should rank the prizes (e.g., ``from highest to lowest value''). 
We do this for three reasons.
First, our purpose is to investigate, and try to improve, participants' \emph{understanding} (rather than, e.g., their reaction to, or trust in, advice; or their tendency to yield to authority). 
In particular, we compare a ``partial menu description,'' i.e., one where the details of how the menu is calculated are not conveyed (Menu-SP), to another SP Property treatment (Textbook-SP) that conveys identical mathematical information---that the mechanism is strategyproof.
Second, our approach relies on weaker assumptions regarding subjects' preferences; in particular, it avoids the conceptual problem of assuming that participants wish to play classically predicted, expected-utility-maximizing strategies. 
Instead, our approach relays a concrete property of the allocation rule; we argue that this approach is worth investigating both experimentally and for potential use in real-world markets as a complement to more conventional, advice-based approaches. 
Third, importantly, our approach may help mitigate an experimenter-demand effect.

\paragraph{Comparisons of treatment descriptions.}

As each of our descriptions conveys different information, it is challenging to make them look and read similarly. 
However, we make an effort to increase the similarity between the two DA Mechanics treatments and between the two SP Property treatments as much as we reasonably can.
In the DA Mechanics treatments, when describing a DA algorithm containing proposals and rejections, we maintain very similar wording in both DA Mechanics descriptions (other than interchanging the role of participants vs.\ prizes, and excluding the participant). 
Still, Menu-DA conveys elements that Trad-DA does not; namely, Menu-DA conveys in addition that the overall process has a menu structure, and also describes exactly how the menu is calculated from the reversed-DA allocation, and how the participant's prize is chosen from the menu. Therefore, Menu-DA is longer.

The two SP Property descriptions are completely different in their wording since they convey strategyproofness via different (though equivalent) definitions, but we design them to be as similar as we can in paragraph structure, sentence lengths, and overall description length.
Still, as discussed above, Menu-SP describes a concrete two-step outline, while Textbook-SP relays the classical mathematical definition of strategyproofness (using as ordinary language as we reasonably can).

\subsection{Training Rounds and Training Score (by Treatment)}
\label{sec:design-training}

After reading the description, the participant completes a set of comprehension exercises that test understanding, while also providing further opportunities to learn. The participant's total scores on these exercises are our first main outcome variable, which we term their Training Score, denoted \% TR.

Specifically, in non-Null treatments, the participant encounters three rich training rounds after the description screen. 
These training rounds differ significantly across treatments, but are designed to be as comparable as possible within the DA Mechanics or SP Property treatment groups.
Unlike in real rounds, in training rounds the prizes are not worth money, the computerized participants' rankings are fixed, and the human participant is instructed which particular ranking of the four prizes to submit.
We then ask the participant specific questions about this allocation scenario, and the participant receives points towards their understanding bonus for each question that they answer correctly on their first attempt.\footnote{
    The most complicated, multi-step questions of the DA Mechanics treatments are counted as 5 regular questions. Thus, the participant earns $5$ points (i.e., the equivalent of five correct answers) toward their understanding bonus for answering correctly on their first attempt. In those questions only, the participant also earns some bonus for answering correctly on their second attempt, of $2$ points.
}
During these screens (and the real-round screens that follow), the participant is able to open a pop-up window to remind them of the text they saw in the description screen.

\paragraph{DA Mechanics training rounds.}
    In the DA Mechanics training rounds, the round is stopped just before the allocation is determined, and the participant is asked to calculate the outcome of DA for themselves with the help of our specially developed allocation-calculation GUI.
    (In Menu-DA, they are also asked to calculate their menu.)
    The first training round includes a detailed step-by-step walk-through, which guides the participant in calculating the allocation, based on the DA Mechanics description screen they previously saw, providing detailed feedback in each step.
    The second training round offers the participant an optional video
    illustrating how to solve the training questions in that round, but the participant must implement the solution without additional hand-holding.
    The third training round asks the participant to use the GUI to calculate the final allocation, without any guidance. 
    Thus, the successive training rounds increase in difficulty and in the independence required from the participant.

\begin{figure}[htbp]
    \centering
    \caption{Main components of DA Mechanics training}
    \label{fig:mechanics-training-UI} 

    \begin{minipage}[t]{0.49\textwidth}
    \centering
        \begin{minipage}{0.8\textwidth}
        \subcaption{Trad-DA allocation-calculation GUI and allocation entry screens}
        \end{minipage}
    \end{minipage}
    \begin{minipage}[t]{0.49\textwidth}
    \centering
        \begin{minipage}{0.9\textwidth}
        \subcaption{Menu-DA allocation-calculation GUI, menu entry screen, and final allocation entry screen}
        \end{minipage}
    \end{minipage}

    \begin{minipage}[t]{0.49\textwidth}
        \vspace{0em}
        \frame{\includegraphics[width=\textwidth]{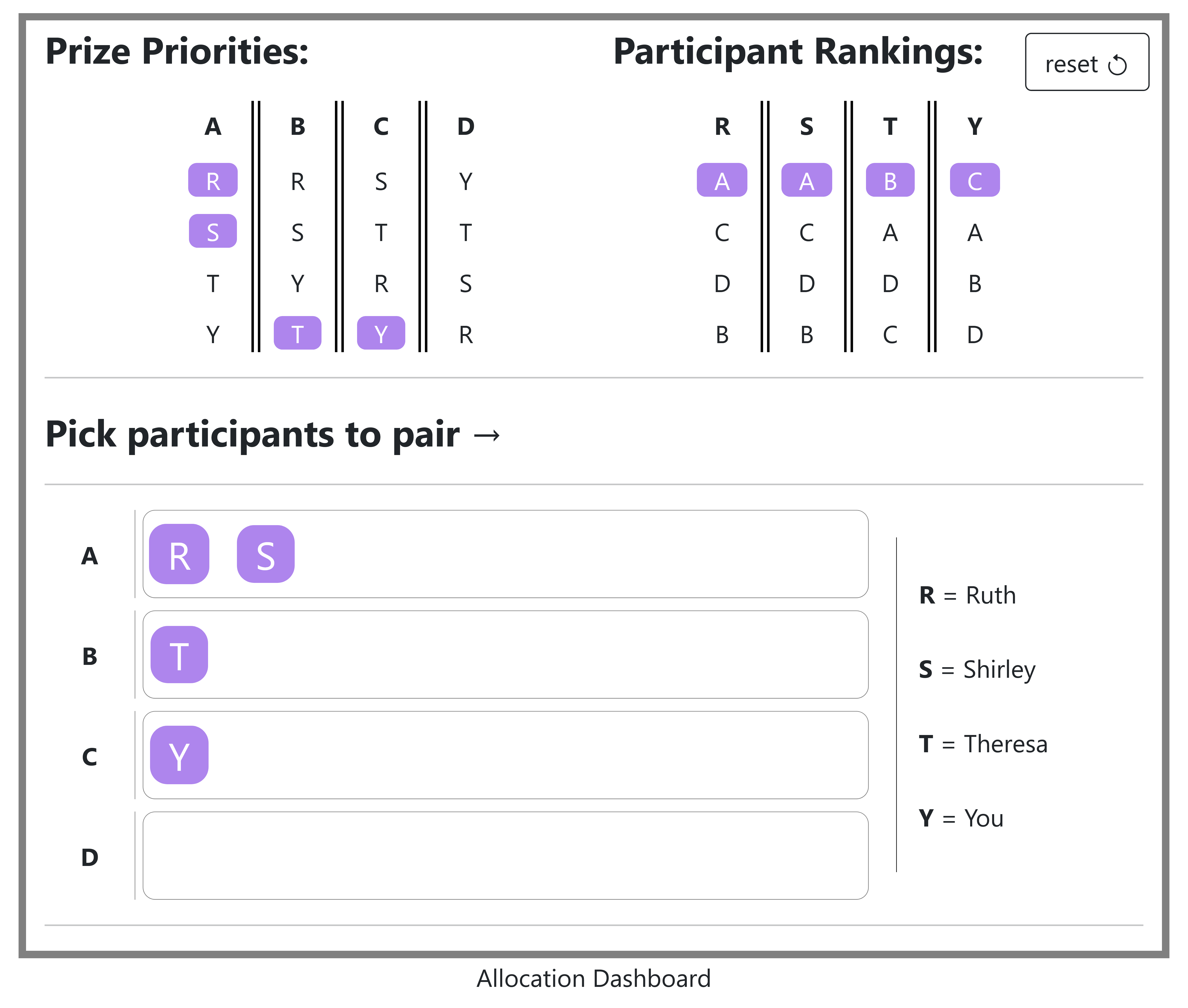}}
    \end{minipage}
    \begin{minipage}[t]{0.49\textwidth}
        \vspace{0em}
        \frame{\includegraphics[width=\textwidth]{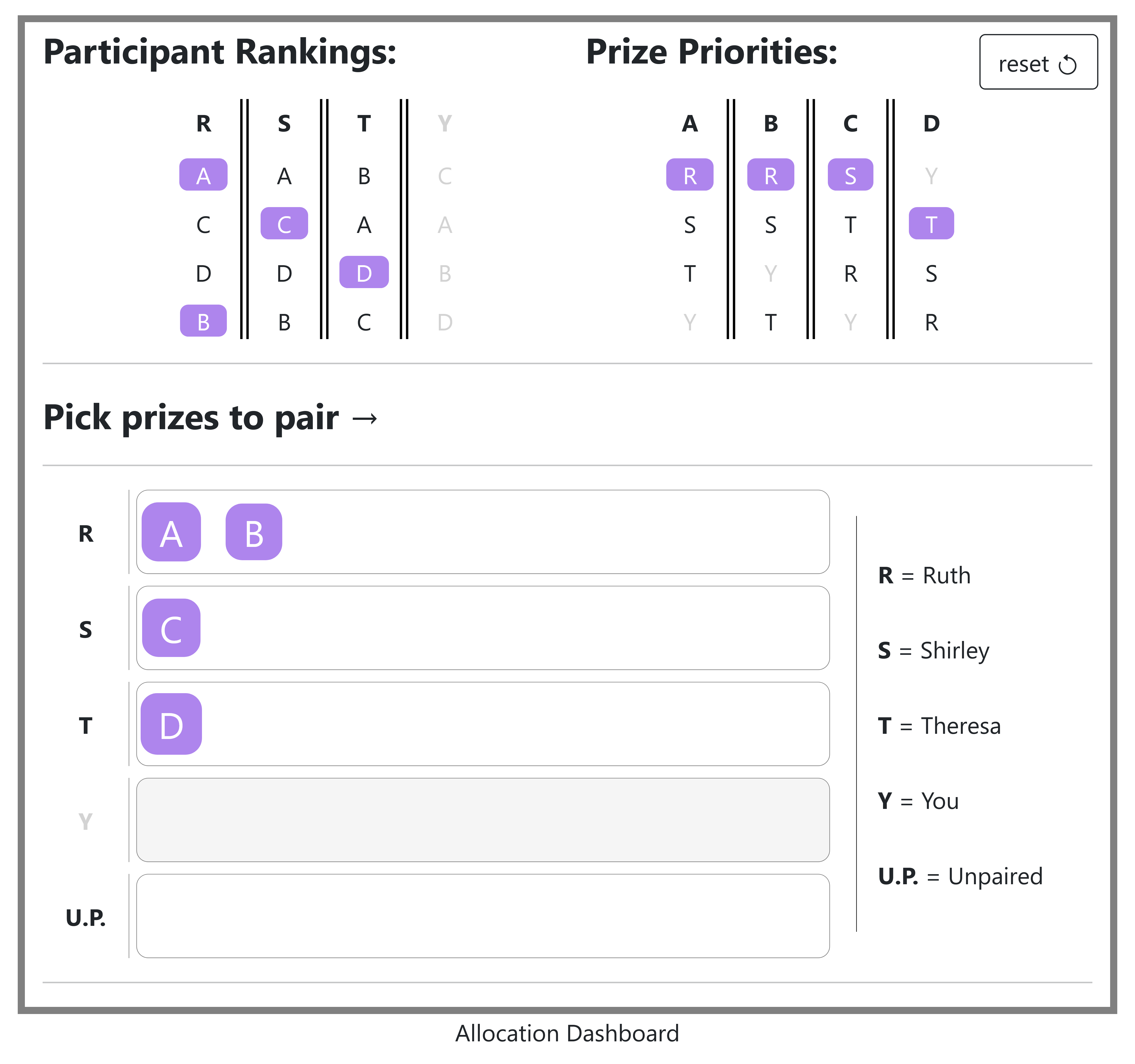}}
    \end{minipage}

    \vspace{0.5em} 
    \begin{minipage}[t]{0.49\textwidth}
        \frame{\includegraphics[width=\textwidth]{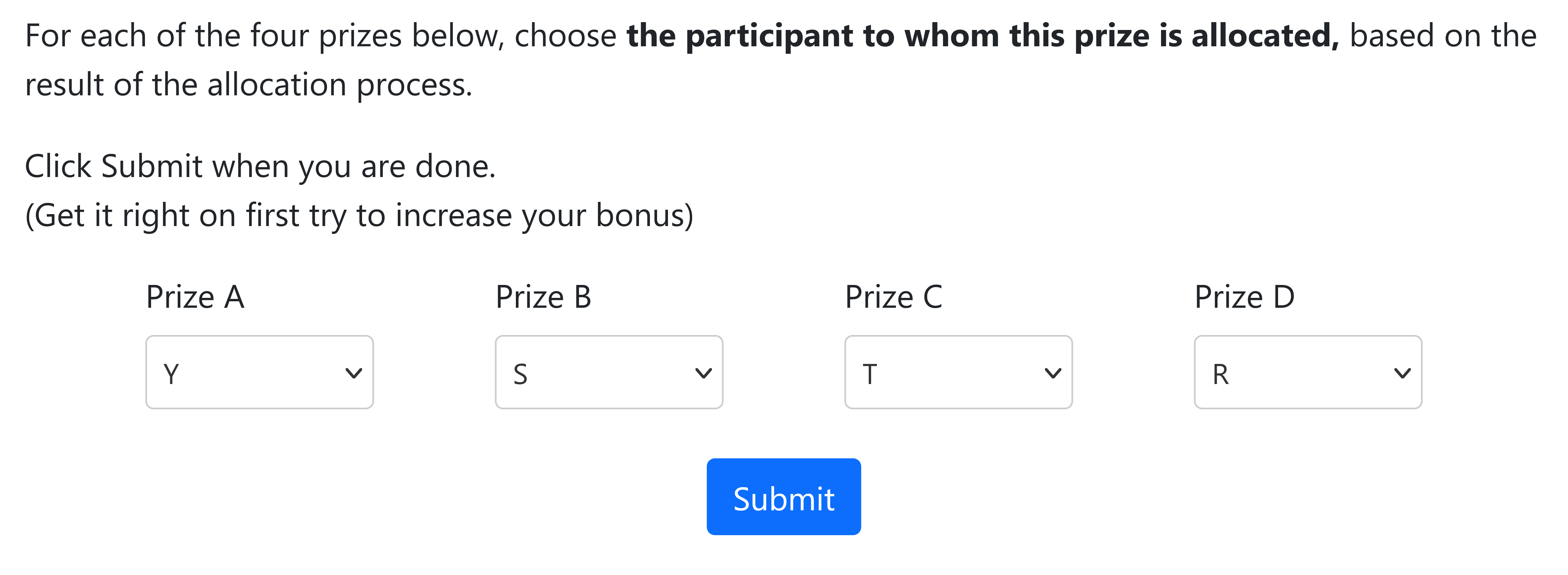}}
    \end{minipage}
    \begin{minipage}[t]{0.49\textwidth}
        \frame{\includegraphics[width=\textwidth]{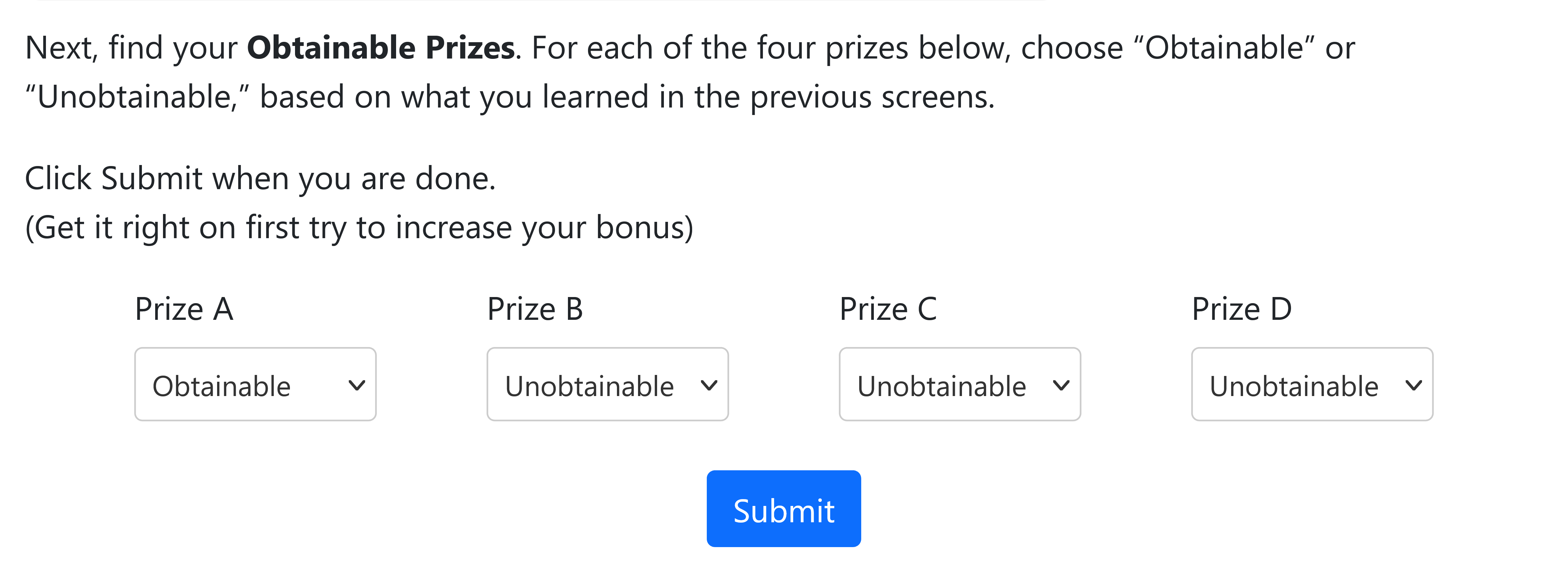}}
    \end{minipage}
    
    \begin{minipage}[t]{0.49\textwidth}
        \vspace{0em}
        \frame{\includegraphics[width=\textwidth]{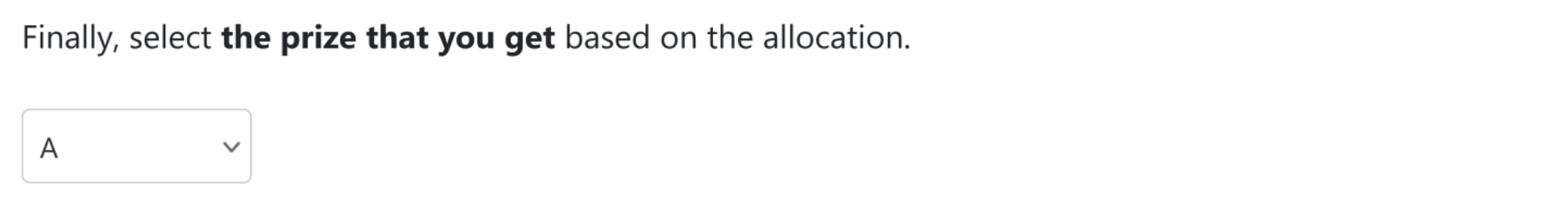}}
    \end{minipage}
    \begin{minipage}[t]{0.49\textwidth}
        \vspace{0em}
        \frame{\includegraphics[width=\textwidth]{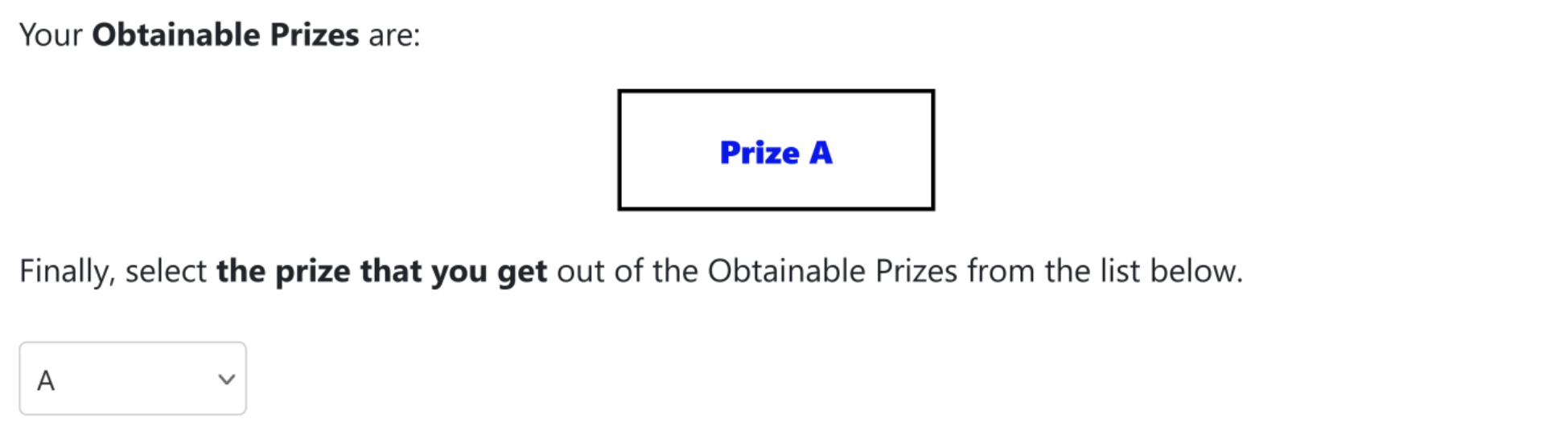}}
    \end{minipage}

    \vspace{0.5em} 
    \begin{minipage}{\textwidth} \footnotesize
        \textbf{Notes:}
        The figure shows each treatment's second training round; the allocation-calculation-GUI snapshots at the top are taken from the optional instructional videos available in this round.
        These training videos are available at
        \url{https://youtu.be/qK9JL32oxJg} (Trad-DA) and
        \url{https://youtu.be/dYePFVdqm5I} (Menu-DA). \emph{Allocation-calculation GUI}: The four prizes are indicated by A, B, C and D, the human participant is indicated by Y (``You'') and the three computerized participants are indicated by their initials R, S, and T. 
        In Menu-DA, Y is grayed out to emphasize that the prize-proposing DA is calculated while excluding the human participant.
        In Trad-DA (Menu-DA), below ``Pick participants (prizes) to pair,'' each row indicates a prize (participant) and the purple boxes in that row indicate the participants (prizes) currently paired to it. The human participant can click the purple boxes to repair prizes (participants) to different participants (prizes). In Menu-DA, the row ``U.P.'' is used to leave one prize unpaired at the end of the prize-proposing DA.
    \end{minipage}
\end{figure}

    \autoref{fig:mechanics-training-UI} shows screenshots of the main components of the DA Mechanics training rounds, including the allocation-calculation GUI.
    The figure also includes links to the optional GUI training video included in the second training round.
    In Trad-DA (respectively, Menu-DA), the participant iteratively clicks on the purple boxes signifying participants (prizes) to tentatively pair them with prizes (participants).

    As \autoref{fig:mechanics-training-UI} shows, in Trad-DA, participants use the GUI to calculate the outcome of participant-proposing DA. 
    In Menu-DA, the training rounds have three steps.
    First, participants use the allocation-calculation GUI to calculate the outcome of prize-proposing DA excluding the human participant; second, they use this to calculate the menu; third, they pick their highest-ranked prize from this menu.
    To somewhat mirror Menu-DA's second and third steps, Trad-DA also includes further questions asking participants to re-state the full allocation and the prize they receive.

    In both DA Mechanics descriptions, if the participant does not correctly answer a question involving the GUI correctly within three attempts (in the first round, a question is one step of the calculation; in the second and third rounds, it is the entire GUI calculation),
    they are given the answer (which, in the second and third rounds, they are able to reference for the remainder of the training round).
    For questions not involving the GUI, the participant must enter the correct answer, regardless of the number of attempts, in order to proceed.

    The same DA scenarios (i.e., the human and computerized participants' rankings and the prize priorities) are presented in both Trad-DA and Menu-DA; see \autoref{app:scenarios} in the Supplementary Materials for the scenarios used in the first and third training rounds (\autoref{fig:mechanics-training-UI} shows the second-round scenario).
    These scenarios are chosen so that running each of these algorithms (whether participant- or prize-proposing) would be of comparable complexity.

\begin{figure}[htbp]
\caption{A sample SP Property training question}
\label{fig:property-training-q}
    \begin{minipage}{0.7\textwidth}
        \vspace{0em}
        \frame{\includegraphics[width=\textwidth]{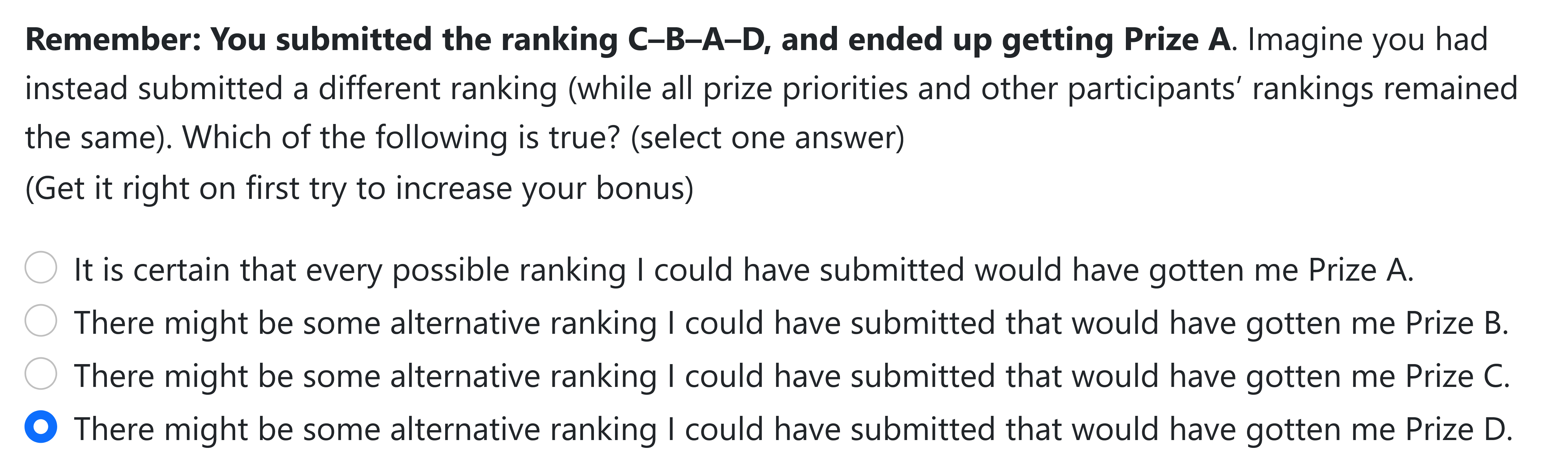}}
        \vspace{0em}
    \end{minipage}
\end{figure}

\paragraph{SP Property training rounds.}
    In contrast to the DA Mechanics treatments, in the SP Property training rounds, the participant experiences none of the above allocation-calculation GUI. Instead, after submitting the particular ranking they were instructed to submit, the allocation is calculated behind the scenes by the computer, and the participant is informed of the calculated allocation.
    Then, the participant is asked a series of multiple-choice questions requiring application of the definition of strategyproofness,
    based on the SP Property description they received. 
    They are asked a few questions aimed to reveal some common misconceptions about strategyproof mechanisms,
    and many questions about counterfactual outcomes, had they submitted other rankings in the context of the current allocation scenario.\footnote{
        The misconceptions which the first questions ask about include viewing rankings as able to ``insure'' against the worst outcome, and viewing a strategy of flipping the first and second prizes in a ranking as increasing the chances of getting the second in case the first is impossible to get.
    } 
    See \autoref{fig:property-training-q} for an example. %

    Following each correct response, the participant gets detailed feedback that explains why this answer is correct. 
    Following each incorrect response to a multiple-choice question, they are asked to try again until they answer correctly.
    With minimal exceptions, the training questions are the same in Menu-SP and Textbook-SP.\footnote{
        The exceptions are: (1) the first training question is different in Menu-SP vs.\ Textbook-SP, because it refers directly to the text of the description; and (2) one later question contains a hint that references the menu in Menu-SP, and references features of the Textbook-SP description in Textbook-SP.
    }
    However, the feedback following correct answers differs across treatments (in both wordings and illustrative examples) in order to explain the answer in terms of the specific Menu-SP vs.\ Textbook-SP framing. 
    The answer to each of the identical questions is completely determined by either SP Property description because these two descriptions are mathematically equivalent \citep{Hammond79}.

\paragraph{Null training rounds.}
    In the Null treatment, recall that the description of the mechanism is replaced with a repeat of the Null description and the two Null training rounds discussed in \autoref{sec:design-setting-and-environment}.
    The participant's training score is calculated based on their performance in this second repetition of the Null training rounds. To maintain three training rounds in all treatments, the Null treatment also includes a third training round that merely provides additional experience with the setting, and does not include training questions.

\subsection{Real Rounds and Ranking Behavior (All Treatments)}
    \label{sec:design-rounds-of-play}
\begin{figure}
        \centering
        \frame{\includegraphics[width=0.7\textwidth]{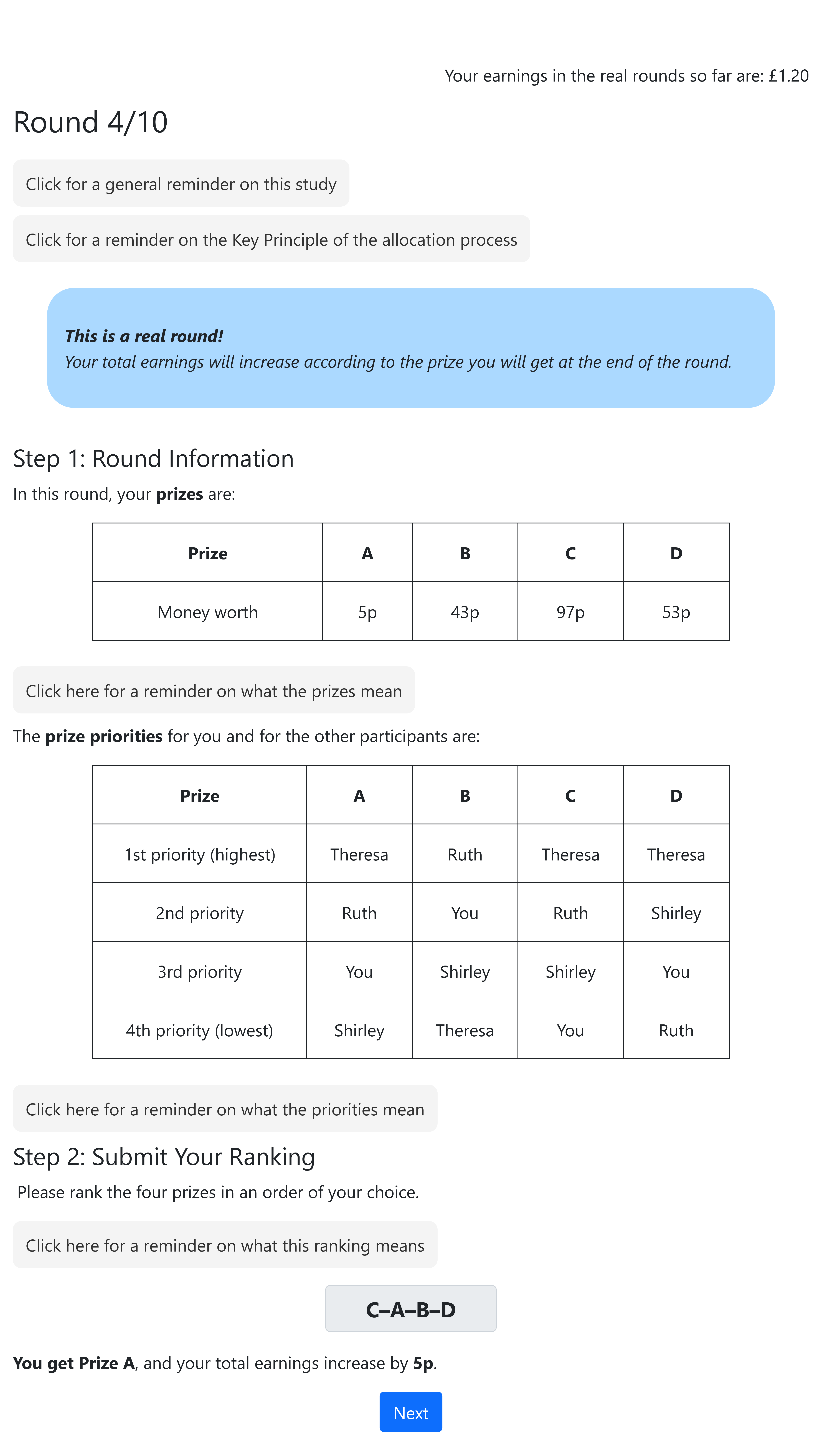}}
        \caption{A real round}
        \label{fig:real-round-example}
        \begin{minipage}{\textwidth} \footnotesize
        \vspace{0.3em}
            \textbf{Notes:}
             The participant entered the ranking C-A-B-D and received Prize A.
        \end{minipage}
    \end{figure}

    After reading the descriptions and completing all training questions, the participant plays ten rounds of the DA mechanism. \autoref{fig:real-round-example} provides a screenshot of a completed round.
    Our second main outcome measure is the fraction of rounds in which  the participant ranks straightforwardly, i.e., in highest-earning to lowest-earning order, denoted \% SF.

    Prize values for the human participant, prize priorities, and computerized participants' rankings are drawn from a specially-tailored joint  distribution (for full details, see \autoref{app:randomization-setting} in the Supplementary Materials). 
    This distribution is designed to encourage an ample fraction of participants---particularly those who fail to understand that the mechanism is strategyproof, or who have non-classical preferences---to carefully consider their ranking and thus play NSF.
    For example, the distribution is meant to make the setting feel competitive by giving the human participant a lower-than-uniform probability of having a high priority for getting their highest-earning prize, and by making it more likely that the computerized participants rank the human participant's highest-earning prize first.
    Additionally, the distribution we use often induces large differences between prize values, so that allocation outcomes meaningfully affect a participant's earnings.

    \paragraph{Information and feedback.}
    As discussed in \autoref{sec:design-setting-and-environment}, at the beginning of each round, the participant observes each prize's value for them and all prize priorities.
    We present the prize priorities both to imitate a real-world context in which the participant has an idea about their likelihood to get different outcomes (e.g., in a school-choice context, how well-positioned they are relative to other students), and to leave the participant with the possibility to strategize based on this information if they so choose.
    At the end of each round, the participant receives limited feedback: they only learn their own outcome (i.e., their assigned prize), with no information on the computerized participants' rankings or outcomes. This is done to minimize learning from experience (through feedback), and hence maximize the effect our different descriptions may have on behavior.

\subsection{Strategyproofness-Understanding Test (All Treatments)}
\begin{figure}
        \centering
        
        \caption{Samples of the strategyproofness understanding test} 
        \label{fig:sp-u-screenshots}

        \subcaption{Abstract}
        \label{fig:sp-u-screenshots-abstract}
        \includegraphics[width=0.9\textwidth]{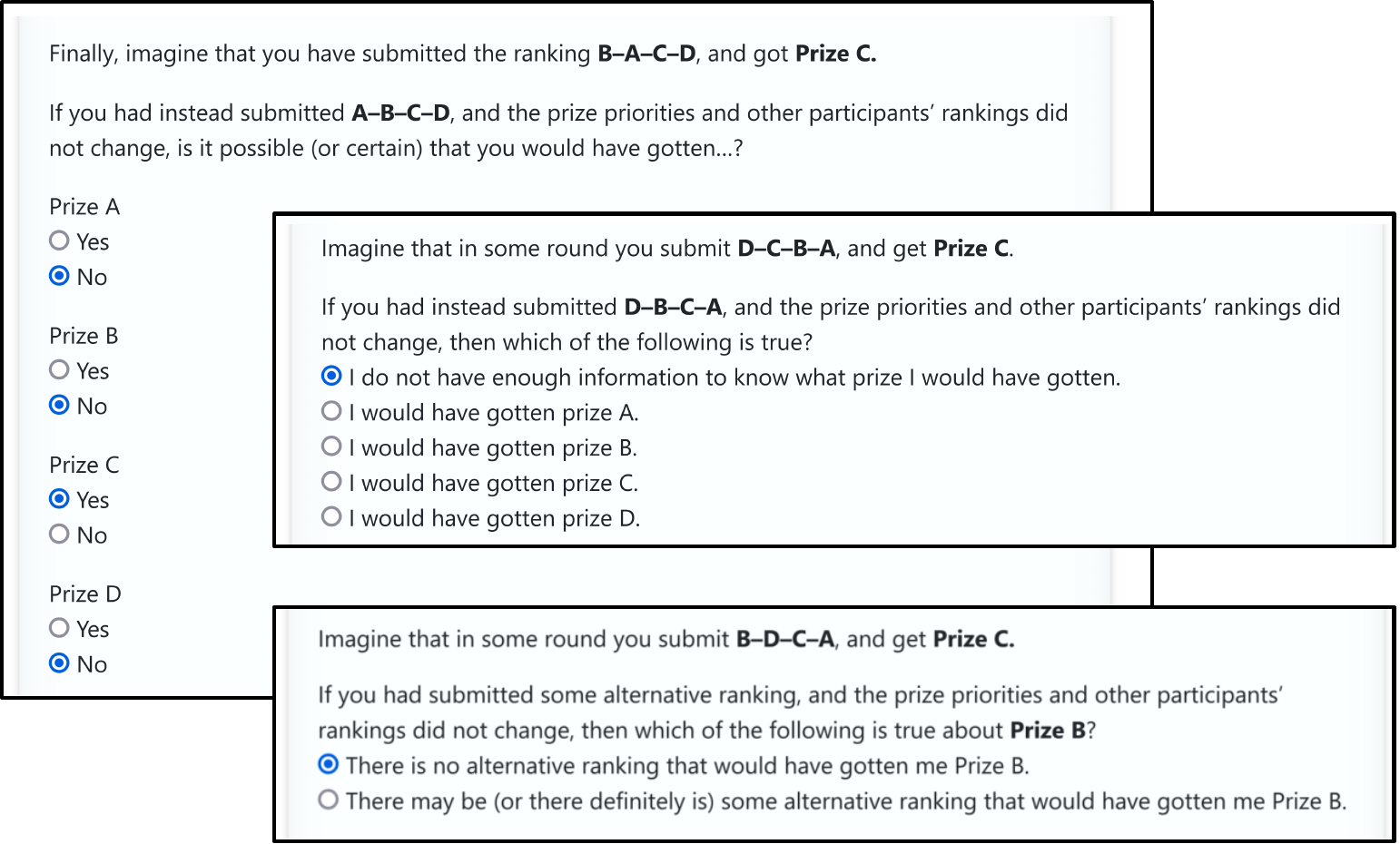}
        
        \vspace{0.4em}
        
        \subcaption{Practical}
        \label{fig:sp-u-screenshots-practical}
        \includegraphics[width=0.9\textwidth]{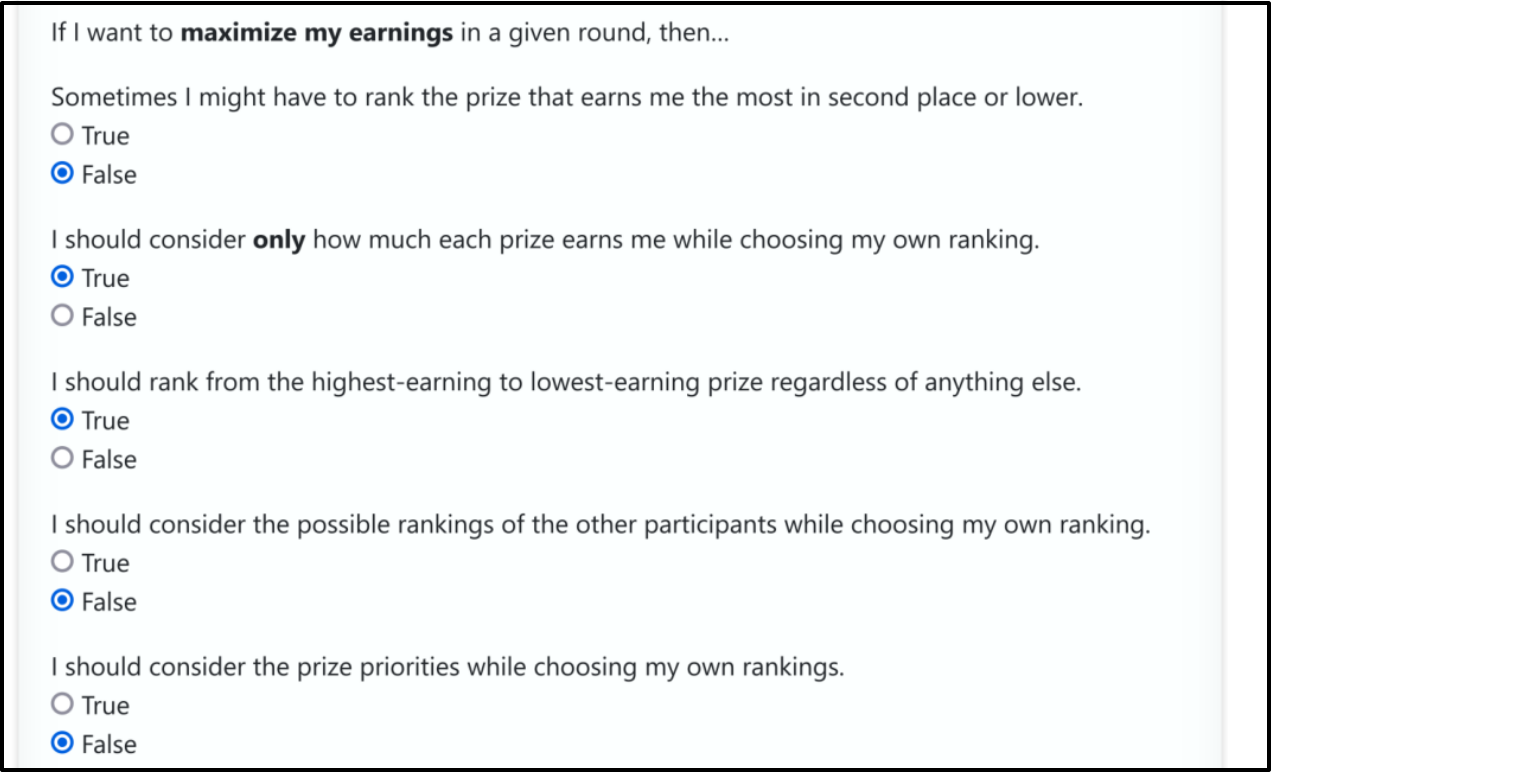}
        
        \begin{minipage}{\textwidth} \footnotesize
        \vspace{1.1em}
            \textbf{Notes:}
            The content of each bordered box appears on a separate screen. 
            \emph{Panel (a)}: questions directly concerning the abstract logical properties of strategyproofness (``Abstract'').
            The first screen contains another similar question, and  an attention-check question.
            The second and third screens also contain, respectively, one and two other similar questions.
            \emph{Panel (b)}: questions on practical implications, i.e., on how participants can maximize earnings (``Practical'').
            This panel shows the entire set of questions of that screen.
        \end{minipage}
    \end{figure}

    After the ten real DA rounds, the participant completes a newly designed strategyproofness-understanding test; see \autoref{fig:sp-u-screenshots} for samples of questions.
    Their test score is our final main outcome variable, denoted \% SP-U. 
    The test consists of four screens, each with its own set of questions.
    For each question the participant answers correctly on the first attempt, they receive $2$ points (i.e., the equivalent of two correct answers) towards their understanding bonus.

    The first three screens of the strategyproofness understanding test are on abstract logical properties of strategyproofness, and we refer to their thirteen questions jointly as Abstract; see \autoref{fig:sp-u-screenshots-abstract} for examples. 
    These questions are similar in nature to the comprehension questions asked during the SP Property training rounds,
    but broader in scope, and common across all five treatments.
    For SP Property treatments, these can be thought of partially as further measures of participants' understanding of the description (similar to those asked during training rounds), and partially as tests of whether the participant can apply their knowledge in different, novel scenarios.
    For DA Mechanics treatments, these can be thought of as measures of whether the participant correctly infers (or guesses) the features relevant to the strategyproofness properties from their mechanical description.
    The first screen also contains an attention-check question.

    The final screen of the strategyproofness understanding test contains questions on practical implications of strategyproofness, and specifically, how one could maximize their earnings. We refer to these five questions jointly as Practical; see \autoref{fig:sp-u-screenshots-practical} for a complete screenshot of this screen.

\subsection{Exit Questions {and Cognitive Score (All Treatments)}}
\label{sec:design-exit}
\label{sec:design-cognitive}
        Finally, additional questions elicit reflections on, and perceptions of, the mechanism; measure cognitive abilities and numeracy; and collect demographic and other data.

        The reflection questionnaire asks the participant several questions about the strategies they played and about how they perceived the mechanism.
        Some of these question are adapted from from previous studies \citep[e.g.,][]{ReesJonesS18} and some are new to this study. 
        The cognitive-abilities questionnaire uses the three questions of the Cognitive Reflection Task \citep[CRT;][]{Frederick05}, and measures numeracy using one question from the Berlin Numeracy Test \citep[][]{Cokely_Galesic_Schulz_Ghazal_Garcia-Retamero_2012}. %
        We use this questionnaire to give participants a cognitive score between $0$ and $4$. 
        We also similarly define an attention score outcome based on two attention-check questions planted in the Null training and SP-understanding test, discussed above.
        For a full list of the demographics and social, economic, and political leanings that we collect, see \autoref{app:demographics} in the Supplementary Materials. 
        The final questionnaire elicits general feedback on the experiment.

\section{Results}\label{sec:results}

\subsection{Sample}
\label{sec:sample}

Our data includes a total of $542$ participants, collected from two sources. $255$ participants were recruited on Prolific---a crowd-sourcing platform designed for scientific use---from August 3 to August 8, 2023. $287$ participants were recruited at the Cornell Johnson Business Simulation Lab (BSL) from September 26 to November 17, 2023.\footnote{
    According to the pre-registration, we stopped collection at each platform after obtaining at least $50$ responses per treatment, resulting in at least $100$ responses per treatment overall. 
    See \url{https://aspredicted.org/7eq7e.pdf} for our pre-registration, and see \autoref{app:prereg} in the Supplementary Materials for a summary and comparison with our analysis and findings.
} 
We use these two different sources to increase the variation of factors such as engagement, education level and cognitive skill within our sample, as they may all play a key role in how our descriptions affect understanding and behavior.
See \autoref{app:sample-basics} for information on recruitment (which was as similar as possible across samples), earnings (which are on average \pounds15.0 and \$25.5 on Prolific and at Cornell, respectively), attrition levels (which do not vary by treatment), and duration.\footnote{
  The overall median completion time is $4$ minutes higher on Prolific than at Cornell.
  In addition, median completion times differ across treatments:
  Trad-DA and Menu-DA take $48$ minutes and $53$ minutes, respectively, Menu-SP and Textbook-SP take $40$ minutes and $39$ minutes, respectively, and Null takes $35$ minutes.
}
See \autoref{app:demographics} in the Supplementary Materials for demographic characteristics of our participants.

Our main results are qualitatively similar across Prolific and Cornell; hence, we pool the two samples for our main analysis (Sections~\ref{sec:results-training-score} through~\ref{sec:results-ranking}).
We report the cross-sample differences 
in \autoref{sec:results-samples-and-controls}.
The two main differences are in baseline levels of all our main outcome variables, and in the magnitude of treatment effects; both are lower on Prolific.

\subsection{Main Result \#1: We Can Teach DA}
\label{sec:results-training-score}

We begin by studying participants' training score (\% TR), elicited through our incentivized training rounds (which, in DA Mechanics treatments, use our training GUI).
Our first main result, and our main result regarding \%~TR, is that the Traditional DA Mechanics description is effective at teaching participants the mechanics of DA.
We begin by showing this result, and then investigate participants' training performance in other treatments.

\paragraph{Training score in Trad-DA.}

\begin{figure}[htb]
    \centering
    \caption{Training score (\% TR) by treatment}
    \label{fig:mean-tr-by-treatment}
    
    \begin{minipage}[t]{0.49\textwidth}
    \centering
        \begin{minipage}{0.6\textwidth}
        \subcaption{Mean overall score}
        \label{fig:mean-tr-by-treatment-actually-the-mean}
        \end{minipage}
    \end{minipage}
    \begin{minipage}[t]{0.49\textwidth}
    \centering
        \begin{minipage}{0.8\textwidth}
        \subcaption{Perfect-solution rates in calculating DA outcome independently}
        \label{fig:sub-measure-tr-by-treatment}
        \end{minipage}
    \end{minipage}

        \includegraphics[width=0.49\figscale\textwidth{}]{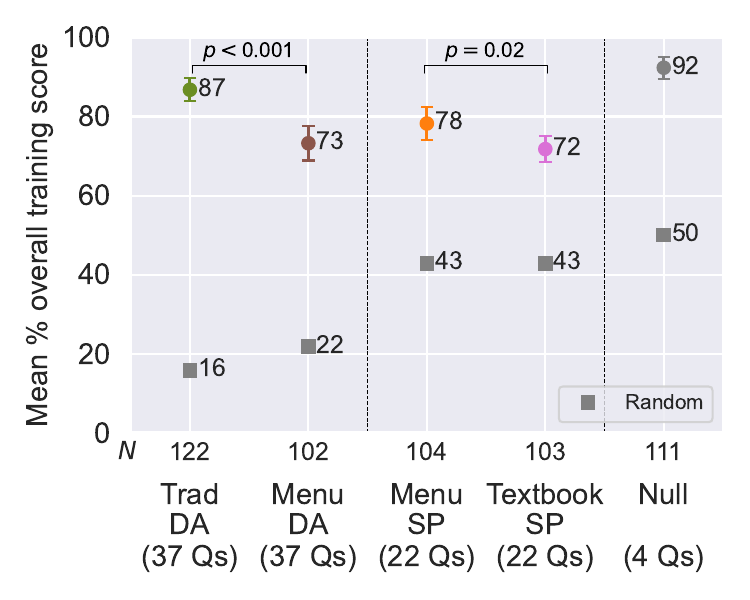}
        \includegraphics[width=0.49\figscale\textwidth{}]{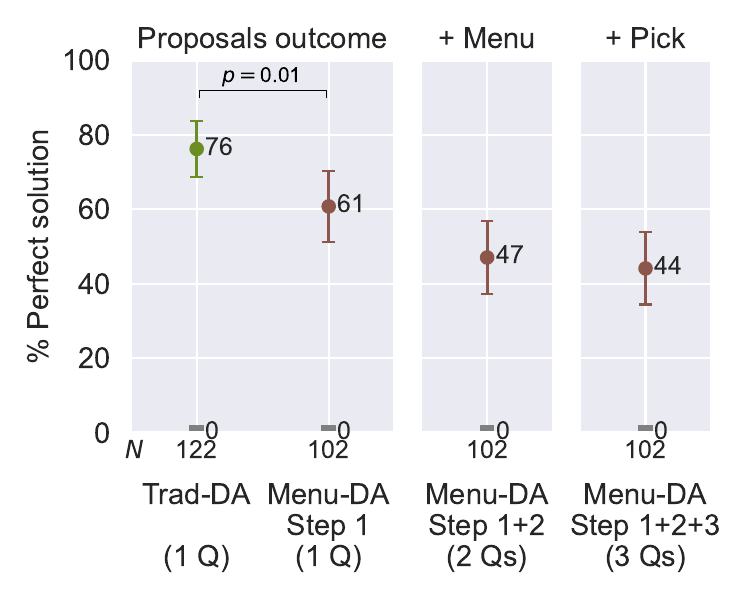}

        \begin{minipage}{\textwidth}  \footnotesize
        \textbf{Notes:}
        \emph{Panel (a)}:
        Mean overall \% TR by treatment.
        \emph{Panel (b)}  (relevant to DA Mechanics treatments only): Rates of perfectly  calculating (parts of) a DA outcome using our training GUI in the final training round, on the first attempt. 
        ``Proposals outcome'': the rate of participants perfectly indicating the DA algorithm proposal-sequence outcome in our GUI. 
        In Menu-DA, this is only the first step;
        ``+ Menu'': the rate of participants who also perfectly calculate their menu based on it;
        ``+ Pick'': the rate of participants who also perfectly indicate that their highest ranked prize is the one picked from the menu.
        \emph{Both Panels:}
        Error bars: 95\% confidence intervals. 
        $p$-values: two-sample $t$-tests.
        ``$N$'': number of observations per treatment.
        ``(\# Qs)'': the number of questions that the relevant score is based on.
        ``Random'': the expected score of answering every question uniformly at random.
        Averages are not comparable across vertical dotted lines.
        \end{minipage}
    \end{figure}

    \autoref{fig:mean-tr-by-treatment} shows participants' average training-score performance.
    \autoref{fig:mean-tr-by-treatment-actually-the-mean} shows average overall \% TR. 
    In Trad-DA, mean \% TR is 87\% (SE $=$ 1\%), i.e., overall very high performance in the 37 questions and problems that participants face. %

    The training modules are specific to each treatment, and in particular differ dramatically between DA Mechanics treatments on the one hand and SP Property treatments on the other hand.
    Hence, in \autoref{fig:mean-tr-by-treatment}, \% TR is only meaningfully comparable within treatment pairs (separated by the dotted lines), but it differs somewhat even within pairs.\footnote{
        As discussed in \autoref{sec:design-training}, while the DA Mechanics training modules in Trad-DA and Menu-DA are designed to be as similar as possible, they still must differ significantly due to the different algorithms involved, with the Menu-DA description being longer and having more steps.
        The SP Property training is much closer across the Menu-SP and Textbook-SP treatments.
    } 
    The panel, along with others like it throughout this section, includes a ``Random'' benchmark that indicates what results would have looked like had participants answered questions uniformly at random. 
    Participants' scores vary somewhat, but they are much higher than the Random benchmark in all treatments. For the full training-score distributions by treatment, %
    see \autoref{app:tr-distributions}, and for more details on question-level performance, see \autoref{app:tr-more-details} in the Supplementary Materials.

   \autoref{fig:sub-measure-tr-by-treatment} shows DA Mechanics participants' performance when focusing only on the final, most-challenging, and most externally valid training problem, 
    which asks participants to correctly calculate a DA allocation outcome completely by themselves.
    Participants use our interactive GUI to perform the entire proposal sequence, which consists of 10 total proposals when executed correctly.
    In Trad-DA, 76\% (SE $=$ 4\%) of participants succeed in solving this problem perfectly and on their first attempt
    (shown in the mini-figure labeled ``Proposals outcome''). 
    This fraction increases to 84\% (SE $=$ 3\%) when including second-attempt successes.

    These findings suggest that with sufficient care and coaching, many participants can be taught (the mechanical step-by-step method of calculating) DA using the traditional description of DA.
    At the same time, given that a quarter of our participants still do not accurately calculate DA on first attempt, our findings cast doubt on the assumption---often implicit in existing work---that most participants perfectly understand DA.

\paragraph{Training score in Menu-DA.}    

    \autoref{fig:mean-tr-by-treatment} also shows that training scores are lower in Menu-DA compared to Trad-DA, suggesting that of the two, Trad-DA is the simpler description.
    \autoref{fig:mean-tr-by-treatment-actually-the-mean} shows that \% TR in Menu-DA is 73\% (SE $=$ 2\%), compared to 87\% in Trad-DA.
    \autoref{fig:sub-measure-tr-by-treatment} then focuses on the hardest training problem, which has three steps in Menu-DA.
    The outcome variables indicating whether participants perfectly solve these steps together on the first attempt are labeled ``Proposals outcome'' (Step 1), ``+ Menu'' (both Steps 1 and 2), and ``+ Pick'' (Steps 1, 2, and 3).
    Participants in Menu-DA find (all steps of) the training more challenging than in Trad-DA. 
    44\% (SE $=$ 5\%) of participants are able to perfectly complete the entire three-step process on their first attempt (which increases to 56\% when including participants who succeed on the first or second attempt in all steps of the process; SE $=$ 5\%).

\paragraph{Training score in SP Property treatments and Null.} 

    Finally, \autoref{fig:mean-tr-by-treatment-actually-the-mean} also shows that training scores in the two SP Property treatments are fairly high, and are more comparable in mean levels compared to the wider gap between Trad-DA and Menu-DA.
    Null training scores are near perfect, suggesting that repeating the understanding questions on the basic components of our setting is indeed simple and straightforward for most participants.

\subsection{Main Results \#2 and \#3: Understanding DA \texorpdfstring{$\neq$}{Is Not} Understanding SP, But We Can Teach SP}
\label{sec:results-strategyproofness-understanding}

Perhaps our most important and novel outcome variable is \% SP-U---participants' score on the strategyproofness-understanding test. This incentivized eighteen-question test, identical across treatments, is conducted after training and playing ten real rounds of DA, and underlies our second and third main results.

Our second main result is that understanding the mechanics of DA does not imply understanding of its SP property, as indicated by low \% SP-U levels in Trad-DA (comparable to those in the Null treatment). 
Our third main result is that SP Property descriptions, %
and particularly Menu-SP, are more successful in teaching SP to participants. This subsection presents these results, as well as a more detailed investigation that separates the full \% SP-U measure into its two main components: Abstract and Practical.

\subsubsection{Overall Measure of Strategyproofness Understanding}

\begin{figure}[htbp]
    \centering
    \caption{Strategyproofness understanding test score (\% SP-U) by treatment}
    \label{fig:mean-sp-u-by-treatment}
    
    \begin{minipage}{\textwidth}
    \subcaption{Mean overall score}
    \label{fig:sp-u-means}
    \end{minipage}
    \includegraphics[width=0.5\figscale\textwidth{}]{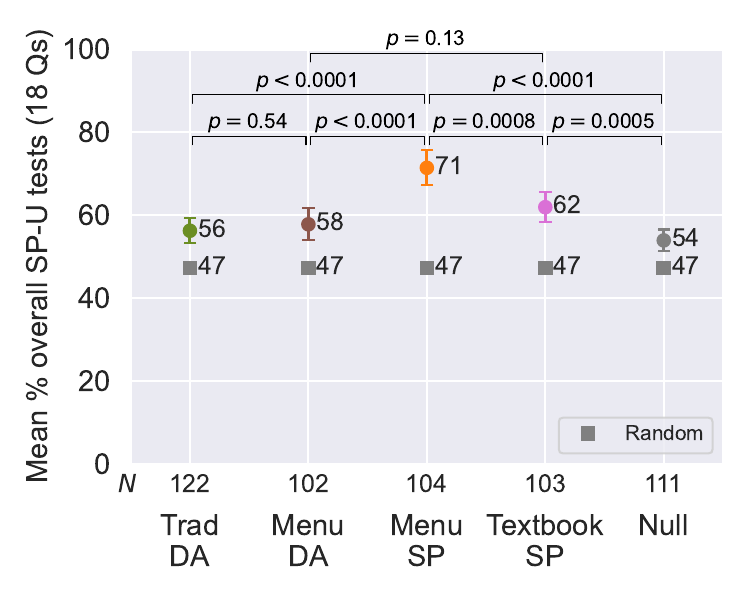}

    \vspace{0.1in}
    \begin{minipage}{\textwidth}
    \subcaption{Distribution of overall score}
    \label{fig:sp-u-hist}
    \end{minipage}

    \includegraphics[width=\figscale\textwidth{}]{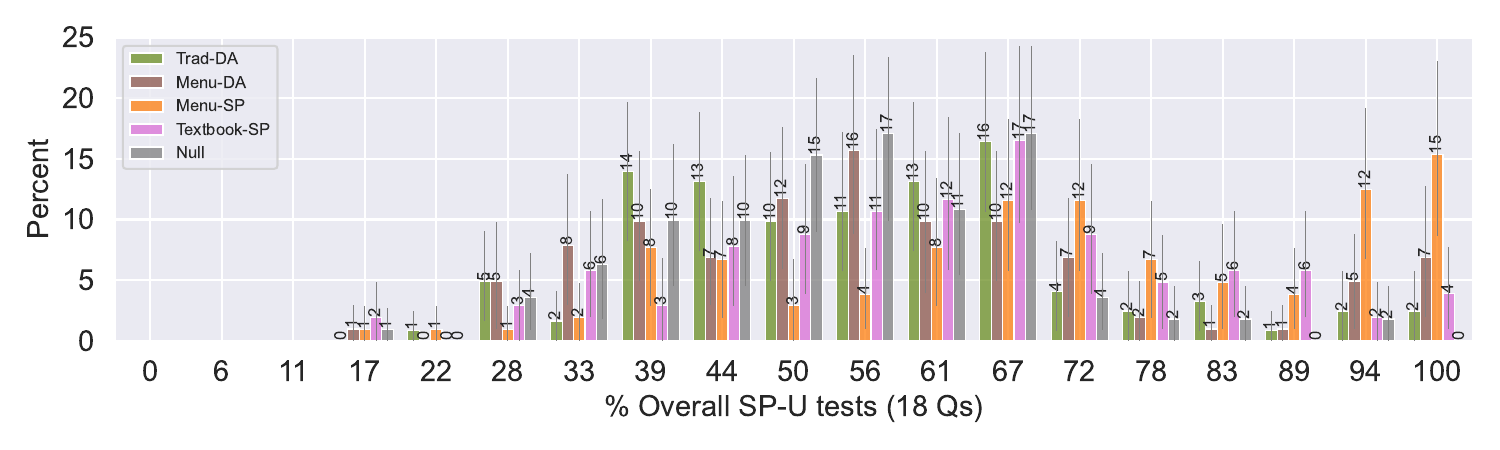}

    \begin{minipage}{\textwidth}  \footnotesize
        \textbf{Notes:}
        \emph{Panel (a):} Mean of \% SP-U by treatment.
        $p$-values: two-sample $t$-tests.
        ``$N$'': number of observations per treatment.
        ``Random'': the expected score of answering every question uniformly at random.
        \emph{Panel (b):} Distribution of \% SP-U by treatment.
        \emph{Both Panels:} Error bars: 95\% confidence intervals. 
        ``(\# Qs)'':  the number of questions that the test consists of.
    \end{minipage}
    \end{figure}

    \autoref{fig:sp-u-means} shows average \% SP-U---the percent of the eighteen questions that a participant answered correctly---across treatments.
    The figure first shows that both DA Mechanics treatments are indistinguishable from each other in \% SP-U, and are essentially indistinguishable from the baseline of 54\% ($\text{SE}=1\%$) Null treatment. 
    Trad-DA's high training scores but low strategyproofness-understanding scores underlies our second main result.
    In particular, in spite of three quarters of Trad-DA participants learning to fully calculate the final DA allocation on their own with no mistakes 
    (see \autoref{sec:results-training-score}), they do not understand that the mechanism is strategyproof more than Null participants, who know essentially nothing about the mechanism!

    In contrast, Menu-SP, and to a much lesser extent Textbook-SP, lead to a noticeable increase in \% SP-U, with mean levels  71\% and 62\% (SE $=$ 2\% in both treatments), respectively---underlying our third main result.%
    \footnote{
        There are ten possible comparisons of \% SP-U across pairs of treatments. Applying an arguably over-conservative Bonferroni correction to address multiple comparisons would multiply all $p$-values in \autoref{fig:sp-u-means} by 10, maintaining the difference between Menu-SP and Textbook-SP with $p<0.01$, and the difference between Menu-SP and any other treatment with $p<0.001$.
    }
   We speculate that the difference in \% SP-U between Menu-DA and Menu-SP---where the former is merely a more detailed version of the latter---suggests that the additional complex details in Menu-DA make the important features of menu descriptions insufficiently salient.

    \autoref{fig:sp-u-hist} presents these results in more detail, reporting the distribution of \% SP-U across treatments. A noticeably larger fraction of participants get perfect or near-perfect SP-U scores in the Menu-SP treatment, with much smaller differences across other treatments (including Null). For example, 28\% (SE $=$ 4\%) of Menu-SP participants get at least seventeen of the eighteen questions correct. This figure drops to 12\% (SE $=$ 3\%) in Menu-DA, 6\% (SE $=$ 2\%) in Textbook-SP, 5\% (SE $=$ 2\%) in Trad-DA, and 2\% (SE $=$ 1\%) in Null.

    \subsubsection{Sub-Measures of Strategyproofness Understanding}
    \label{sec:sp-u-sub-measures}

\begin{figure}[htbp]
    \centering
    \caption{Abstract and Practical sub-measures of strategyproofness understanding test score (\% SP-U) by treatment}
    \label{fig:mean-sp-u-abstract-practical-by-treatment}

    \vspace{0.1in}
    \begin{minipage}[t]{0.49\textwidth}
        \centering
        \begin{minipage}{0.6\textwidth}
        \subcaption{Mean Abstract sub-measure score}
        \label{fig:sp-u-abstract-means}
        \end{minipage}
    \end{minipage}
    \begin{minipage}[t]{0.49\textwidth}
        \centering
        \begin{minipage}{0.6\textwidth}
        \subcaption{Mean Practical sub-measure score}
        \label{fig:sp-u-practical-means}
        \end{minipage}
    \end{minipage}
    \includegraphics[width=0.49\figscale\textwidth{}]{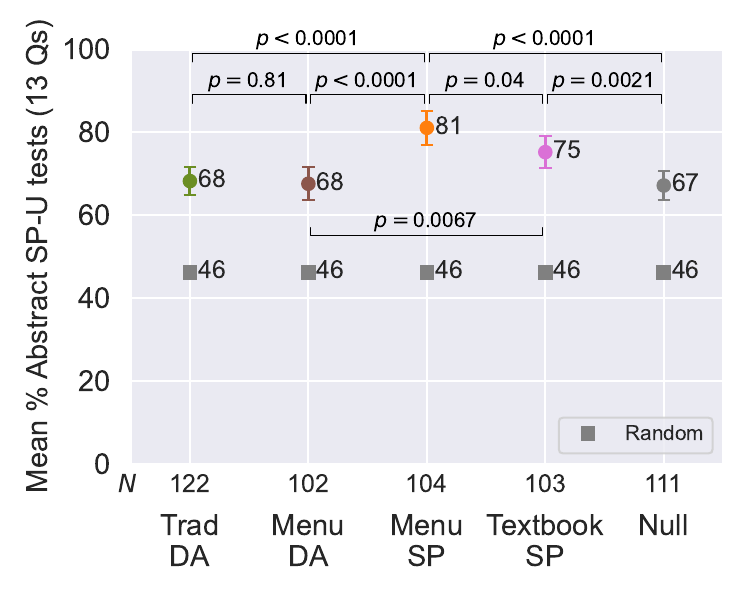}
    \includegraphics[width=0.49\figscale\textwidth{}]{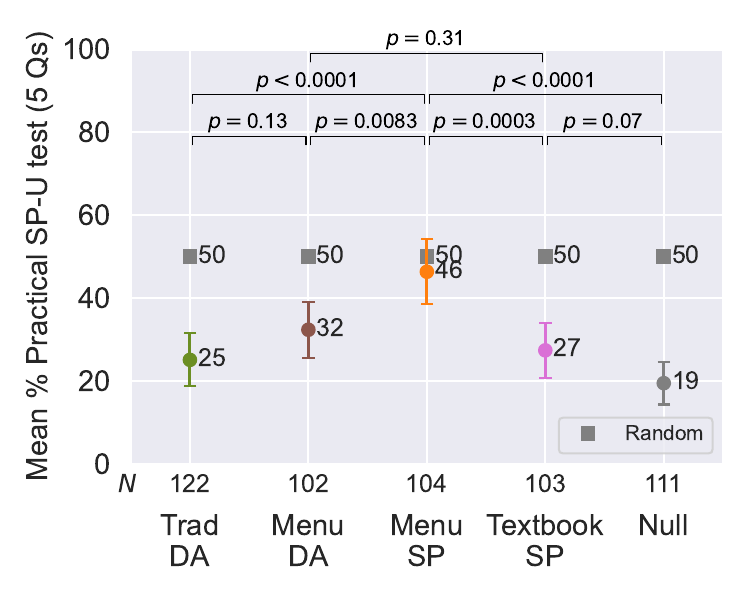}

    \vspace{0.1in}
    \begin{minipage}[t]{0.66\textwidth}
        \centering
        \begin{minipage}{0.46\textwidth}
        \subcaption{Distribution of Abstract sub-measure score}
        \label{fig:sp-u-abstract-hist}
        \end{minipage}
    \end{minipage}
    \begin{minipage}[t]{0.33\textwidth}
        \centering
        \begin{minipage}{0.9\textwidth}
        \subcaption{Distribution of Practical sub-measure score}
        \label{fig:sp-u-practical-hist}
        \end{minipage}
    \end{minipage}
    \includegraphics[width=\figscale\textwidth{}]{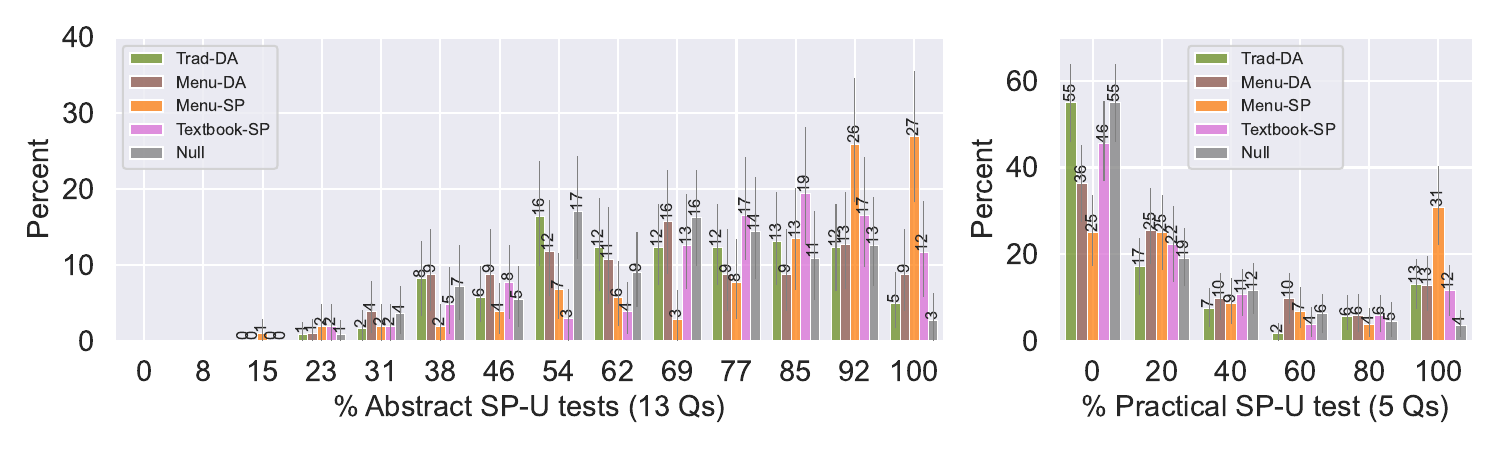}

    \begin{minipage}{\textwidth}  \footnotesize
        \textbf{Notes:}
        \emph{Panel (a):} Mean of Abstract sub-measure of \% SP-U, i.e.,  the first three screens of the test (13 questions), which ask about  abstract logical properties of strategyproofness.
        \emph{Panel (b):} Mean of Practical sub-measure of \% SP-U, i.e.,  the last screen of the test (5 questions), which ask about  practical implications of strategyproofness (specifically, how participants could maximize earnings).
        $p$-values: two-sample $t$-tests.
        ``$N$'': number of observations per treatment.
        ``Random'': the expected score of answering every question uniformly at random.
        \emph{All Panels:} Error bars: 95\% confidence intervals. 
        ``(\# Qs)'': the number of questions that the relevant part of the strategyproofness understanding test consists of.
    \end{minipage}
    \end{figure}

    \autoref{fig:mean-sp-u-abstract-practical-by-treatment} shows participants' performance separately on the Abstract (consisting of 13 questions) and Practical (5 questions) parts of the SP-U test.\footnote{
        See \autoref{app:sp-more-details} for participants' performance at the question level, along with a summary of each question.
    }
Figures~\ref{fig:sp-u-abstract-means} and~\ref{fig:sp-u-practical-means} support our second main result---by showing that Trad-DA participants score low (statistically, as low as Null) on both sub-measures---and our third main result---by showing that Menu-SP participants score high compared to all other treatments on both sub-measures. 
Remarkably, \autoref{fig:sp-u-practical-means} also shows that with the exception of Menu-SP, participants' Practical SP-U scores in all other treatments are significantly worse than Random. 
We return to this point momentarily.

    \autoref{fig:sp-u-abstract-hist} shows the Abstract SP-U score distributions  within each treatment. They seem very roughly bell-shaped, with a mode that moves right across treatments, from Trad-DA and Null at 54\% correct (i.e., seven out of thirteen questions) to Menu-SP at 100\% correct (with 27\% of participants getting this score, and another 26\% getting the next possible highest score).

    In contrast, \autoref{fig:sp-u-practical-hist}, which
    shows the Practical SP-U score distributions, suggests that they are bimodal, with many participants scoring close to zero while a group of others get a perfect 100\% score (i.e., five out of five questions).
    This bimodality may suggest that participants typically have one of two mental models of how one can maximize earnings: One in which they perfectly perceive strategyproofness, and one in which they perceive exactly the opposite of strategyproofness (e.g., that one should consider the prize priorities and use them to strategically report an NSF ranking).

    Additionally, \autoref{fig:sp-u-practical-hist} shows that in the Trad-DA treatment---which is based upon the traditional description used to explain DA in real-world markets---more than \emph{half} of our participants score zero---i.e., answer according to an ``opposite-SP'' perception of DA---a rate similar to that among Null participants, who are told almost nothing about the mechanism. 
    In contrast, in Menu-SP, 
    many fewer participants have this ``opposite-SP'' perception while many more participants compared to other treatments correctly perceive strategyproofness.

    In \autoref{app:sp-more-details}, we further investigate the joint distribution of Abstract and Practical, and how the two sum up to \% SP-U.
    Our main finding is that the variation in \% SP-U is largely driven by the different sub-measures for different ranges of the \% SP-U distribution.
    Among observations with \% SP-U $<$ 75\%, Practical is typically low---approximately at its ``opposite-SP'' perception mode (with an average of 17\%; SE $<1\%$)---and variation in \% SP-U is mostly determined by Abstract. In contrast, when \% SP-U $\ge$ 75\%, Abstract is close to its maximal value and changes only by little, while the bimodality in Practical drives the weaker form of bimodality seen in the full distribution of \% SP-U, namely, the ``dip'' between 78\% and 89\% in \autoref{fig:sp-u-hist}. 
    At higher levels of \% SP-U, Practical has already shifted to its ``correct SP'' perception mode; for instance, Practical's average in the \% SP-U $\ge$ 75\% range is 85\% (SE $<1\%$).\footnote{
        The specific value of (approximately) 75\% seems largely due to the relative weight of the sub-measures in the overall measure---13/18 questions for Abstract, and 5/18 for Practical.
    } 
    This pattern suggests a potentially more general feature of participants' understanding of SP---that reaching an intermediate level of SP understanding does not mean ``partially'' understanding how to maximize earnings; rather, (some) participants realize how to maximize their earnings only when they understand SP quite well.

    Overall, the joint distribution of Abstract and Practical suggests that in our setting, in order for participants to understand the practical implications of SP on earnings maximization, they need to understand its abstract logical properties.
    However, understanding the abstract logical properties is far from implying understanding of the practical implications. 
    Even in Menu-SP, where many participants score perfectly or nearly perfectly on Abstract, 
    a majority of
    participants are still closer to perfect opposite-SP than to perfect SP perception in Practical.

\subsection{Main Result \#4: Strategyproofness Understanding Predicts Straightforward Behavior}
\label{sec:results-ranking}

Our final main outcome measure is the rate of straightforward behavior (\% SF) in standard incentivized DA rounds. 
We first find that our treatments' effects on \% SF are overall small. 
However, when investigating the relation between \% SP-U and \% SF, we find our fourth main result: Participants with high-enough SP-understanding play SF at very high rates.

\subsubsection{Mean Levels of Straightforward Behavior}

\begin{figure}[htbp]
    \centering
    \caption{Straightforward ranking behavior (\% SF) by treatment}
    \label{fig:mean-sf-by-treatment}

    \begin{minipage}[t]{0.49\textwidth}
    \centering
        \begin{minipage}{0.6\textwidth}
        \subcaption{Mean \% SF}
        \label{fig:mean-sf-by-treatment-actually-the-mean}
        \end{minipage}
    \end{minipage}
    \begin{minipage}[t]{0.49\textwidth}
    \centering
        \begin{minipage}{0.8\textwidth}
        \subcaption{Distribution of \% SF}
        \label{fig:sf-histogram-by-treatment}
        \end{minipage}
    \end{minipage}

    \includegraphics[width=\figscale\textwidth{}]{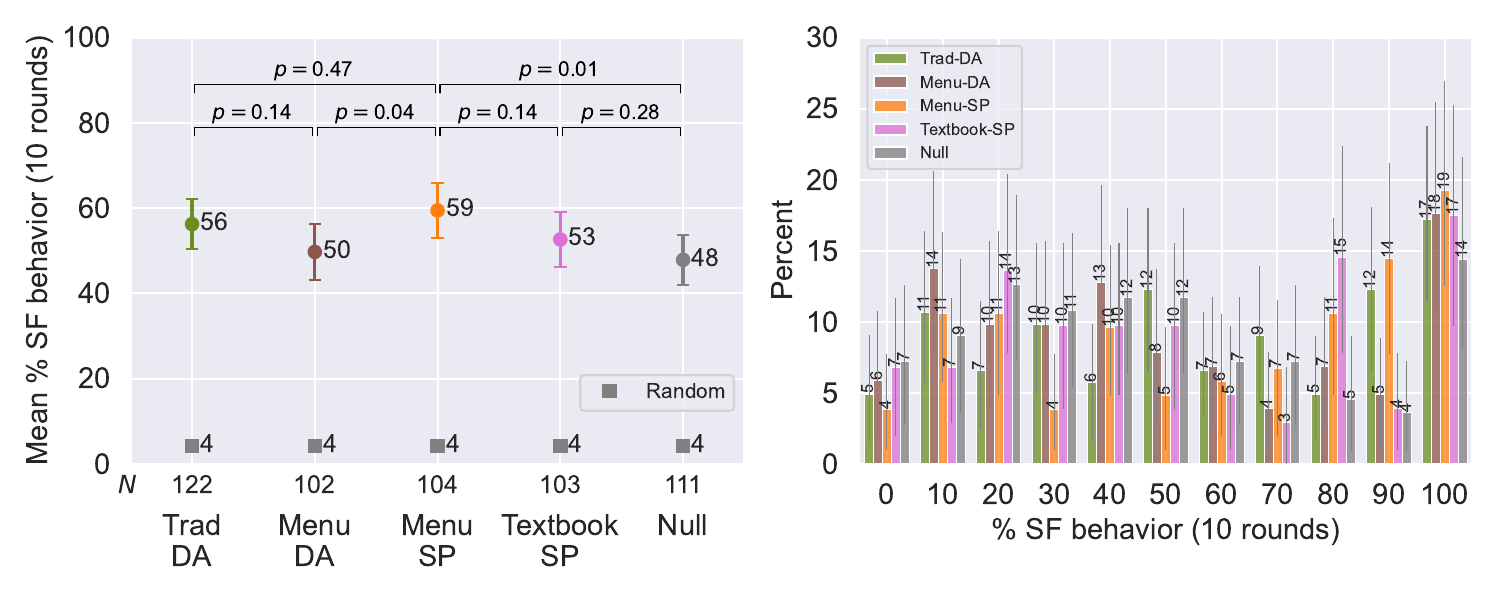}
    \begin{minipage}{\textwidth}  \footnotesize
        \textbf{Notes:}
        \emph{Panel (a)}:
        $p$-values: two-sample $t$-tests.
        ``$N$'': number of observations per treatment.
        ``Random'': the expected \% SF from ranking the prizes uniformly at random.
        \emph{Both Panels}:
        Error bars: 95\% confidence intervals. 
    \end{minipage}
    \end{figure}

   \autoref{fig:mean-sf-by-treatment} shows means and distributions  of \% SF across treatments.
   \% SF  is never particularly high on average.
   Its baseline rate is 48\% in Null.
   In Menu-SP, in spite of the significant increase in \% SP-U we observe in \autoref{sec:results-strategyproofness-understanding}, \% SF is still only 59\%, and only slightly above Trad-DA's 56\%.
   Menu-DA and Textbook-SP remain even lower, at 50\% and 53\% respectively---statistically indistinguishable from Null.
   ($\text{SE}=3\%$ for \% SF in all treatments.)
   Turning to distributions, \% SF seems widely spread in all treatments, and again (at least somewhat) bimodal, with a plateau around 20-40\%, and a peak at 100\%.

    In \autoref{app:ranking-patterns}, we look for patterns of \% SF over the 10 played rounds, as well as patterns in participants' ranking behavior beyond \% SF, i.e., beyond the binary classification of  SF vs.\ NSF play.
    By far the most common NSF ranking flips the first and second prizes, but we do not find additional strong trends in ranking behavior beyond those conveyed by \% SF.
    In particular, while most participants play SF in some rounds and NSF in others, the specific choice of when to play NSF seems to depend only weakly on parameters of the specific round of DA.\footnote{
      For example, we find that across all treatments, participants, and rounds, the fraction of times a participant \emph{does not} rank their highest-earning prize first is 40\%.
      In rounds where the participant has highest priority of getting their highest-earning prize, this fraction is not much lower, at 34\%,
      and we see little variation in this comparison across treatments. 
      We also find little variation in ranking patterns conditional on the difference in monetary value between the highest-earning prize and the second-highest-earning prize.
    } 
    We do not find strong trends of \% SF over the 10 rounds in most treatments. 
    The exception is Menu-SP, where there is an average increase of 1.6\% (SE $=$ 0.5\%) per round in \%~SF, suggesting potential effects of learning, experience, fatigue, and other dynamic phenomena.

\subsubsection{Relation Between Strategyproofness Understanding and Straightforward Behavior}
\label{sec:sp-sf}

\begin{figure}[t]
    \centering
    \caption{Relationship between \% SP-U and \% SF by treatment}
    \label{fig:sp-sf-detailed-relationship}
    \includegraphics[width=\textwidth]{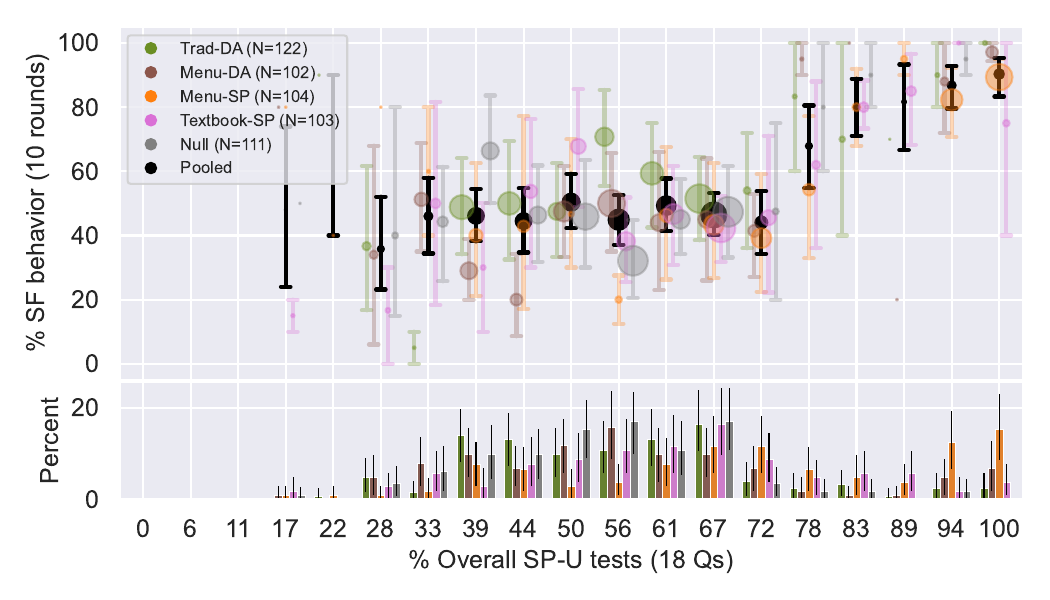}

    \begin{minipage}{\textwidth}  \footnotesize
        \textbf{Notes:}
        \emph{Top:} Mean \% SF by treatment and by \% SP-U score.
        \emph{Bottom:} Distribution of \% SP-U by treatment (replicating \autoref{fig:sp-u-hist}).
        \emph{Both top and bottom}: Circular markers (top) are proportional in diameter to corresponding fraction of participants (bottom). Error bars: 95\% confidence intervals. 
    \end{minipage}
\end{figure}

Finally, we investigate the joint distribution of \% SP-U and \% SF.
While not part of our pre-registered analysis plan, we find strong results, robust  across treatments and sub-samples.
\autoref{fig:sp-sf-detailed-relationship} shows mean \% SF conditioned on (i.e., binned by)  \% SP-U.
At its bottom, the figure reports the distribution of \% SP-U scores by treatment (replicating \autoref{fig:sp-u-hist}).
At its top, it reports mean \% SF among participants with given \% SP-U scores, both by treatment (in semitransparent colors) and pooled (black). The diameter of circle markers is proportional to  bin size (in the histogram at the bottom).\looseness=-1

\autoref{fig:sp-sf-detailed-relationship} shows our fourth main result: That those who grasp strategyproofness and its implications sufficiently well play SF significantly more than those who do not.
Indeed, we see that mean \%~SF ``jumps'' to noticeably high levels once participants score sufficiently high on \%~SP-U, with a threshold at roughly 75--80\%. 
While more noisy at the single-treatment level, this relation seems remarkably global. 
That it is not treatment-specific is consistent with the notion that better understanding of SP may increase SF behavior of some participants, regardless of the description or training that improved SP understanding. %
Moreover, the \% SP-U distribution in the lower panel (also indicated by marker sizes in the top panel) suggests that of our five descriptions, Menu-SP is the most effective at pushing participants to high levels of both \% SP-U and \% SF.

This strong relation between \% SP-U and \% SF, and the strong variation in \% SP-U across treatments from \autoref{sec:results-strategyproofness-understanding},
may initially appear somewhat at odds with our findings %
of small overall differences in mean \% SF across treatments.
\autoref{fig:sp-sf-detailed-relationship} suggests at least a partial explanation.
First, it shows that Pooled across all treatments, the mean \% SF among participants with \% SP-U less than 75\% is 46\%, while for participants above 75\% it is 83\%, a difference of 36\% percentage points. 
Second, it also shows that the fraction of participants with \% SP-U above 75\% ranges from 43\% in Menu-SP to 5\% in Null, a difference of 38\% percentage points.
Thus, one would expect a ballpark difference in \% SF of at most $38\%\times36\% = 14\%$ across treatments. 
This is consistent with the (noisy) range of mean \% SF values observed across treatments in \autoref{fig:mean-sf-by-treatment-actually-the-mean}. 
As the $p$-values reported in that figure suggest,
our sample size is slightly under-powered to detect most differences within this range.

In \autoref{app:sp-sf-more-details}, we investigate the non-conditional joint distribution of \% SP-U and \% SF. %
We observe that Menu-SP is most effective at moving participants into a region where they both understand SP well and play SF at high rates. 
For instance, 32\% (SE $=$ 5\%) of participants in Menu-SP have both high \% SP-U and high \% SF, namely, above 75\% in both metrics. 
In all other treatments this fraction is 17\% (SE $=$ 4\%) or less.
This suggests that participants in Menu-SP tend to not only move towards the higher mode of understanding of strategyproofness, but may additionally act on this understanding by moving towards the higher mode of \% SF.

In \autoref{app:sp-sf-more-details}, we additionally investigate how the Abstract and Practical sub-measures of \% SP-U differently contribute to the overall relation between \% SP-U and\ \% SF.
We find that the Practical sub-measure can by itself explain most of the overall relation in \autoref{fig:sp-sf-detailed-relationship} and, in particular, it effectively separates participants into high vs.\ low SF play based on high vs.\ low sub-measure score.
However, we find that including Abstract may increase the sharpness of this separation, especially among participants who score in the middle range of Practical.\footnote{
    For example, among participants whose Practical scores are 40\% or 60\% (two or three correct answers out of five), the average SF rate moves from 49\% (SE $=$ 4\%) to 80\% (SE $=$ 7\%) when Abstract changes from below 90\% to above 90\%. 
}

\subsection{Additional Results and Robustness}
\label{sec:results-samples-and-controls}

In this subsection, we investigate differences in our main results across the Prolific and Cornell samples,  test robustness to controlling for a set of additional variables, and summarize additional findings on the relationship between our three outcome variables.

\subsubsection{Variation Across Samples}

\begin{figure}[htbp]
\centering
\caption{Main results within the Prolific and Cornell samples}
\label{fig:main-results-by-sample}
\begin{minipage}{0.49\textwidth}
    \begin{center}
        \subcaption{Mean outcome variables by sample}
        \label{fig:main-results-by-sample-means}
        \includegraphics[width=\textwidth]{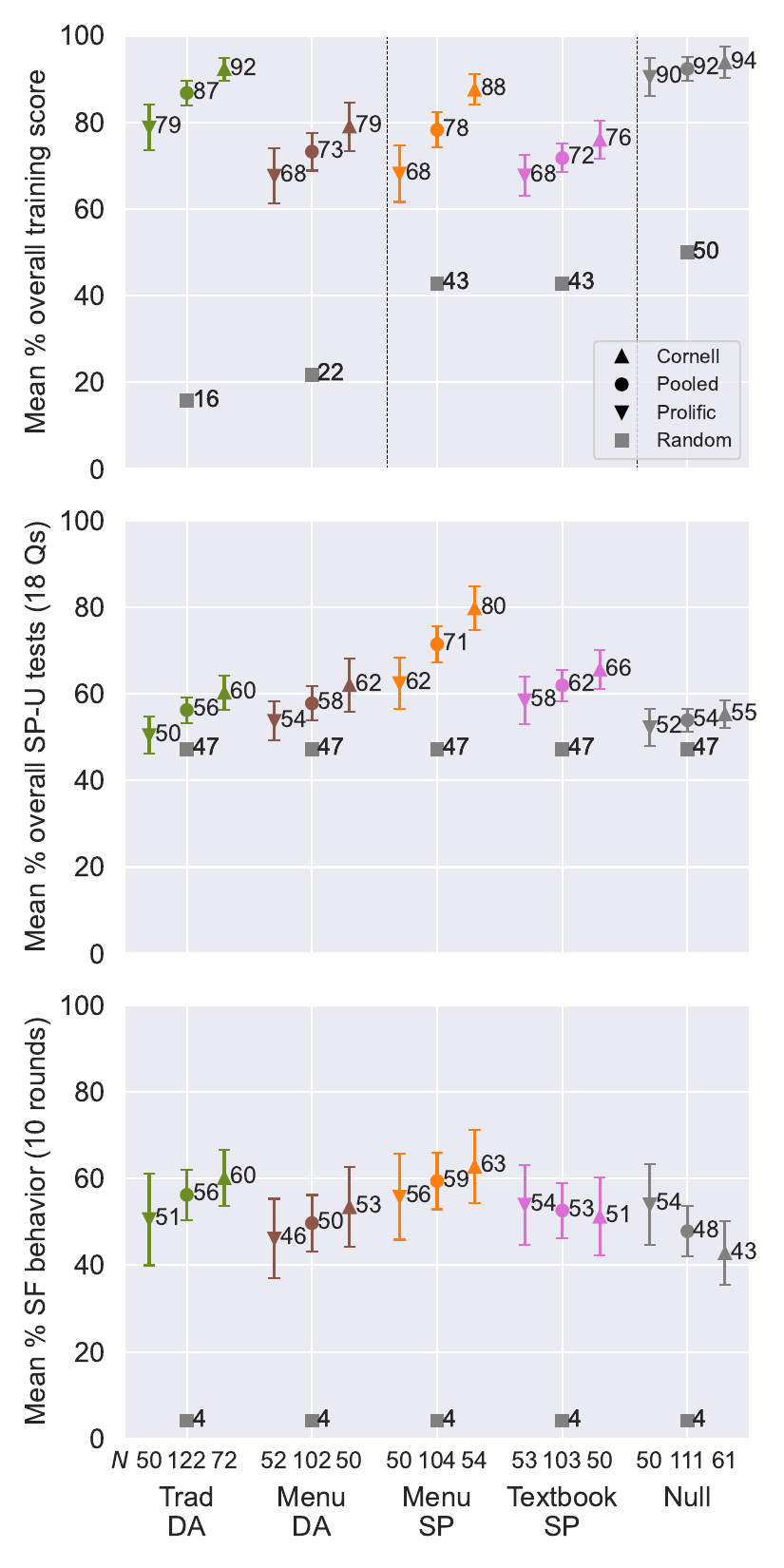}
    \end{center}
\end{minipage}
\hfill
\begin{minipage}{0.49\textwidth}
    \begin{center}
        \subcaption{Relation between \% SF and \% SP-U by sample}
        \label{fig:main-results-by-sample-sf-sp}
        \includegraphics[width=\textwidth]{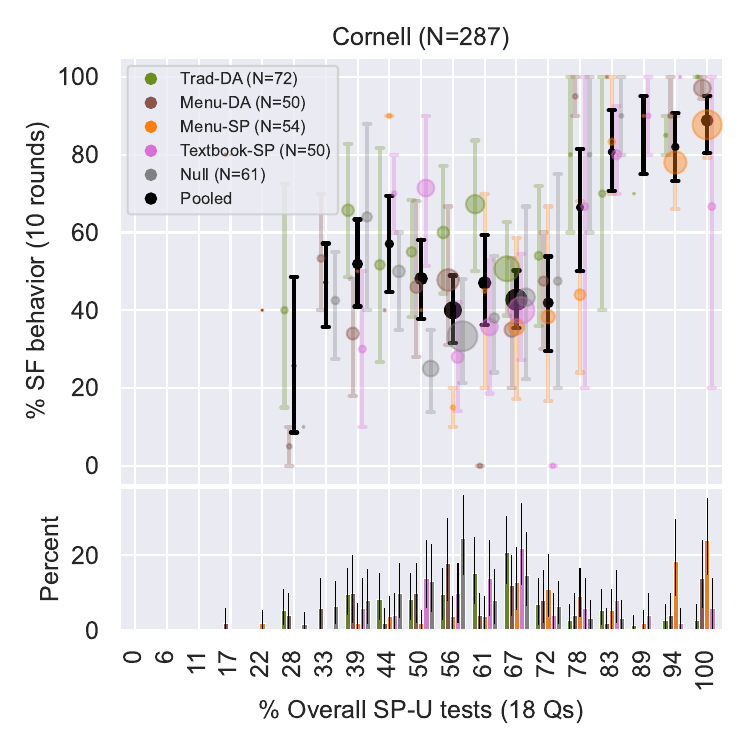}\\
        \includegraphics[width=\textwidth]{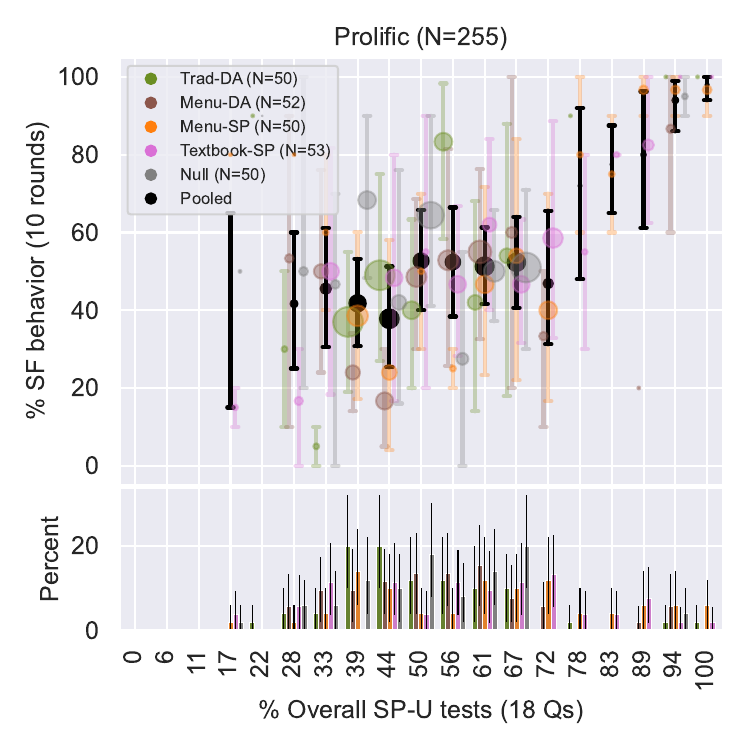}
    \end{center}
\end{minipage}

\begin{minipage}{\textwidth}  \footnotesize
        \textbf{Notes:}               
        \emph{Panel (a)}: The three mini-figures report the mean rates of \% TR (top), \% SP-U (middle), and \% SF (bottom), by treatment. Each mini-figure reports the overall mean and the means among Cornell participants ($N=287$, $\blacktriangle$) and among Prolific participants ($N=255$, $\blacktriangledown$; see \autoref{sec:sample}).
        ``$N$'': number of observations per treatment.
        ``Random'': the expected score of answering every question uniformly at random.
        \emph{Panel (b)}:  \autoref{fig:sp-sf-detailed-relationship} by sample (top: Cornell; bottom: Prolific).
    \end{minipage}
\end{figure}

\autoref{fig:main-results-by-sample} reproduces the main figures of this section---Figures \ref{fig:mean-tr-by-treatment-actually-the-mean}, \ref{fig:sp-u-means}, %
\ref{fig:mean-sf-by-treatment-actually-the-mean}, and \ref{fig:sp-sf-detailed-relationship}---split by sample (Cornell vs.\ Prolific). 
As seen most clearly in \autoref{fig:main-results-by-sample-means}, the baseline levels of outcome variables in non-Null treatments are lower on Prolific ($\blacktriangledown$) than at Cornell ($\blacktriangle$), particularly \% TR (\autoref{fig:main-results-by-sample-means}'s top mini-figure) and \% SP-U (middle mini-figure). And, as seen in both panels, Menu-SP's effect of moving participants to both high \% SP-U and high \% SF is less pronounced (in accordance with fewer participants achieving high \% SP-U in the Prolific sample).

We draw three conclusions. First, these results support the idea that our novel metrics, \%~TR and \% SP-U, are meaningfully measuring understanding levels. In non-Null treatments, those levels should in principle increase with  education, cognitive ability, attention, and ability to focus in a physical lab (vs.\ remote participation); we indeed find \% TR and \%~SP-U to increase within each non-Null treatment for Cornell vs.\ Prolific.
Our new metrics thus pass an important sanity check.

Second, these results suggest that despite the sample differences, cross-treatment differences are at least qualitatively similar across the two samples. This conclusion applies to both the average cross-treatment differences in mean outcome variables (\autoref{fig:main-results-by-sample-means}) and the sharp change in \% SF when \% SP-U increases from low to high levels (\autoref{fig:main-results-by-sample-sf-sp}).

Third, importantly, these cross-sample differences highlight a potential limitation of Menu-SP: Its effect in improving SP understanding appears stronger among those of our participants who are more educated, more highly cognitively skilled, more attentive, or participate onsite (vs.\ remotely). Each of these factors may imply that real-world implementations of Menu-SP may have significant distributional consequences, to be investigated in future work.

In \autoref{app:cognitive}, we conduct a similar comparison of participants with high vs.\ low cognitive scores, and of high vs.\ low attention scores, and see similar results to the Cornell vs.\ Prolific comparison (for more details on the cognitive and attention scores, see \autoref{app:cognitive-attention} in the Supplementary Materials).

\subsubsection{Robustness to Controls}

In \autoref{app:treatment-comparison-controls}, we use the demographic control variables (elicited at the end of the experiment) to conduct two additional robustness tests to our main results. First, we verify that the mean cross-treatment differences of \% TR, \% SP-U, and \% SF (shown in Figures~\ref{fig:mean-tr-by-treatment}, \ref{fig:mean-sp-u-by-treatment}, and~\ref{fig:mean-sf-by-treatment}) change little quantitatively when controlling for demographic characteristics, cognitive and attention scores, and session-date fixed effects.
Second, we find that the relation of \% SP-U and \% SF (shown in \autoref{fig:sp-sf-detailed-relationship}) maintains its step-function shape  %
when controlling for the above variables, across treatments.

\subsubsection{Other Relations Between Outcome Variables}

\autoref{sec:sp-sf} focused on the relation between \% SP-U and \% SF. Among the three pairwise relations between our three main outcome variables, this is the only one that significantly contributes to our understanding of our data beyond that conveyed by mean levels.

For completeness, in \autoref{app:new-joint-distrs} in the Supplementary Materials, we investigate the other joint relationships between our main outcome variables.
We make three main observations. 
First, average relations between \% TR, \% SP-U and \% SF using linear regressions and correlation coefficients
give further evidence that the relation between \% SP-U and \% SF is fairly consistent across treatments.
Second, we investigate the relation between \% TR and \% SP-U in more detail.
In both DA Mechanics treatments, we find within-treatment support for our second main result, that teaching participants (the mechanics of) DA does not imply teaching them strategyproofness, even for those participants who perform very well in \% TR.
In the SP Property treatments, we find a strong monotone relationship between \% TR and \% SP-U, suggesting that the SP Property treatments and the standardized SP-U tests are well-aligned in the way they measure strategyproofness understanding.
Finally, 
we investigate the relation between \% TR and \% SF, which looks like a noisy combination of the two other relations. This is consistent with the idea that \% TR is mostly related to \% SF through its relation with \% SP-U.

\section{Related Literature}
\label{sec:related}

Our paper sits most directly within the literature on behavioral mechanism and market design.
Excellent reviews can be found in \citet{HakimovK19, Rees-JonesS23}.
Early works within this literature compare behavior across different social choice rules, e.g., \citet{KagelL93, ChenS06, PaisP08, FeatherstoneN16}, among others.
More recent papers (surveyed below, e.g., \cite{Li17}) often consider a fixed social choice rule, and vary some feature of how the rule is implemented.
Our paper considers a fixed implementation---i.e., we keep the rounds of play identical across treatments---and we vary the framing and information provided to participants.
We now discuss, and compare our paper with, the most directly related prior work and themes in the literature.

\paragraph{Menu descriptions.}
As discussed, our Menu DA Mechanics treatment is directly based on the mechanism design theory paper GHT~(\citeyear{GonczarowskiHT23}), which also indirectly inspired our Menu SP Property treatment.

Other than GHT, the most closely related paper to ours is %
the experimental paper by \citet{KatuscakK2020}.
Like our study, they conduct an experiment with a menu description of a matching mechanism, Top Trading Cycles (henceforth TTC).\footnote{
    Phrased in the language of our paper, one treatment of \citet{KatuscakK2020} is a Traditional (TTC) Mechanics treatment, and one is a Menu SP Property treatment with an optional supplemental appendix given to participants, which specifies the details of one form of Menu (TTC) Mechanics.} 
Beyond differing in the mechanism used, their experiment has only behavior as an outcome variable, while our focus %
is on the interplay between training score, SP understanding, and behavior. Moreover, they do not conduct training, and do not test SP understanding separately from behavior. 
They find higher rates of SF play in their menu treatment of TTC, especially for participants with higher cognitive-ability scores.

\paragraph{Training.}
Many papers on behavioral market design conduct pen-and-paper training quizzes that ask participants to solve simple instances of DA, but either do not record participants' performance on such quizzes, or do not report any analysis based on such quizzes.
All exceptions of which we are aware give a single pen-and-paper task without providing feedback or coaching on the descriptions.\footnote{
    These exceptions include \citet{GuillenH14, BoH20, GuillenV21}. Each of them reports some analysis based on participants' performance in a single-question training quiz.
    \citet{GuillenH14} study TTC, and report that roughly half of participants complete the quiz correctly, but that performance on the quiz does not correlate with behavior.
    \citet{BoH20} study DA, and report a positive correlation between answering the quiz correctly and SF play.
    \citet{GuillenV21} study TTC and DA, and report that performance on the quiz does not seem to strongly change under the framing change of reversing the order in which each participant's submitted ranking is considered by the algorithm.
} 
Our training questions and allocation-calculation GUI extend these papers' descriptions and their training tasks, and formalizes them into a computerized interface with hand-held coaching and feedback.
Recently, \citet{SerizawaST24} conduct an experiment in a strategyproof auction setting with two types of quizzes depending on the treatment; in only one treatment, their quiz asks participants how they can maximize their earnings, constituting a ``hint'' towards strategyproofness. 
They find this quiz ``with hint'' increases SF play substantially above the baseline.

\paragraph{Advice.}
Prior work that explores strategic advice in matching mechanisms includes \citet{GuillenH14, GuillenH18}.
\citet{MasudaMSSW20} conduct similar experiments in a second-price auction, while aiming to reduce the experimenter-demand effect (by prefacing the advice with a disclaimer indicating that the advice may or may not be true, and the participant should freely choose whether or not to follow the advice).
These papers generally find that advice is effective at increasing SF behavior, but does not completely eliminate NSF behavior. Our SP Property treatments describe the strategyproofness property, and avoid providing any explicit strategic advice to choose the SF ranking.

\paragraph{Framing and information provision changes.}
Various papers investigate framing effects in mechanism design settings in ways different from menu descriptions.
\citet{BreitmoserS19} study ascending-price-based framings of both static and dynamic auctions, and find that some framings of static auctions in terms of dynamic ones can influence behavior nearly as much as actually implementing the mechanism via dynamic auctions.
\citet{GuillenV21} study the theoretically-trivial change to DA of reversing the order in which participants submit their rankings, and find that a lower fraction of participants submit (reversed) straightforward rankings.
\citet{DanzVW22} and \citet{GuillenH14} show that, in different strategyproof environments, providing participants with more information on the environment's mechanics can \emph{decrease} straightforward reporting. Related to such results, we find that Menu-DA is less effective at teaching SP relative to Menu-SP---despite merely providing more information.

\paragraph{Behavioral mechanism design more broadly.}
Other papers propose behavioral models of behavior or confusion in strategyproof mechanisms, and investigate these models empirically or theoretically.
\citet{Li17} theoretically and experimentally studies failures to grasp that a mechanism is strategyproof due to limited contingent-reasoning skills, and defines and proposes switching to obviously strategyproof (OSP) implementations of strategyproof choice rules. \citet{AshlagiG18} and \citet{Thomas21} show that, except under highly restrictive sets of priorities, DA does not have an OSP implementation.\footnote{
    While classifying OSP matching  mechanisms is often an intricate theoretical task, the main finding of this literature is that obviously strategyproof matching mechanisms do not exist in many settings.
}
\citet{PyciaT23} theoretically study a hierarchy of strategic simplicity notions strengthening OSP.

\citet{DreyfussHR22, DreyfusGHR22, MeisnerW23} study loss aversion in matching mechanisms and DA in particular, and find that it can explain NSF behavior patterns well. 
The designs of these papers typically give participants more information on their chances of getting different prizes, since their behavioral theories concern participants' beliefs about the expected earnings.
Since our work focuses on participants' misunderstandings of strategyproofness, we do not provide explicit information on probabilities of getting different prizes, and instead focus on explicit information about the DA algorithm.
\citet{BoH20, BoH23} study interactive mechanisms that ask participants to repeatedly indicate their favorite from some set of still-available options, and find that such mechanism can increase SF play.

Our work is also indirectly inspired by a vast literature %
of empirical papers that study real-world implementations of matching mechanisms. 
See \citet{Pathak17} for a review.

\section{Conclusion}
\label{sec:conclusion}

Describing designed markets, such as those using DA, to participants is a challenging task. 
One reason is that calculating matches in DA and in various other popular mechanisms requires a detailed combinatorial algorithm.
We view the encouraging training scores in all our treatments as a successful indication that both the mechanics of DA, as well as the strategyproofness property, can be taught to participants with careful instruction.
Building on the tools we have developed, such as our experimental flow, training GUI, and strategyproofness understanding tests, future work may further refine methods designed to help participants better understand DA and other strategyproof mechanisms.

We view the relation we find between participants' understanding of strategyproofness and straightforward play as an important contribution of our work.
Other recent works have suggested that NSF play may be partly intentional, for example, as a strategy played by loss-averse participants in order to avoid disappointment \citep{DreyfussHR22, DreyfusGHR22, MeisnerW23}. 
We believe it is \emph{also} important to investigate the extent to which NSF play may be the result of simple misunderstandings of strategyproofness.
The step-function-like behavior illustrated in \autoref{fig:sp-sf-detailed-relationship} may shed some light in this direction: Above a certain sufficiently high level of strategyproofness understanding, participants indeed play SF at rather high rates.

Our main results suggest that the traditional description of DA can be taught to participants without teaching them strategyproofness. At the same time, participants \emph{can} (to some extent) be taught strategyproofness via our novel Menu-SP description. 
While it was our Menu-\emph{DA} description---which we based on the main theoretical result in \citep{GonczarowskiHT23}---that originally inspired us to embark on our experimental investigation into how participants respond to different descriptions of DA, our results suggest that mechanical descriptions may remain too complicated to effectively convey strategyproofness, even in their menu form.
In contrast, our Menu-SP treatment---a stripped-down version of the full Menu-DA that focuses on the main principle of menu descriptions---is found to be much more effective in our data.
Its simplicity may be the key to success. 
It simply informs participants of a concrete property satisfied by the matching algorithm:
Each participant's allocation will always be their highest-ranked choice among some set of possibilities that their ranking cannot influence.

Our study suggests many promising directions.
Future work may develop SP Property descriptions further,
explore the interaction between conveying SP Property and providing concrete ranking advice,
investigate real-world deployments,
and, importantly, pay careful attention to potential redistributional consequences. 
Future work may also study broadly different new approaches to conveying other important properties of mechanisms, such as those related to different notions of fairness.

{

\begin{refcontext}[sorting=nyt]
\printbibliography
\end{refcontext}

}

\newpage

\appendix

\setcounter{page}{1} \renewcommand{\thepage}{A\arabic{page}}
\numberwithin{figure}{section}
\numberwithin{table}{section}

\clearpage

\section{Additional Results}
\label{app:additional-analysis}

In this appendix, we present additional results and analyses of our data.

\subsection{Sample Collection, Earnings, and Duration}
\label{app:sample-basics}

We provide further details on our sample collection, including attrition rates by sample and treatment, and on earnings and completion times.

At Prolific, $291$ participants started the experiment and $257$ completed it. Of the $34$ who dropped out, $27$ dropped before reaching any treatment-specific parts. Specifically, they dropped out at or before an optional exit point, right after the common Null training, where they could quit for partial payment; we included this salient exit point to reduce attrition conditional on treatment. 
We had to drop another $2$ observations that included corrupt DA round data due to technical issues, leaving a final sample of $N = 255$. At Cornell, $295$ participants started the experiment and $293$ completed it; of the two incompletes, one voluntarily dropped out, and one stopped due to a computer crash. We had to drop another $6$ observations due to other technical data-saving errors unrelated to the experiment content, leaving a final sample of $N = 287$.

In the Prolific sample, of the $9$ participants who dropped out after allocation into treatment or whose data was corrupt, $3$ were allocated to Trad-DA, $3$ to Menu-DA, $2$ to Menu-SP and $1$ to Textbook-SP. In the Cornell sample, of the $2$ participants who dropped out after allocation into treatment or whose computer crashed, $1$ was allocated to Trad-DA and $1$ to Textbook-SP. Treatment data was not saved for the additional $6$ participants who had data-saving errors.

Recruitment and experiment texts were as similar as possible across platforms. Changes to experiment text were limited to differences in currency---British pounds (\pounds) on Prolific and US dollars (\$) at Cornell; differences in the fixed participation fee; and that the optional early exit point for partial payment was enabled for Prolific participants only. 
Prolific prescreening included only participants from the US, with a past Prolific approval rating of at least $99\%$ and at least $50$ approved past tasks. Prolific participants were required to participate in a camera-on video-conference meeting with the experimenter for the duration of the experiment. Cornell prescreening included only students. Cornell participants participated in physical lab sessions. 
For recruiting materials, a detailed description of the experiment protocol, and more details on the sessions, see \autoref{app:experiment-procedures} in the Supplementary Materials. For demographics characteristics of our sample, see \autoref{app:demographics} in the Supplementary Materials.

Prolific participants earned an average of \pounds15.0 (a fixed \pounds7 participation fee; an average of \pounds3.0 for TR and SP-U questions; and an average of \pounds5.0 for incentivized rounds).
Cornell participants earned an average of \$25.5 (a fixed \$17 participation fee, an average of \$3.4 for TR and SP-U questions, and an average of \$5.1 for incentivized rounds; participants could choose to receive one credit for their studies at Cornell instead of the \$17 participation fee; all monetary earnings were paid through Amazon gift cards).

Median completion times differ across treatments, where Trad-DA and Menu-DA take $48$ minutes and $53$ minutes, respectively, Menu-SP and Textbook-SP take $40$ minutes and $39$ minutes, respectively, and Null takes $35$ minutes. 
The overall median time on Prolific is $4$ minutes higher than at Cornell.

\begin{figure}[htbp]
    \centering
    \includegraphics[width=\textwidth]{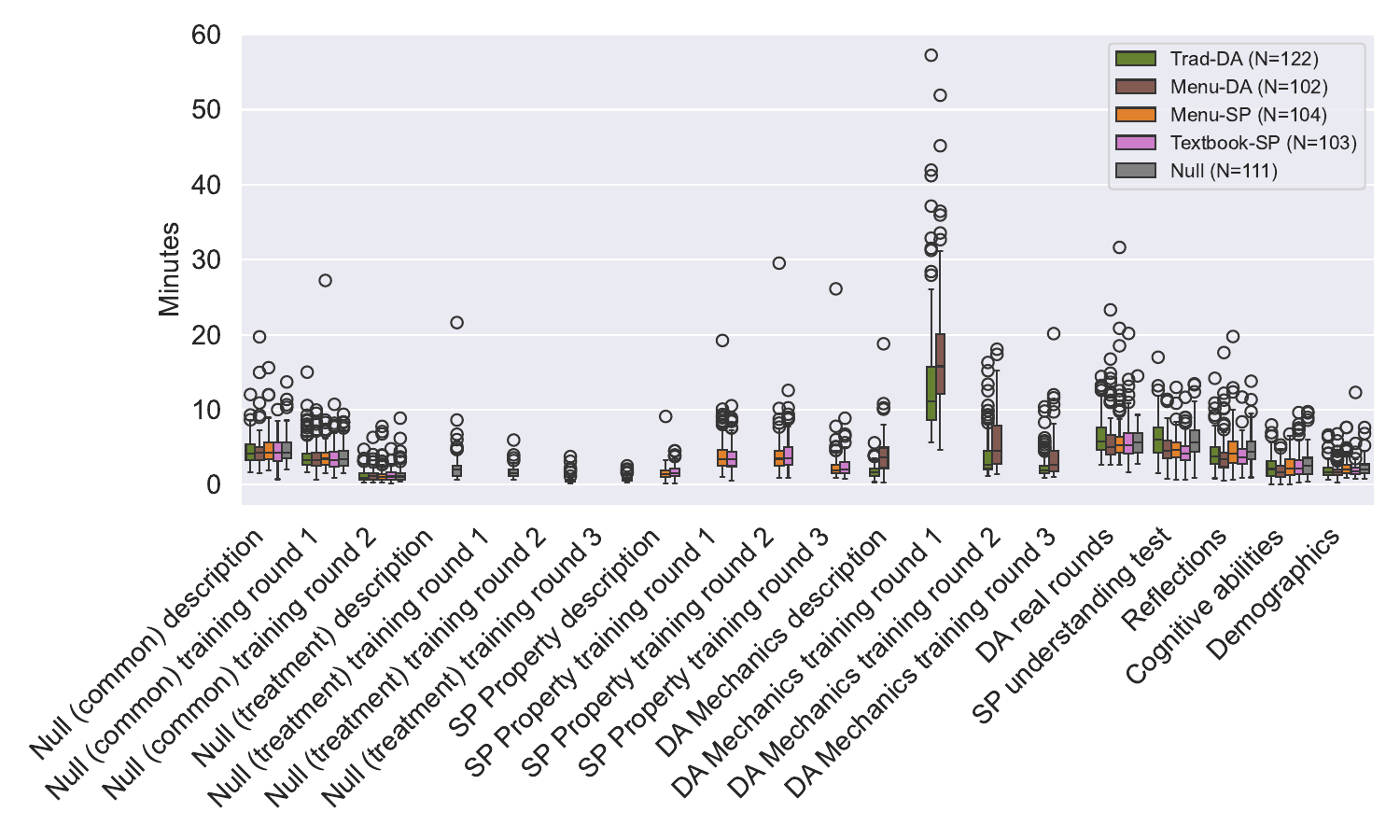}
    \caption{Experiment duration by treatment and part}
    \label{fig:experiment-duration}
\end{figure}

\autoref{fig:experiment-duration} shows the distribution of experiment duration across treatments and the different main parts of the experiment.

\subsection{Full Training Score Distribution}
\label{app:tr-distributions}

We present more details on participants' performance in our training modules, by showing the overall training-score distributions.

\begin{figure}[htbp]
    \centering
    \caption{Distribution of training scores and the Null-description training score}
    \includegraphics[width=\textwidth]{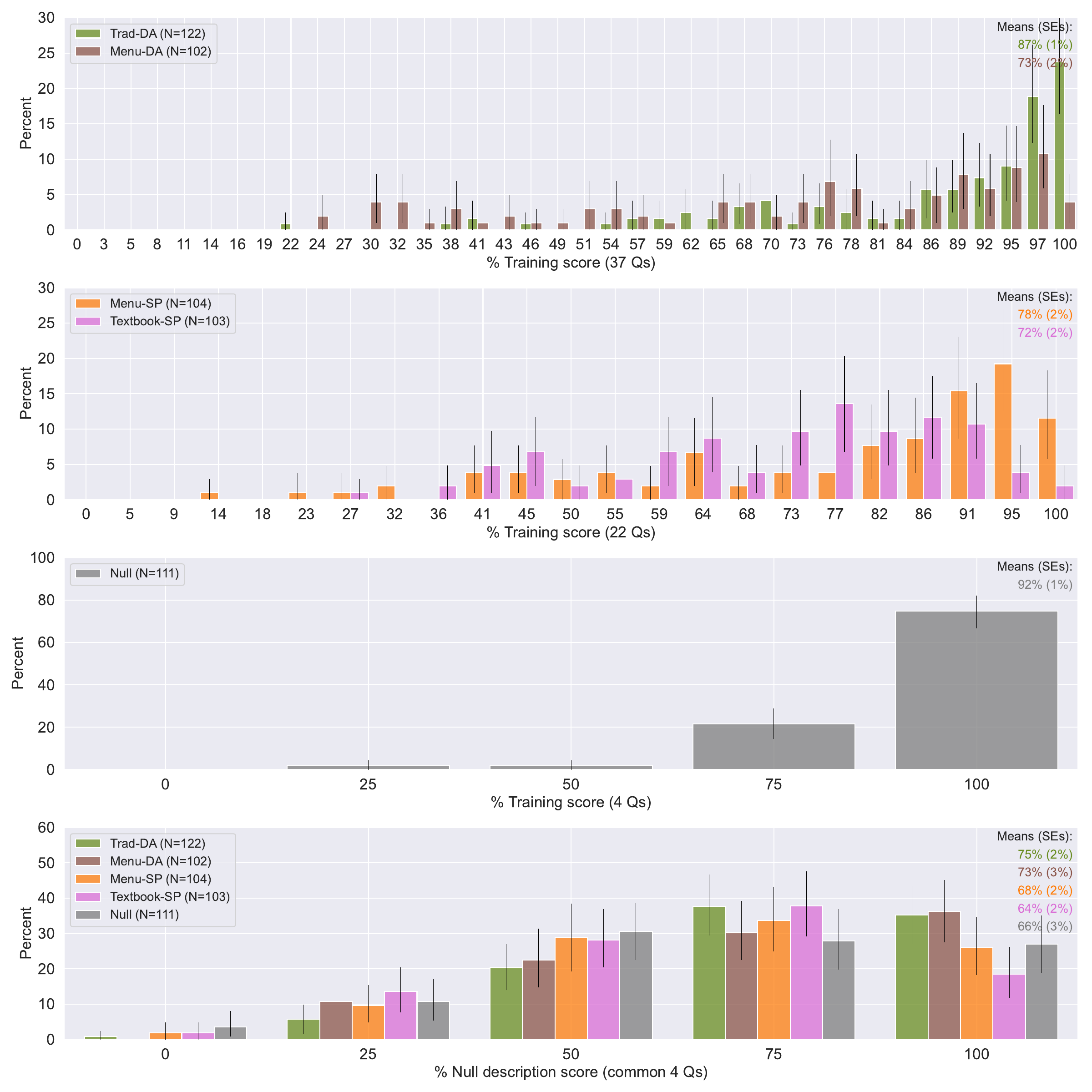}
    \label{fig:all-training-hists}
    \begin{minipage}{\textwidth}  \footnotesize
    \textbf{Notes:}
    \emph{First row}: The two DA Mechanics treatments.
    \emph{Second row}: The two SP Property treatments.
    \emph{Third row}: The Null treatment.
    \emph{Fourth row}: The four Null training round questions, common to all treatments (and not included in their training scores). 
    \emph{All rows}: Mean values (SEs) are included in the upper right corner. ``(\# Qs)'': the number of questions that the relevant score is based on. Error bars: 95\% confidence intervals.
    \end{minipage}
\end{figure}

\autoref{fig:all-training-hists} shows the overall distribution of participants' performance in all training modules. From top to bottom, the figure rows focus on: (1) the two DA Mechanics treatments, (2) the two SP Property treatments, (3) the Null treatment, and (4) participants' performance in the training round following the Null description, which is common to all participants. These are the only training questions in the experiment which are \emph{not} a part of \% TR measure. 

For details on the mean question-level performance within each treatment, see \autoref{app:tr-more-details} in the Supplementary Materials.

\subsection{Sub-Measures of \% SP-U}
\label{app:sp-more-details}

We study the distributions of sub-measures of \% SP-U, particularly the joint distribution of the Abstract and Practical sub-measures.

\begin{figure}[htbp]
    \centering
    \caption{Success rates in all SP-U questions}
    \includegraphics[width=\textwidth]{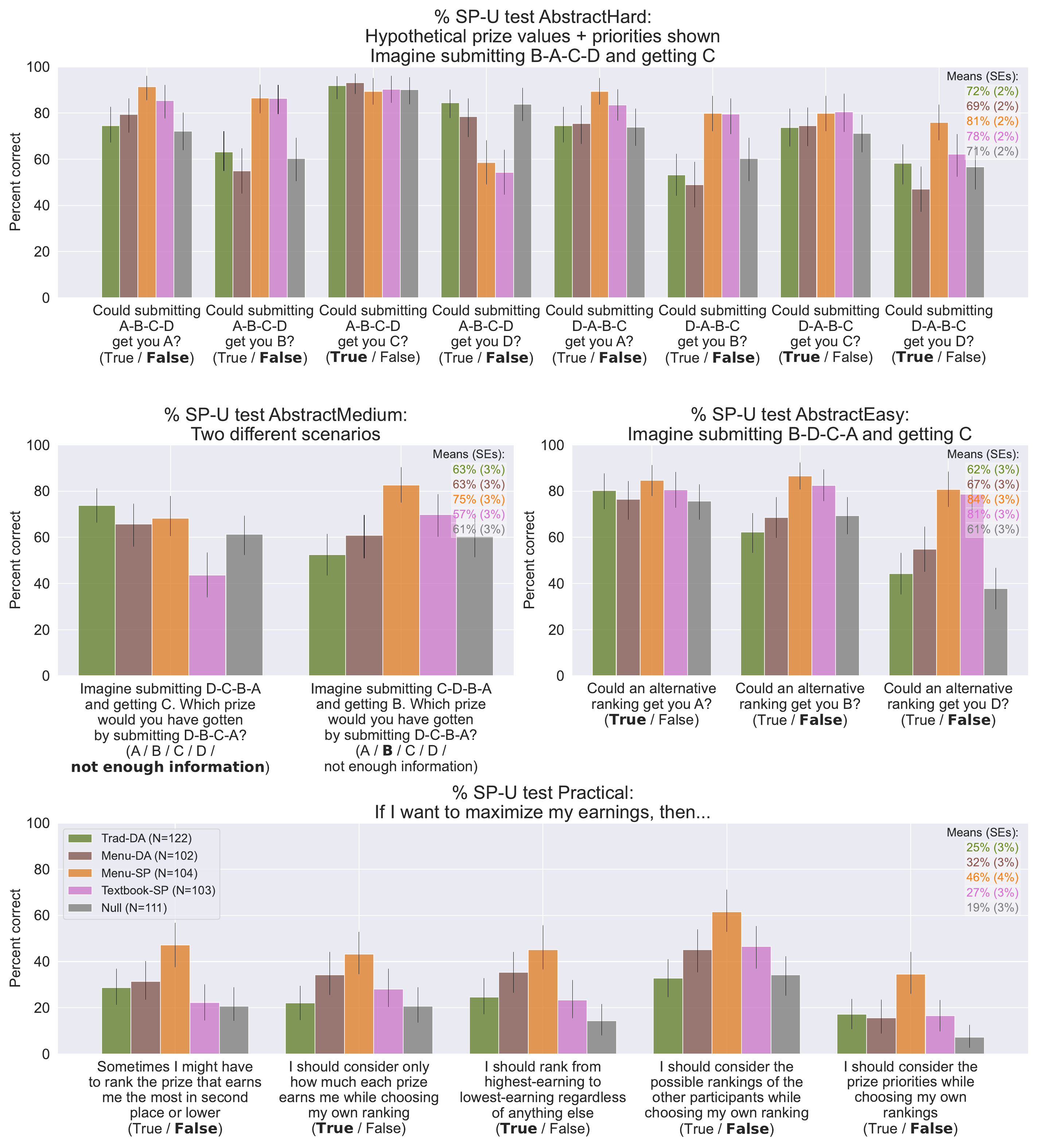}
    \label{fig:sp-u-all-questions}
    \begin{minipage}{\textwidth}  \footnotesize
    \textbf{Notes:} Each mini-figure shows the question-level success rates in one of the four SP-U sub-measures across treatments. Error bars: 95\% confidence intervals.
    \end{minipage}
\end{figure}

First, 
\autoref{fig:sp-u-all-questions} shows the average success rates by treatment in \emph{all} individual SP-U questions, along with summaries of the questions.

\begin{figure}[htbp]
    \centering
    \caption{Joint distribution of the Abstract and Practical \% SP-U sub-measures by treatment}
    
    \includegraphics[width=\textwidth]{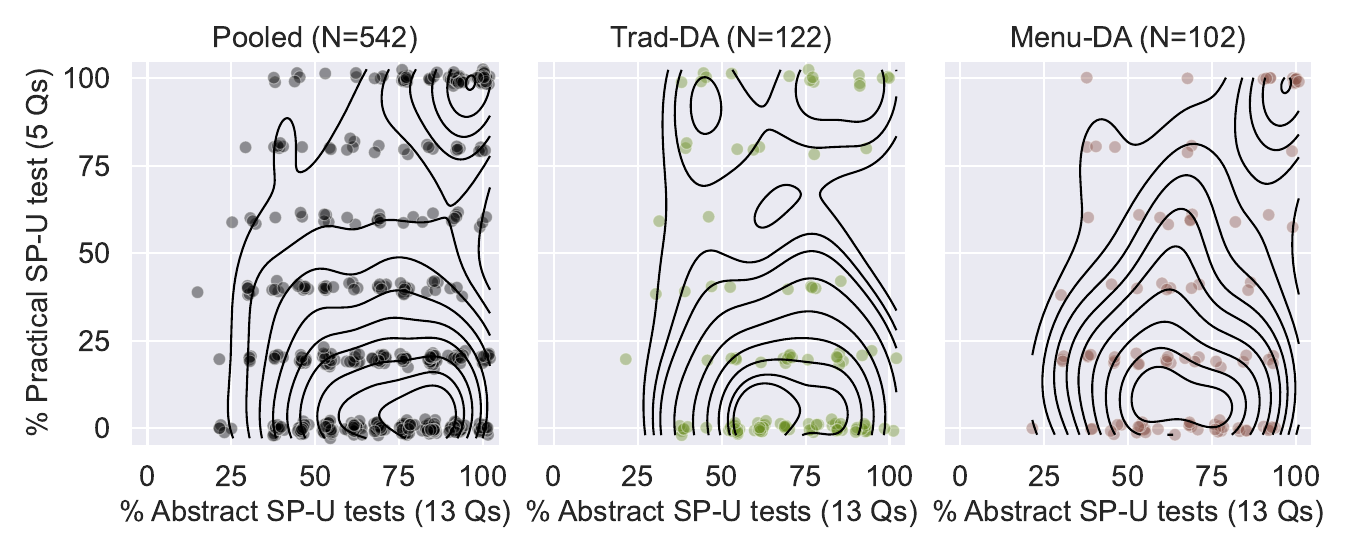}
    \includegraphics[width=\textwidth]{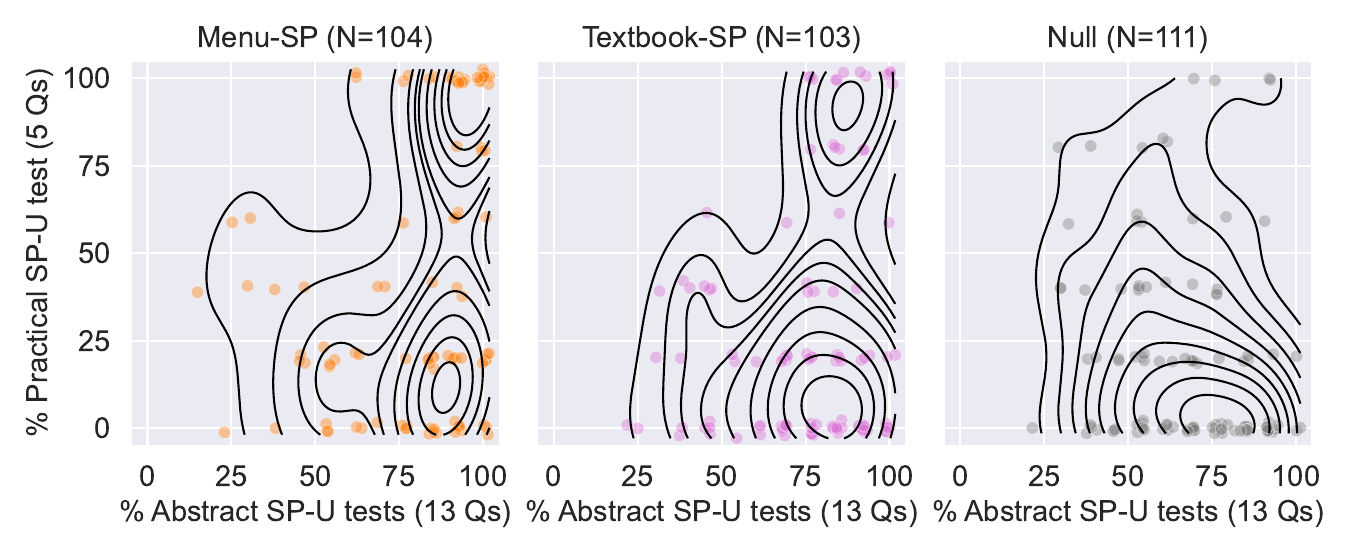}

    \label{fig:sp_u_sub_tests_joint_distribution}
    \begin{minipage}{\textwidth}  \footnotesize
    \textbf{Notes:}
    Each mini-figure contains a jittered scatter plot and estimated contours smoothing the two-dimensional distribution. 
    Smoothing is performed using a Gaussian kernel with a bandwidth chosen according to Scott's Rule (\cite{scott2015multivariate}).    
    \end{minipage}
\end{figure}

Next, in \autoref{fig:sp_u_sub_tests_joint_distribution},
we investigate the joint distribution between the main sub-measures of \% SP-U, Abstract and Practical. 
The figure shows that high scores in Practical are predictive of high scores in Abstract; for example, among all participants with Practical $\geq$ 75\%, the average Abstract score is 83\%.
In the other direction, the Abstract score is not as predictive regarding Practical; for example, even among participants with Abstract $\geq$ 90\%, the average Practical score is 47\%. 
This suggests that a high Abstract score is nearly a necessary condition---albeit an insufficient one---for a high Practical score.

\begin{figure}[htbp]
    \centering
    \caption{Relationship between \% SP-U and its Abstract and Practical sub-measures}
    
    \begin{minipage}{0.48\textwidth}
    \includegraphics[width=\textwidth]{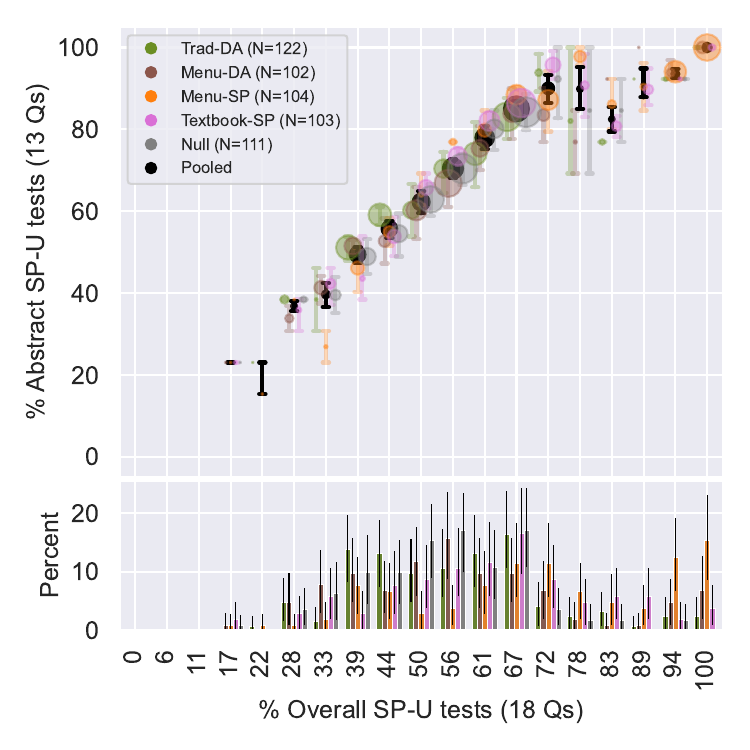}
    \end{minipage}    
    \begin{minipage}{0.48\textwidth}
    \includegraphics[width=\textwidth]{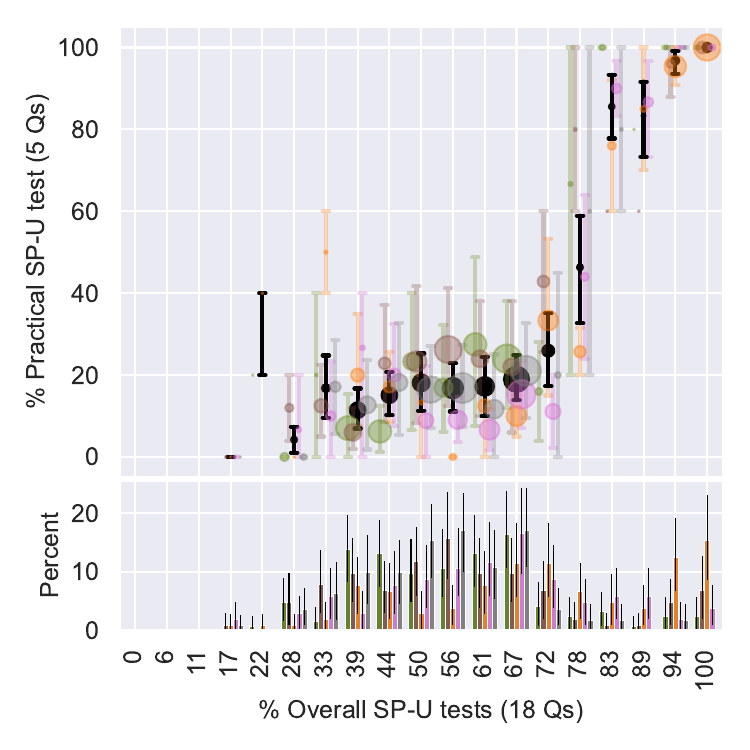}
    \end{minipage}
    
    \label{fig:sp_u_and_its_sub_tests}
    \begin{minipage}{\textwidth}  \footnotesize
    \textbf{Notes:}
    \emph{Left}: Histogram of \% SP-U at the bottom and the conditional means of the Abstract \% SP-U sub-measure given all possible \% SP-U scores at the top (see \autoref{fig:sp-sf-detailed-relationship}, which is similar in structure, for more details).
    \emph{Right}: a similar figure with the Practical sub-measure instead of Abstract. 
    \end{minipage}
\end{figure}

Next, \autoref{fig:sp_u_and_its_sub_tests} suggests that \% SP-U can be understood as having two main ranges. The figure shows the means of Abstract and Practical conditional on different total \% SP-U scores. 
The range of \% SP-U $<$ 75\% is governed mostly by the Abstract score, where Practical is approximately fixed in its low mode. The range of \% SP-U $\geq$ 75\% is governed mostly by the Practical score, where Abstract is approximately fixed near-perfect, and the weak bimodality observed in \% SP-U in this range reflects Practical's bimodality.

\subsection{Ranking Patterns Beyond \% SF}
\label{app:ranking-patterns}

We investigate the details of participants' ranking behavior.

\begin{figure}[htbp]
    \centering
    \caption{Distribution of submitted rankings: overall and by SP understanding score}

    \includegraphics[width=\textwidth]{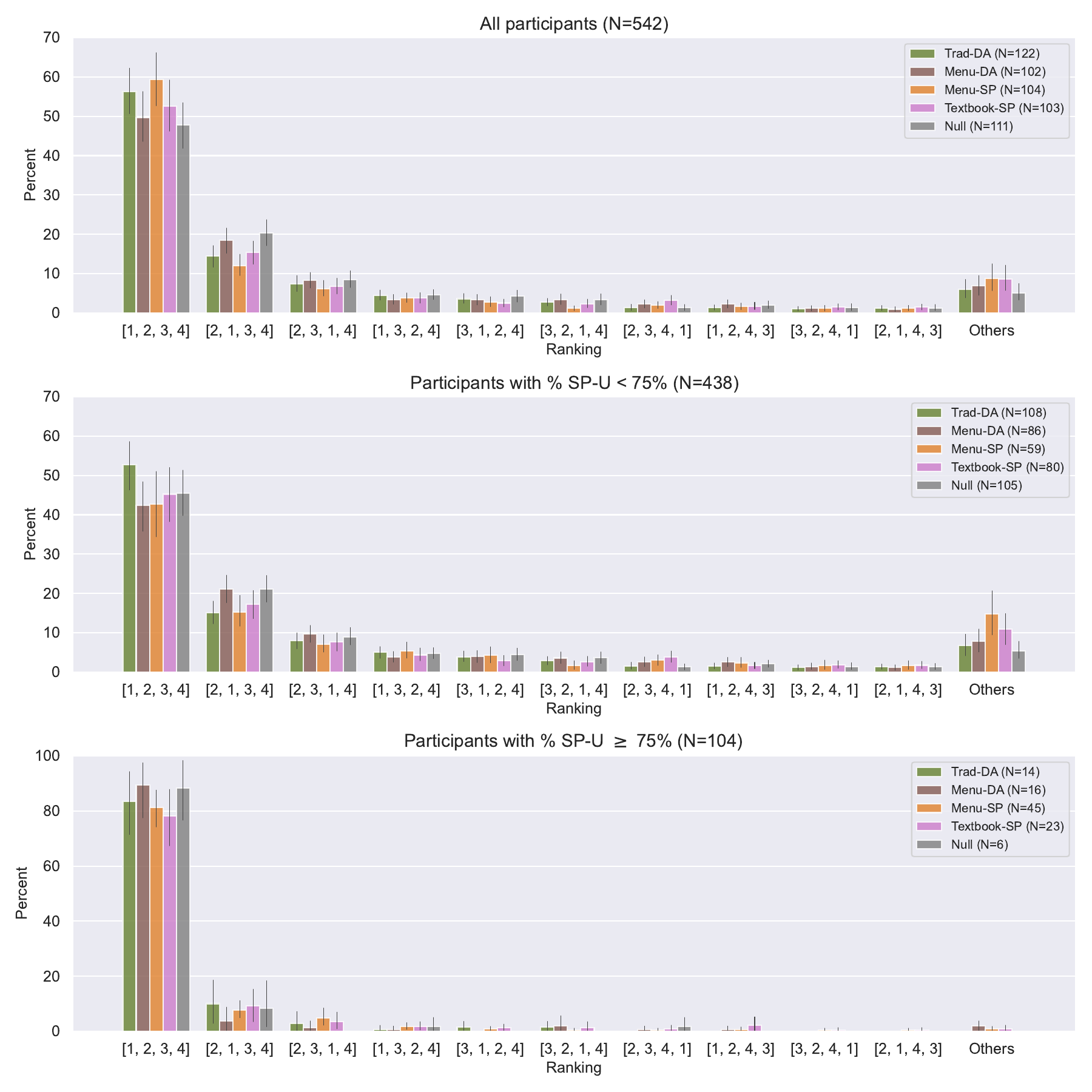}
    \label{fig:most-common-submitted-lists-by-sp-u}
    
    \begin{minipage}{\textwidth}  \footnotesize
    \textbf{Notes:}
    Each $k\in \{1,2,3,4\}$ denotes the subject's $k$th highest-earning prize, and rankings are written from highest to lowest.
    For example, ranking $[1,3,2,4]$ flips the second and third highest-earning prizes relative to the SF ranking ($[1,2,3,4]$). Error bars: 95\% confidence intervals.
    \end{minipage}
\end{figure}

\autoref{fig:most-common-submitted-lists-by-sp-u}, top row, shows the distribution of the most common ranking patterns in the real rounds of playing DA. The SF behavior---ranking $[1,2,3,4]$ where numbers reflect ordinal rank of prize value from highest- to lowest-earning---is the most common strategy across all treatments. Additionally, the most common NSF ranking is $[2,1,3,4]$, i.e., a ranking that flips the highest- and second-highest-earning prizes. We see little variation in types of NSF behavior across treatments. The figure's middle and bottom rows show this distribution among participants with different levels of SP understanding. Participants with low \% SP-U (below 75\%), who play NSF in higher rates, still mostly use the $[2,1,3,4]$ strategy among NSF ones.

\begin{figure}[htbp]
    \centering
    \caption{The frequency of different strategies conditional on round parameters}
    \includegraphics[width=0.85\textwidth]{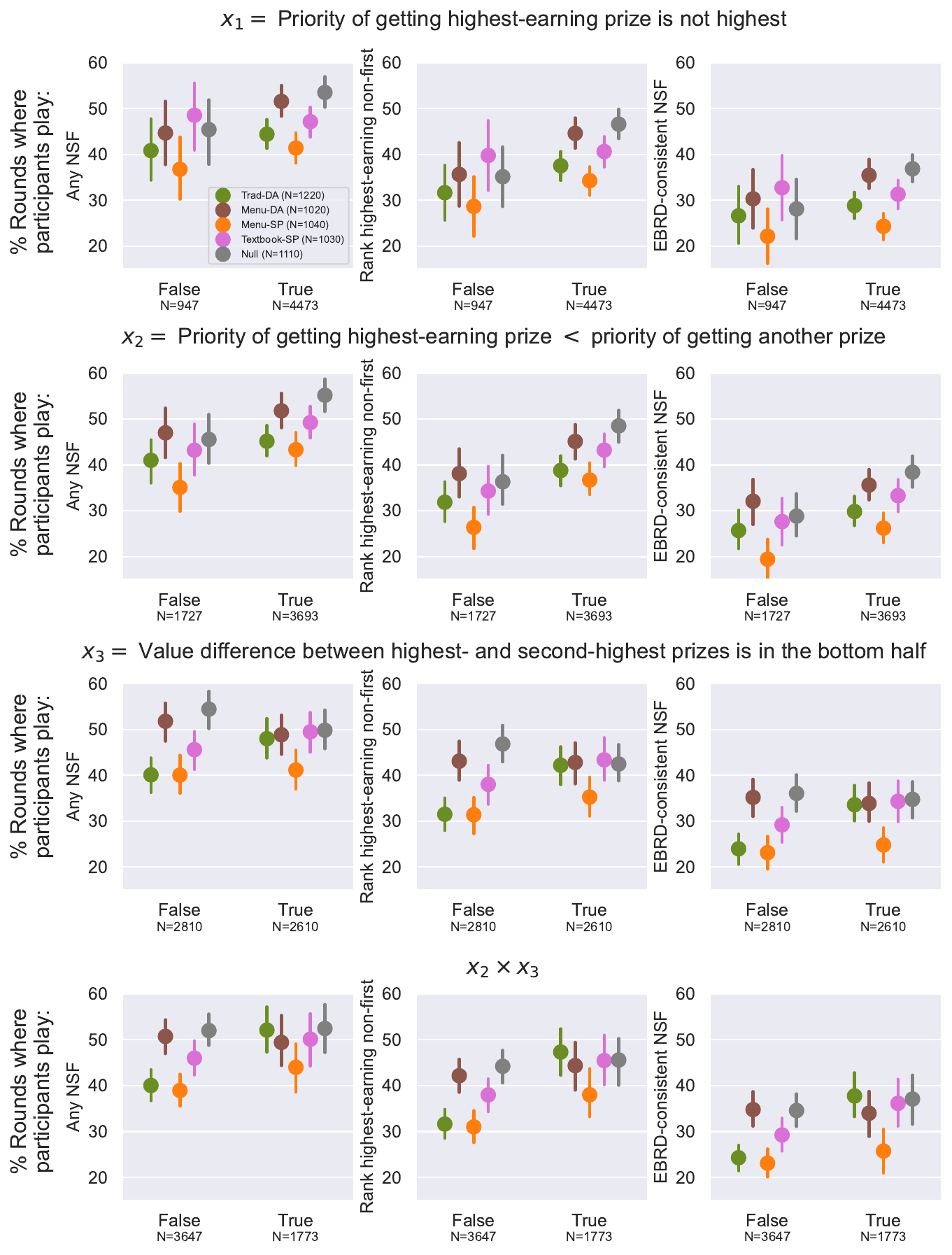}
    \label{fig:nsf-round-level-patterns}
    \begin{minipage}{\textwidth}  \footnotesize
    \textbf{Notes:}
    Each row in the grid investigates a different binary classification of rounds according to their prize priorities and prize monetary values (resulting in a binary variable $x_i$ coded for each round).
    Each column investigates the frequency of a group of strategies conditional on the row's classification.
    In each panel the five treatments are shown separately for comparison. Error bars: 95\% confidence intervals.
    \end{minipage}
\end{figure}

Next, \autoref{fig:nsf-round-level-patterns} explores whether ranking behavior depends on key  parameters of the DA rounds. 
The figure shows the frequency of three different categories of NSF ranking behavior, and compares these behaviors across different categories of DA rounds.
The categories of ranking behavior are 
(1) any NSF, 
(2) any NSF where the highest-earning prize is not ranked first, and
(3) any NSF strategy consistent with ranking according to expectations-based reference-dependent (EBRD) preferences under some beliefs (according to a characterization given by \cite{MeisnerW23}).\footnote{
    EBRD is a model shown to explain NSF patterns well in some settings \citep[see][]{DreyfusGHR22,DreyfussHR22}.
    Unlike these papers, we cannot make sharp EBRD predictions in our settting since it was impossible for participants to calculate their probabilities of getting different prizes
    (the computerized participants rankings' distribution were unknown). 
    We instead use the property ``top-choice monotonicity'' from \citet{MeisnerW23}, that classifies strategies which are EBRD-consistent under some beliefs vs.\ strategies which are never EBRD-consistent. 
}
The figure compares behavior across different binary conditions of the DA rounds.
Each condition is such that, when the condition is true, we expect a higher rate of NSF play (under hypothesized common misconceptions, e.g., that prizes where the participant has low priority, or which do not offer a large increase in the money reward, are ``not worth'' ranking first).
The conditions are: (1)  the participant's priority of getting the highest-earning prize is not highest,
(2) the participant's priority of getting that prize is strictly less than that of getting some other prize, 
(3) the difference between the monetary values of the highest- and second-highest-earning prizes is in the lower half of the distribution of such differences, 
and (4) a combination of the previous two categories.

Overall, we find that, in line with our expectations, NSF play generally increases when the conditions are true.
However, that increase is fairly mild (generally around 10 percentage points), and we do not find any strong pattern across treatments beyond those conveyed by overall differences in mean \% SF.

\begin{figure}[htbp]
    \centering
    \caption{The frequency of different strategies conditional on round parameters: participants with \% SP-U $<$ 75\% vs.\ \% SP-U $\ge$ 75\%}
    \includegraphics[width=0.85\textwidth]{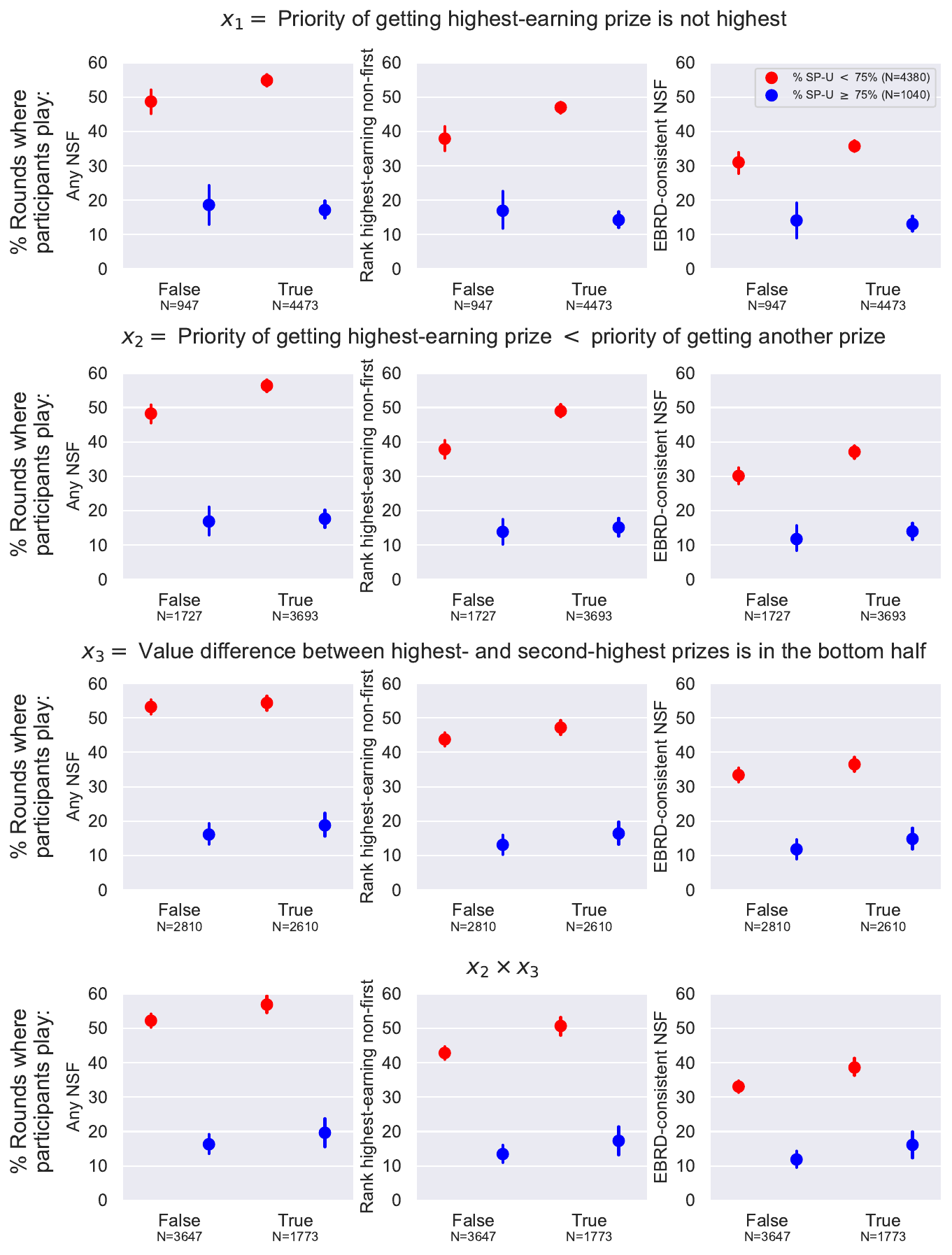}
    \label{fig:nsf-round-level-patterns-by-sp-u}
    \begin{minipage}{\textwidth}  \footnotesize
    \textbf{Notes:}
    See under \autoref{fig:nsf-round-level-patterns}, except that instead of comparing the five treatments, each panel compares two sub-samples of participants with \% SP-U below and above 75\%. 
    \end{minipage}
\end{figure}

\autoref{fig:nsf-round-level-patterns-by-sp-u} shows that the weak patterns we find may be somewhat stronger for participants with \% SP-U $<$ 75\%, and are non-distinguishable from zero for participants with \% SP-U $\ge$ 75\%.

Next, we investigate whether ranking behavior changes over the 10 rounds that each participant play. \autoref{tab:first-five-last-five} shows the difference in participants ranking behavior in rounds 1--5 vs.\ rounds 6--10, and also shows the coefficient from an OLS regression of whether or not the participant played SF that round on the round number.
Rates of SF play also seems stable across the 10 rounds, except perhaps in Menu-SP, where there is a slight trend towards more SF play in later rounds, with an average increase of a 1.6\% (SE $=$ 0.5\%) per round in the fraction of participants playing SF in that round.

\newcommand{\mytabhspace}{0em}
\newcommand{\firstFiveHeader}{1--5}
\newcommand{\lastFiveHeader}{6--10}
\newcommand{\myTwoLineCell}[2]{\makecell{{#1} \\ {\footnotesize(#2)}}}

\begin{table}[htbp]
    \begin{center}
    \caption{
    SF play by round number and treatment
    }
    \label{tab:first-five-last-five}
    \bgroup
    \small
    \def\arraystretch{1.5}%
    \begin{tabular}{lcccccccccccccccccccc}
      \toprule
      & 
      & \multicolumn{2}{c}{{Trad-DA}}
      & \hspace{\mytabhspace} 
      & \multicolumn{2}{c}{{Menu-DA}}
      & \hspace{\mytabhspace} 
      & \multicolumn{2}{c}{{Menu-SP}}
      & \hspace{\mytabhspace} 
      & \multicolumn{2}{c}{{Textbook-SP}}
      & \hspace{\mytabhspace} 
      & \multicolumn{2}{c}{{Null}}
      \\ \cmidrule{3-4} \cmidrule{6-7} \cmidrule{9-10} \cmidrule{12-13} \cmidrule{15-16}
      Rounds &  &  \firstFiveHeader & \lastFiveHeader
      &  &  \firstFiveHeader & \lastFiveHeader
      &  &  \firstFiveHeader & \lastFiveHeader
      &  &  \firstFiveHeader & \lastFiveHeader
      &  &  \firstFiveHeader & \lastFiveHeader
      \\ \cmidrule{3-4} \cmidrule{6-7} \cmidrule{9-10} \cmidrule{12-13} \cmidrule{15-16}
      \% SF
      & & \myTwoLineCell{ 57 }{3} & \myTwoLineCell{ 56 }{3} 
      & & \myTwoLineCell{ 49 }{3} & \myTwoLineCell{ 50 }{4} 
      & & \myTwoLineCell{ 56 }{4} & \myTwoLineCell{ 63 }{4} 
      & & \myTwoLineCell{ 54 }{3} & \myTwoLineCell{ 52 }{4} 
      & & \myTwoLineCell{ 47 }{3} & \myTwoLineCell{ 48 }{3}
      \\[1.5em]
      \% All SF 
      & & \myTwoLineCell{ 26 }{4} & \myTwoLineCell{ 26 }{4} 
      & & \myTwoLineCell{ 22 }{4} & \myTwoLineCell{ 28 }{4} 
      & & \myTwoLineCell{ 28 }{4} & \myTwoLineCell{ 36 }{5} 
      & & \myTwoLineCell{ 23 }{4} & \myTwoLineCell{ 25 }{4} 
      & & \myTwoLineCell{ 20 }{4} & \myTwoLineCell{ 21 }{4}
      \\[1.0em]
      Slope
      & & \multicolumn{2}{c}{0.2 {\footnotesize (0.5)}}
      & & \multicolumn{2}{c}{0.1 {\footnotesize (0.5)}}
      & & \multicolumn{2}{c}{1.6 {\footnotesize (0.5)}}
      & & \multicolumn{2}{c}{0.1 {\footnotesize (0.5)}}
      & & \multicolumn{2}{c}{0.4 {\footnotesize (0.5)}}
      \\ \bottomrule
    \end{tabular}
    \egroup
    \end{center}
    \begin{minipage}{\textwidth}  \footnotesize
    \textbf{Notes:}
    ``\% All SF'': fraction of participants with straightforward play
     in all relevant rounds (1--5 or 6--10).
     ``Slope'': coefficient from a OLS regression of an indicator for SF play (for an observation of participant $\times$ round) on round number.
     Robust standard errors in parentheses.
    \end{minipage}
\end{table}

\subsection{Relationship Between \% SF and (Sub-Measures of) \% SP-U}
\label{app:sp-sf-more-details}

We first investigate the relationship between SP understanding and SF behavior shown in \autoref{fig:sp-sf-detailed-relationship} in more detail. Second, we investigate how the different sub-measures of \% SP-U contribute to the overall relation.

\begin{figure}[htbp]
    \caption{Joint distributions of \% SP-U and \% SF by treatment%
    }
    \label{fig:spu-sf-histogram-by-sp-u-groups}
    
    \vspace{-0.1in}
    
    \includegraphics[width=\textwidth]{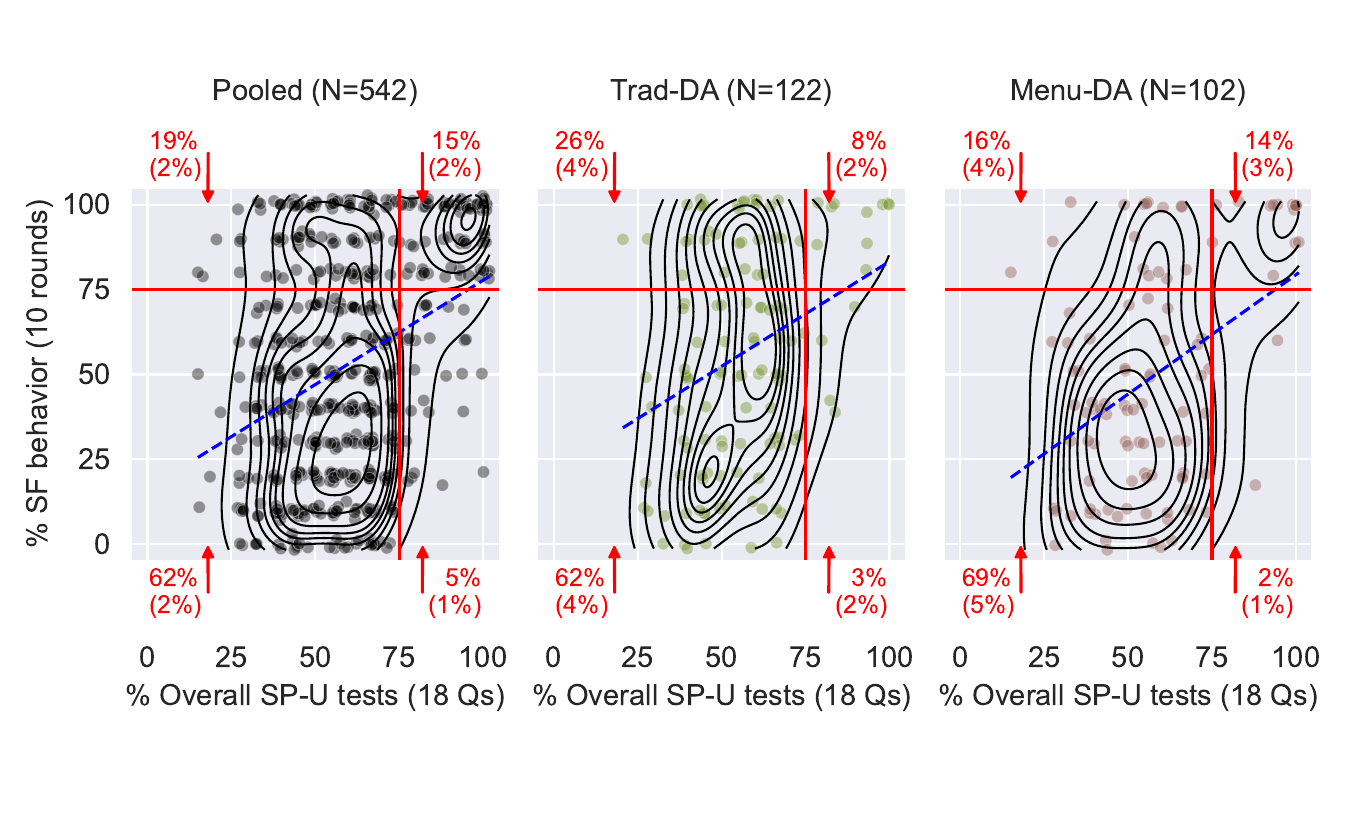}

    \vspace{-0.5in}

    \includegraphics[width=\textwidth]{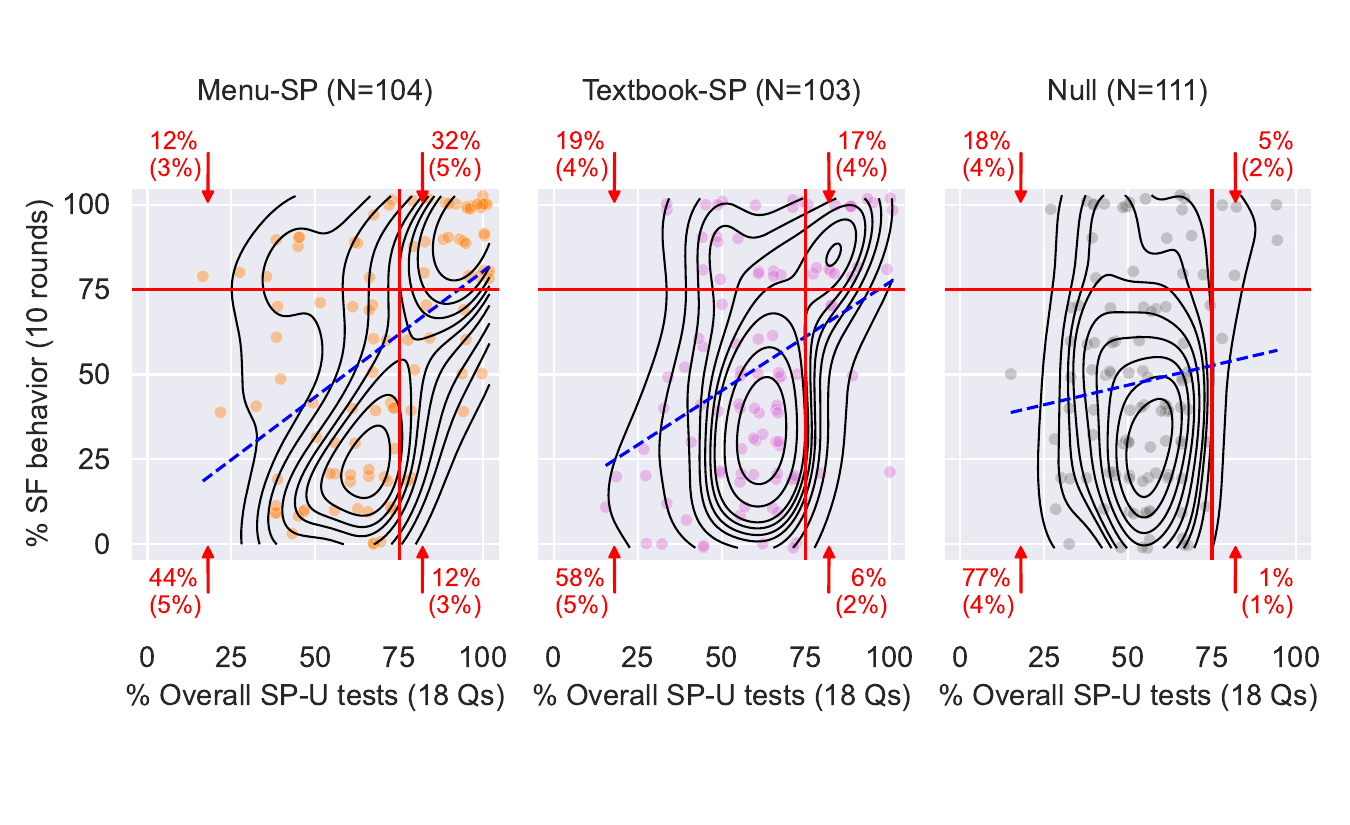}

    \vspace{-0.3in}
    
    \begin{minipage}{\textwidth}  \footnotesize
        \textbf{Notes:}
        Each mini-figure contains a jittered scatter plot and estimated contours smoothing the two-dimensional distribution. 
        Smoothing is performed using a Gaussian kernel with a bandwidth chosen according to Scott's Rule (\cite{scott2015multivariate}).
        \emph{Dashed blue lines}: Predicted \% SF values according to OLS regressions on \% SP-U (see \autoref{tab:basic-regressions} in \autoref{app:univariate-regressions} in the Supplementary Materials for full regression results and more details).
        \emph{Solid Red lines}: \% SP-U = 75\% and \% SF = 75\% lines. 
       \emph{Red numbers}: fractions of participants in the four quadrants separated by the solid red lines (with SEs in parentheses). 
    \end{minipage}
\end{figure}

\autoref{fig:spu-sf-histogram-by-sp-u-groups} shows the (non-conditional) joint distribution of \% SP-U and \% SF.
The figure also displays the fractions of participants falling into four quadrants in the full joint distribution of \% SP-U and \% SF based on the threshold of 75\% for both measures.
The 75\% cutoff of \% SP-U and \% SF is motivated by the 
location of the modes %
in \autoref{fig:sp-u-hist} and \autoref{fig:sf-histogram-by-treatment}, and %
of the jump in \autoref{fig:sp-sf-detailed-relationship}.

We make two observations regarding \autoref{fig:spu-sf-histogram-by-sp-u-groups}, which add to the findings in \autoref{fig:sp-sf-detailed-relationship}.
First, 32\% (SE $=$ 5\%) of participants in Menu-SP are in the upper right quadrant of high SP-understanding and SF rates, while in all other treatments this fraction is 17\% (SE $=$ 4\%) or less. Thus, Menu-SP is most effective in moving participants into the upper quadrant of the \% SP-U, \% SF distribution.
Second, the figure  suggests a \emph{two}-dimensional bimodality of \% SP-U and \% SF in Menu-SP  (and perhaps also in Menu-DA and Textbook-SP, albeit a weaker one).

Next, we investigate how the different sub-measures of \% SP-U contribute to its overall relation with \% SF shown in \autoref{fig:sp-sf-detailed-relationship}.

\begin{figure}[hbtp]
    \centering
    \caption{Relationship between \% SF and the Abstract and Practical sub-measures of \% SP-U}
    \label{fig:define-maxEarn-sf-detailed-relationship}
    
    \begin{minipage}{0.48\textwidth}
    \includegraphics[width=\textwidth]{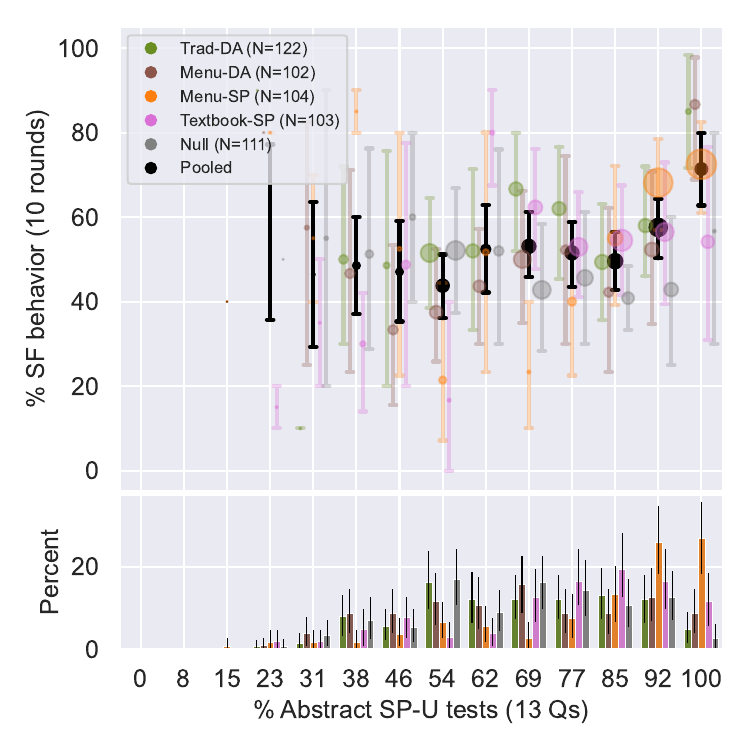}
    \end{minipage}    
    \begin{minipage}{0.48\textwidth}
    \includegraphics[width=\textwidth]{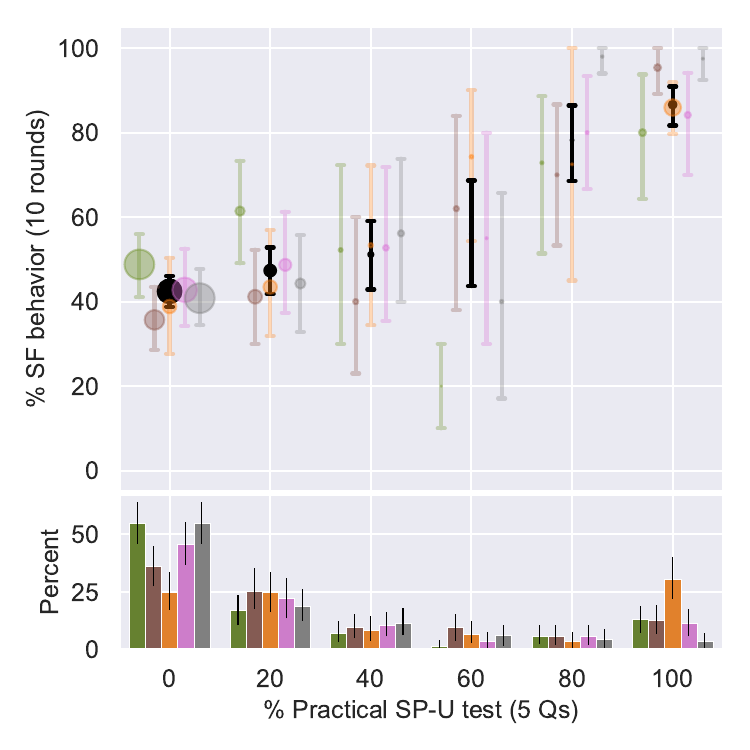}
    \end{minipage}
    
    \begin{minipage}{\textwidth}  \footnotesize
    \textbf{Notes:}
    \emph{Left}: Histogram of Abstract \% SP-U sub-measure at the bottom and the conditional means of \% SF given all possible Abstract scores at the top (see \autoref{fig:sp-sf-detailed-relationship}, which is similar in structure, for more details).
    \emph{Right}: a similar figure with the Practical sub-measure instead of Abstract. 
    \end{minipage}
    
\end{figure}

\autoref{fig:define-maxEarn-sf-detailed-relationship} shows two versions of \autoref{fig:sp-sf-detailed-relationship} focusing on Abstract and Practical \emph{separately}. 
While Abstract is able to explain a small amount of \autoref{fig:sp-sf-detailed-relationship}'s increase in \% SF as \% SP-U increases, Practical seems to explain the full increase.
In contrast to the sharp increase of \% SF in the high \% SP-U end in \autoref{fig:sp-sf-detailed-relationship}, the \% SF increase as a function of Practical alone is more gradual. 
The \% SF estimates at the mid Practical range are noisy and statistically indistinguishable from \% SF rates of low-end Practical scores.

Finally, we investigate the extent to which Abstract and Practical are \emph{jointly} required to explain the increase in \% SF rates. \autoref{tab:sf-sp123-joint-separate} summarizes OLS regressions showing the average relation between \% SF, Abstract and Practical, and total \% SP-U.
(For a breakdown of Abstract into its three sub-measures AbstractHard, AbstractMedium, and AbstractEasy, see \autoref{app:sp-u-sf-more} in the Supplementary Materials.)
The regression results suggest that Practical is much more strongly related to \% SF than Abstract or its sub-measures.
They also suggest that the relation between Abstract alone and \% SF is mostly due to the correlation between Abstract and Practical, 
since the coefficient on Abstract becomes close to zero when adding Practical as another independent variable. These findings are robust to the inclusion of control variables and treatment indicators.

\begin{table}[hbtp]
    \centering
    Dependent variable: \% SF
    {
\def\sym#1{\ifmmode^{#1}\else\(^{#1}\)\fi}
\setlength\tabcolsep{2pt}
\begin{tabular}{@{\extracolsep{2pt}}l*{8}{c}@{}}
\hline\hline

 & (1) & (2) & (3) & (4) & (5) & (6) & (7) & (8) \\
\hline
\% Abstract & 0.26 & 0.05 &  &  & 0.11 & $-$0.00 &  &  \\
 & (0.07) & (0.10) &  &  & (0.07) & (0.09) &  &  \\
\% Practical &  &  & 0.43 & 0.41 & 0.41 & 0.41 &  &  \\
 &  &  & (0.03) & (0.04) & (0.03) & (0.04) &  &  \\
\% SP-U &  &  &  &  &  &  & 0.62 & 0.60 \\
 &  &  &  &  &  &  & (0.06) & (0.10) \\
Controls &  & X &  & X &  & X &  & X \\
Treatment &  & X &  & X &  & X &  & X \\

\hline
R² & 0.03 & 0.31 & 0.22 & 0.45 & 0.22 & 0.45 & 0.13 & 0.37 \\
N & 542 & 542 & 542 & 542 & 542 & 542 & 542 & 542 \\
\hline\hline
\end{tabular}
}

    \caption{Relation of Abstract and Practical \% SP-U sub-measures with SF behavior}
    \label{tab:sf-sp123-joint-separate}

    \begin{minipage}{\textwidth}  \footnotesize
    \textbf{Notes:}
    Coefficients from OLS regressions of \% SF on \% SP-U, including a separation of \% SP-U into its sub-measures. Each regression is shown without and with including a full set of controls in the regression, described in \autoref{app:demographics} in the Supplementary Materials. Treatment indicators are also included in the controlled regressions. Robust standard errors in parentheses.
    \end{minipage}
\end{table}

\begin{figure}[hbtp]
    \centering
    \caption{\% SF rates and fractions of participants at different combinations of Abstract and Practical}
    \label{fig:sf_by_def_maxEarn}
    \includegraphics[width=\textwidth]{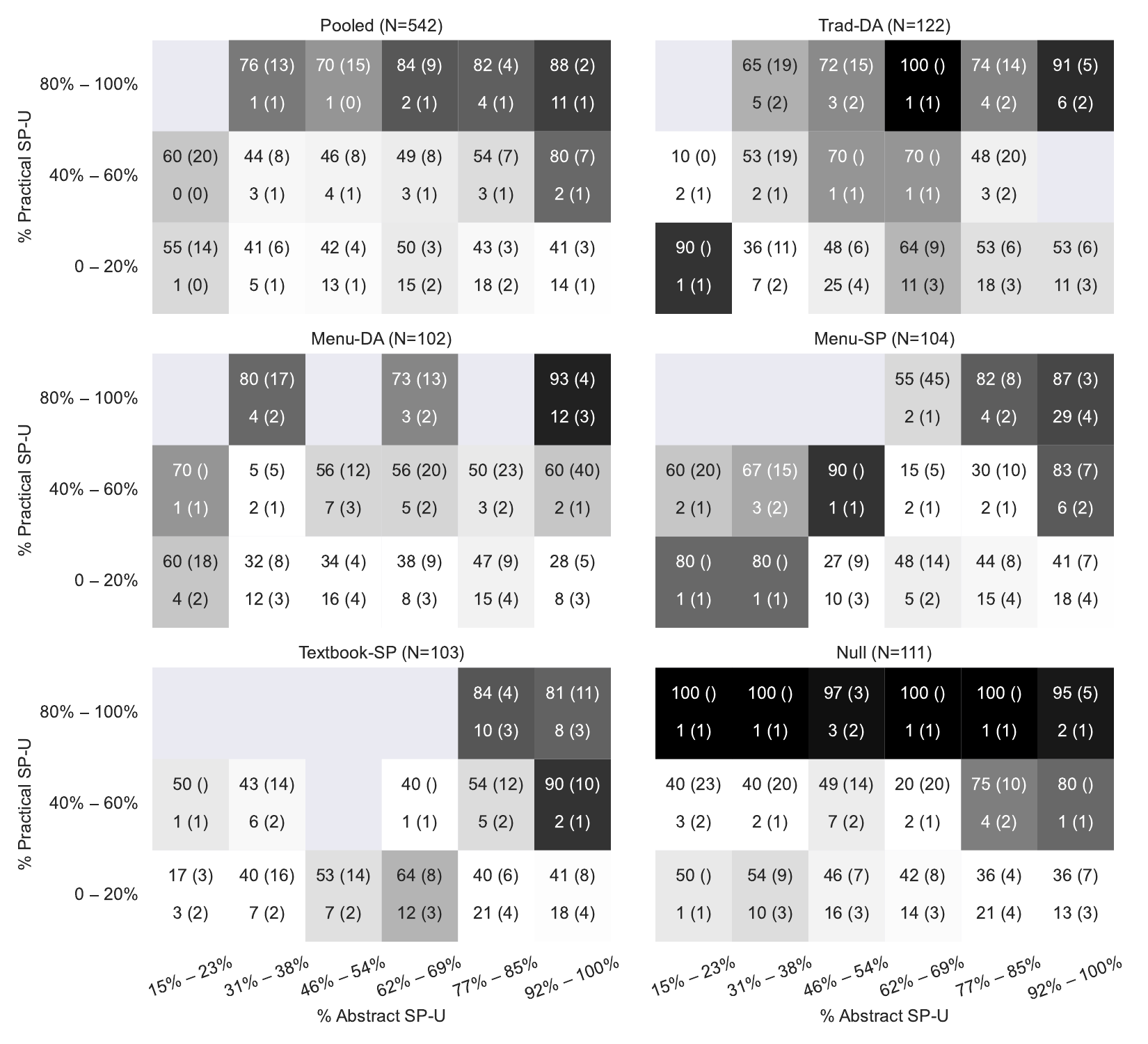}
    \begin{minipage}{\textwidth}  \footnotesize
    \textbf{Notes:}
    In each cell, the top row indicates \% SF (SE) and the bottom row indicates \% participants at this cell, out of the whole panel (SE). A darker cell color corresponds to a higher \% SF rate. Each panel focuses on a different treatment or on the pooled sample (top left). Abstract scores are presented in 6 bins, each bin including 2 adjacent scores out of the total 12 scores which have non-zero numbers of participants. Similarly, Practical scores are presented in 3 bins summarizing the 6 possible scores in this sub-measure. Standard errors are shown in parentheses, where no SE at the top row indicates a single-observation cell.
    \end{minipage}
\end{figure}

Despite that Abstract has an average insignificant relation to \% SF when considered jointly with Practical, it may contribute to the overall relation in more specific cases. \autoref{fig:sf_by_def_maxEarn} non-parametrically investigates \% SF rates at different combinations of Abstract and Practical. Pooling all treatments on the top-left panel, the figure suggests that Abstract is able to separate between low-SF and high-SF participants at a mid level of Practical, between the two modes of its bimodal distribution. For Practical between 40\% and 60\%, the average  \% SF difference between Abstract $<$ 90\% and Abstract $\geq$ 90\% is 31\% (SE $=$ 8\%). Adding a full set of controls, this difference becomes 29\% (SE $=$ 11\%).
Cautiously note, however, that this pattern is estimated on a small sample of 82 participants who have these mid-level Practical scores, and there is insufficient statistical power to test whether it persists within treatments.

\subsection{Prolific vs.\ Cornell Samples, Cognitive Score, Attention Score, and their Mediating Effects on Main Results}
\label{app:cognitive}

Among possible variables mediating our treatment effects, education, cognitive ability, attention, and ability to focus in a physical lab (vs.\ remote participation) seem highly important to effectively learning from any description, since they are related to the ability to learn, process, and comprehend the complicated information conveyed in the experiment.%

We use three key measures to study these effects: an indicator for whether a participant belongs to the Prolific or Cornell sub-sample, a cognitive score measure elicited at the end of the experiment, and an attention score elicited in two attention-check questions planted within the experiment. In addition, and according to our pre-registration, we test whether participants with extreme experiment duration---who take very little or very long time to complete the experiment compared to their treatment distribution---drive our main results. See \autoref{app:cognitive-attention} in the Supplementary Materials for texts of our cognitive questions and attention-check questions, and for the distributions of cognitive and attention scores in our sample.

\begin{figure}[htbp]
\centering
\caption{Main results by top vs.\ bottom cognitive scores}
\label{fig:main-results-by-cognitive}

\begin{minipage}{0.49\textwidth}
    \begin{center}
        \subcaption{Mean outcome variables by cognitive score}
        \label{fig:main-results-by-cognitive-means}        \includegraphics[width=\textwidth]{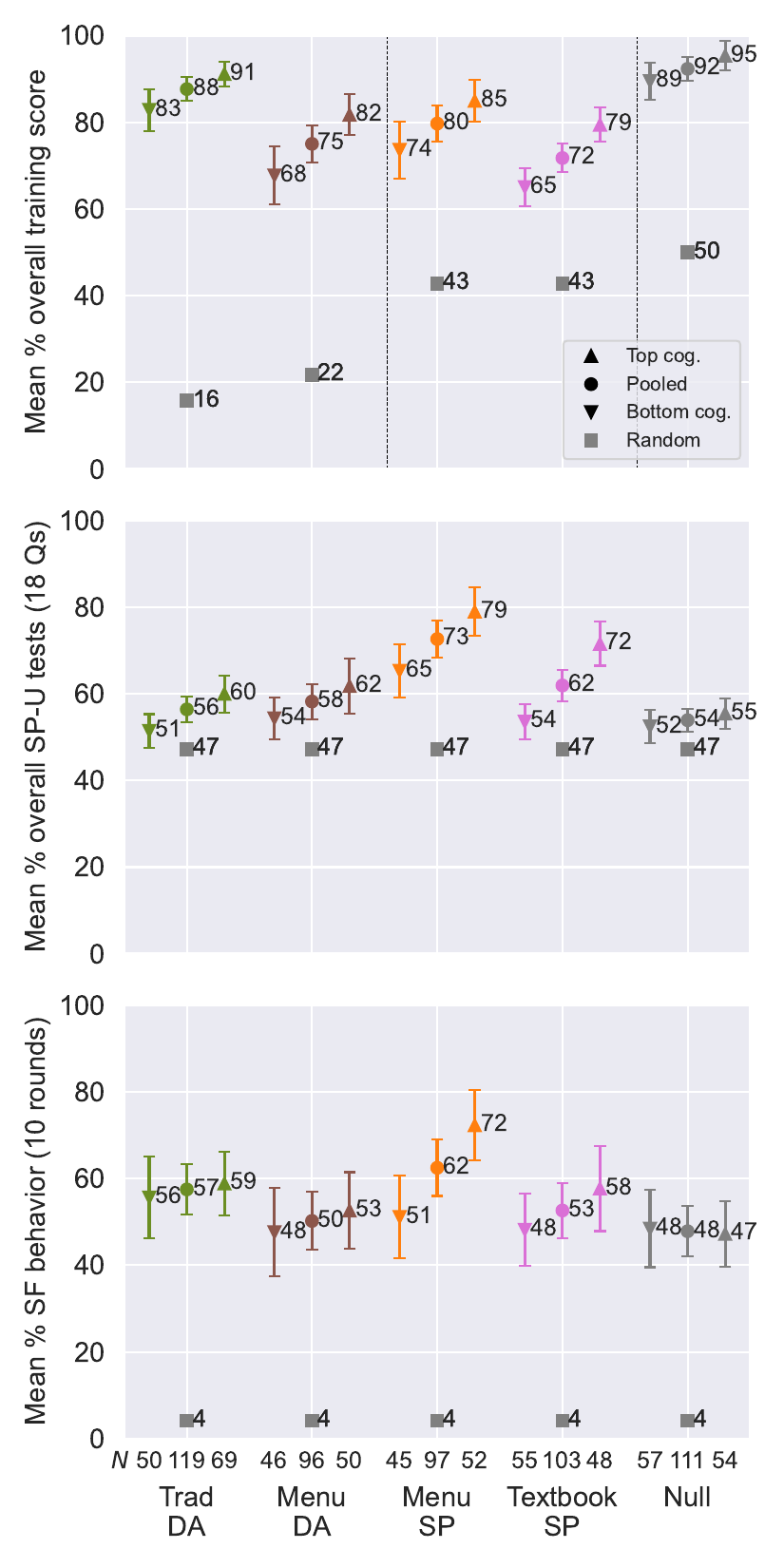}
    \end{center}
\end{minipage}
\hfill
\begin{minipage}{0.49\textwidth}
    \begin{center}
        \subcaption{Relation between \% SF and \% SP-U by cognitive score}
        \label{fig:main-results-by-cognitive-sf-sp}
        \includegraphics[width=\textwidth]{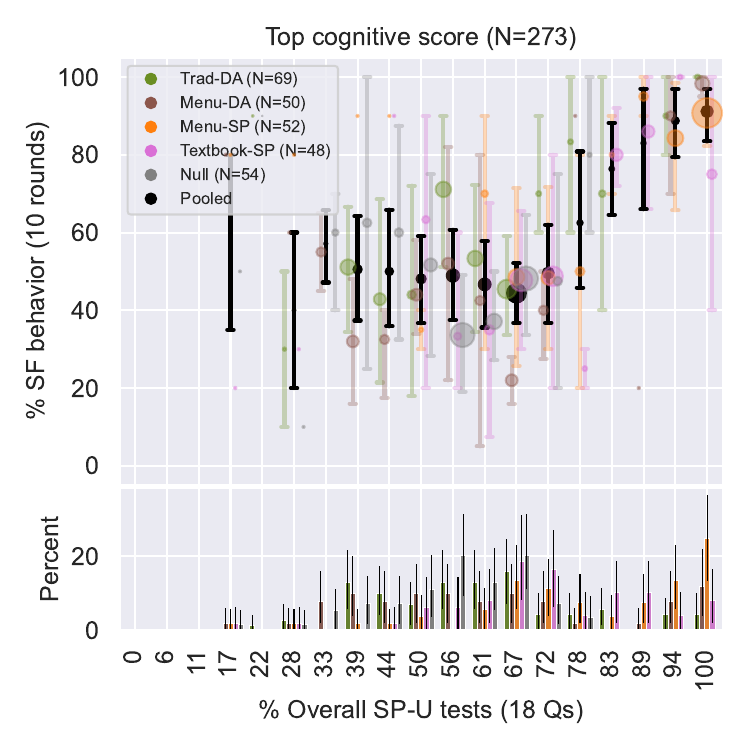}\\
        \includegraphics[width=\textwidth]{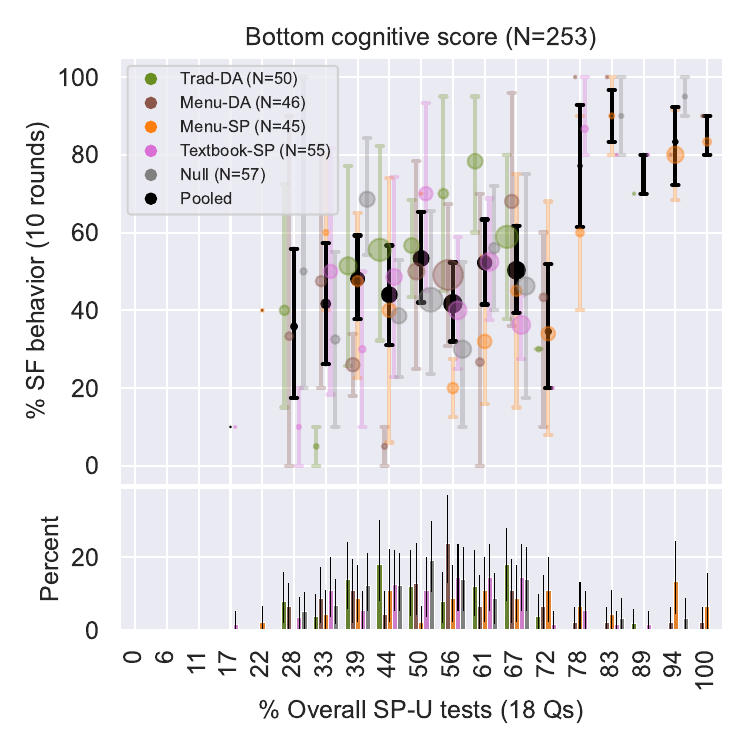}
    \end{center}
\end{minipage}

\begin{minipage}{\textwidth}  \footnotesize
        \textbf{Notes:}               
        See under \autoref{fig:main-results-by-sample}; the sample split is according to top (3--4) vs.\ bottom (0--2) scores in the cognitive-ability test conducted at the end of the experiment.

    \end{minipage}
\end{figure}

\begin{figure}[htbp]
\centering
\caption{Main results by top vs.\ bottom attention scores}
\label{fig:main-results-by-attention}

\begin{minipage}{0.49\textwidth}
    \begin{center}
        \subcaption{Mean outcome variables by attention score}
        \label{fig:main-results-by-attention-means}
        \includegraphics[width=\textwidth]{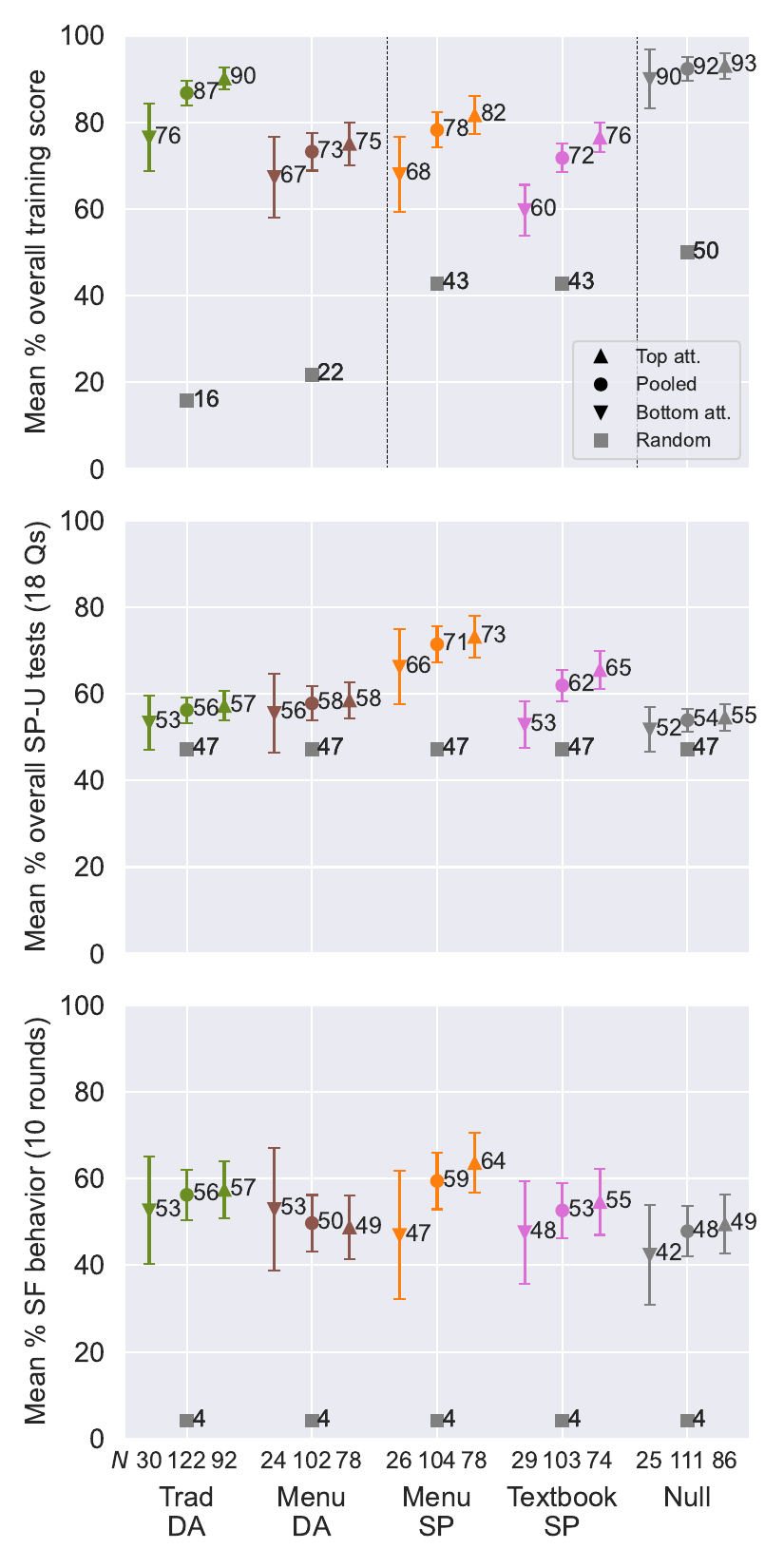}
    \end{center}
\end{minipage}
\hfill
\begin{minipage}{0.49\textwidth}
    \begin{center}
        \subcaption{Relation between \% SF and \% SP-U by attention score}
        \label{fig:main-results-by-attention-sf-sp}
        \includegraphics[width=\textwidth]{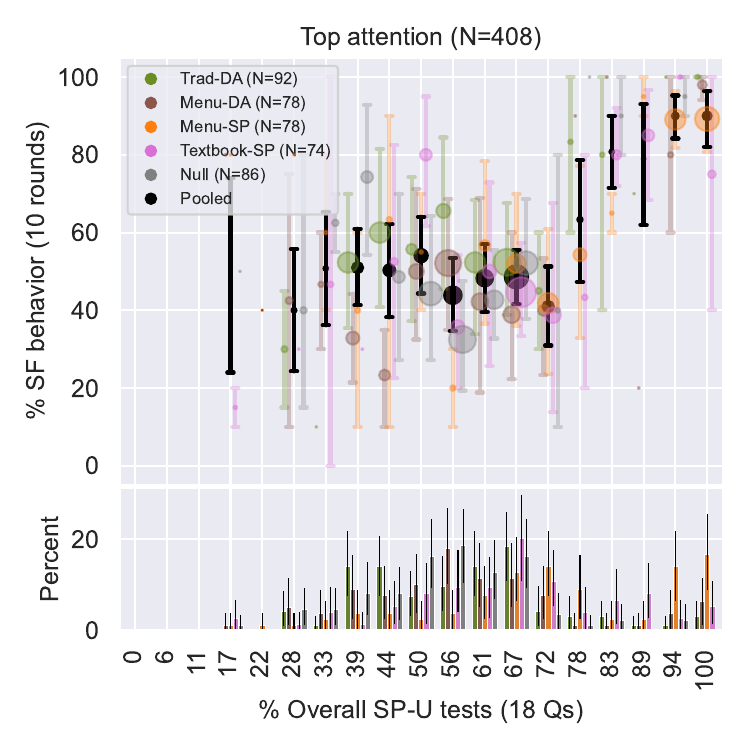}\\
        \includegraphics[width=\textwidth]{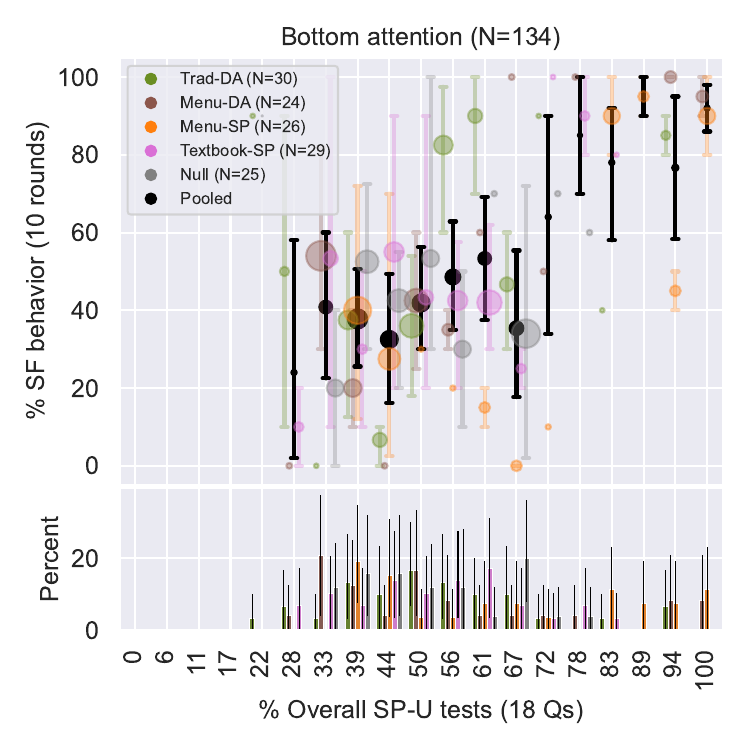}
    \end{center}
\end{minipage}

\begin{minipage}{\textwidth}  \footnotesize
        \textbf{Notes:}               
        See under \autoref{fig:main-results-by-sample}; the sample split is according to top (2) vs.\ bottom (0--1) scores in the attention-check questions planted within the experiment.
    \end{minipage}
\end{figure}

\autoref{fig:main-results-by-sample} in \autoref{sec:results-samples-and-controls} tests the effects of these measures on our main findings, by reporting differences between the Prolific and Cornell samples.  
\autoref{fig:main-results-by-cognitive} similarly adds a comparison of the bottom (roughly) half of cognitive scores (0--2) vs.\ the top half (3--4), and \autoref{fig:main-results-by-attention} adds a comparison of bottom attention scores (0--1) vs.\ a top (perfect) attention score (2). 
Both of these comparisons yield similar insights to the Prolific versus Cornell comparison from \autoref{sec:results-samples-and-controls}.

\begin{figure}[htbp]
\centering
\caption{Main results by non-extreme vs.\ extreme completion time}
\label{fig:main-results-by-duration}

\begin{minipage}{0.49\textwidth}
    \begin{center}
        \subcaption{Mean outcome variables by completion time}
        \label{fig:main-results-by-completion-means}
        \includegraphics[width=\textwidth]{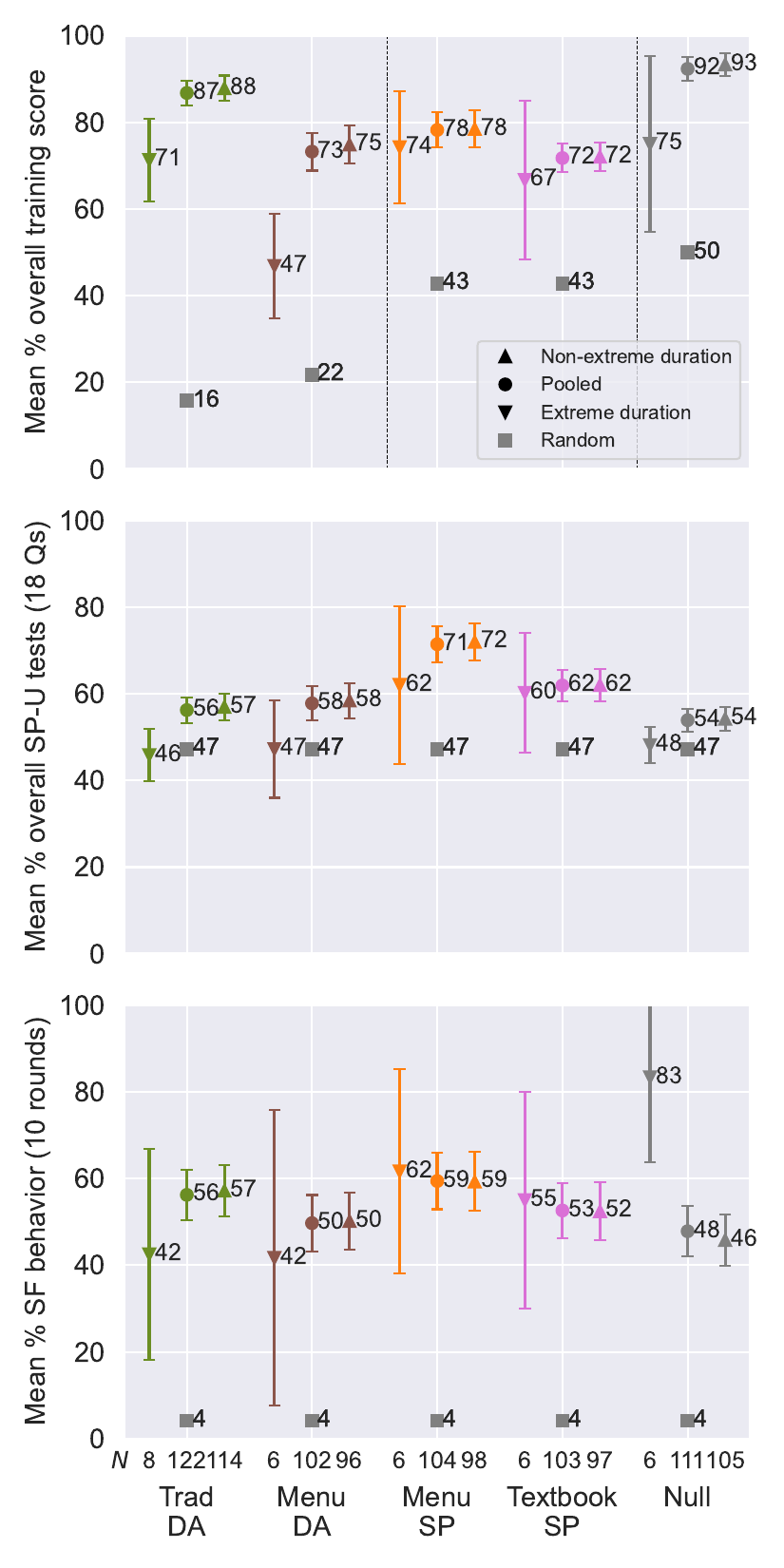}
    \end{center}
\end{minipage}
\hfill
\begin{minipage}{0.49\textwidth}
    \begin{center}
        \subcaption{Relation between \% SF and \% SP-U by completion time}
        \label{fig:main-results-by-completion-sf-sp}
        \includegraphics[width=\textwidth]{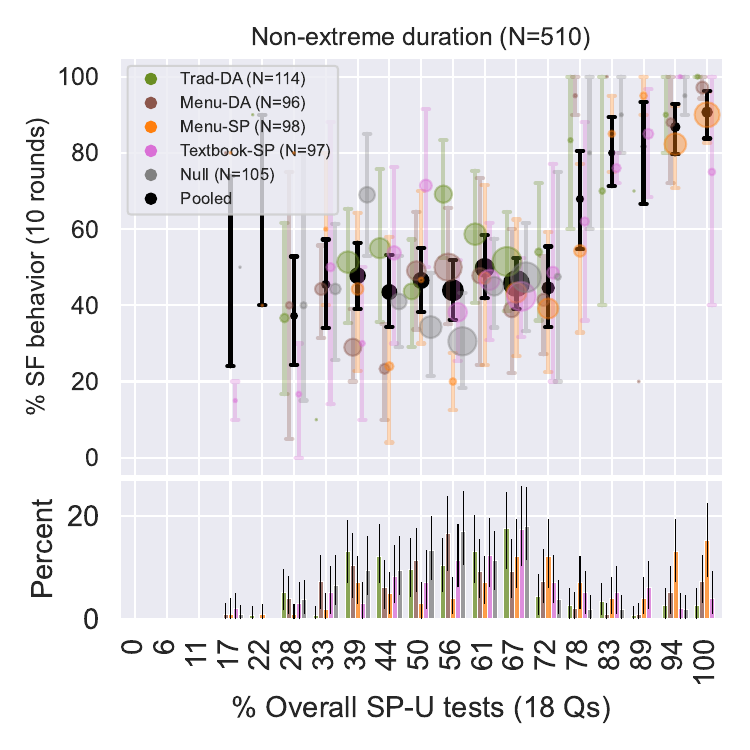}\\
        \includegraphics[width=\textwidth]{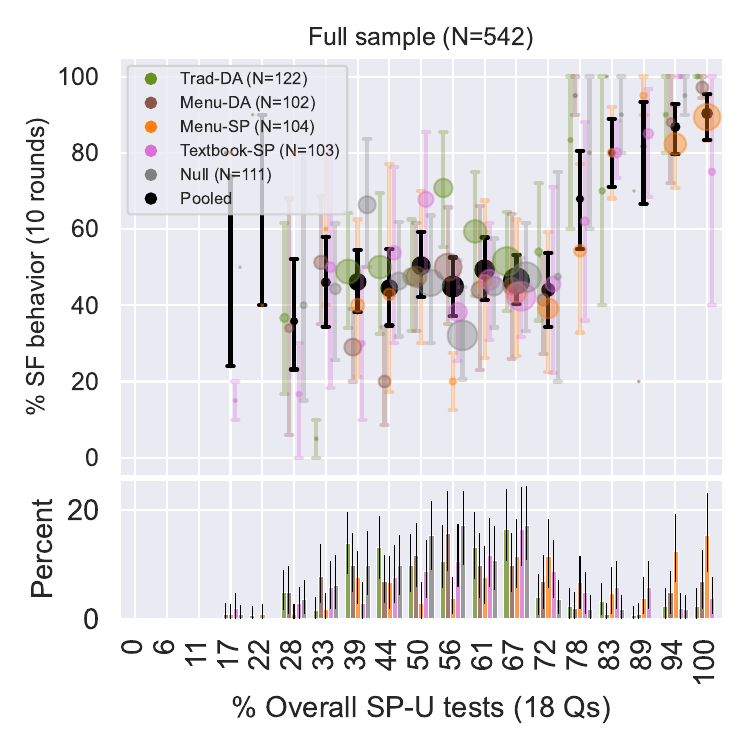}
    \end{center}
\end{minipage}

\begin{minipage}{\textwidth}  \footnotesize
        \textbf{Notes:}               
        See under \autoref{fig:main-results-by-sample}; on the left, the sample split is according to non-extreme vs.\ extreme experiment duration (defined as being $\leq$ the 2.5 percentile or $\geq$ the 97.5 percentile of the per-treatment duration distribution); on the right, the sub-sample with non-extreme duration is compared to the full sample, since the extreme-duration sub-sample is too small for a meaningful comparison.
    \end{minipage}
\end{figure}

Finally, \autoref{fig:main-results-by-duration} reports a comparison of participants with non-extreme per-treatment completion time to participants with an extreme completion time. According to our pre-registration, extreme completion time is defined as belonging to the top or bottom 2.5 percentages of the per-treatment duration distribution. The per-treatment 2.5-percentile, median and 97.5-percentile duration are as follows: (31, 48, 95) minutes in Trad-DA, (30, 53, 101) minutes in Menu-DA, (19, 40, 75) minutes in Menu-SP, (22, 39, 78) minutes in Textbook-SP, and (21, 35, 68) minutes in Null. Our main results (among the full sample) remain largely unaffected when excluding extreme completion times.

\subsection{Robustness of Main Results to Adding Controls}
\label{app:treatment-comparison-controls}

We test whether our four main findings are robust to adding a rich set of controls as explanatory variables.

We first find that adding controls does not change the treatment-means of our three main outcome variables by much. \autoref{tab:outcome-vars-with-controls} shows the treatment effects on the three outcome variables \% TR, \% SP-U and \% SF using OLS regressions on treatment indicators. The columns without controls repeat the mean values shown in Figures~\ref{fig:mean-tr-by-treatment-actually-the-mean},~\ref{fig:sp-u-means}, and~\ref{fig:mean-sf-by-treatment-actually-the-mean} in terms of difference from the Null treatment mean. 
The control variables include (1) a set of demographic indicators, (2) indicators for every possible cognitive and attention scores (described in \autoref{app:cognitive-attention} in the Supplementary Materials) and for every possible score in the Null training questions common to all treatments, and (3) date indicators for all days in which session of the experiment took place. (1) and (2) include a category for missing values in case no response was recorded for a question.

The demographic characteristics (described fully in \autoref{app:demographics} in the Supplementary Materials) include state of residence (not used in regressions), age, number of people in household, number of people in household aged at least 18, gender, race, education, primary education focus, knowledge of the DA mechanism, participation in real-life DA, marital status, employment status, social views, economic views, identification with a political party, vote in the 2020 presidential elections, and combined household income.

\begin{table}[htbp]
    \centering
    {
\def\sym#1{\ifmmode^{#1}\else\(^{#1}\)\fi}
\setlength\tabcolsep{2pt}
\begin{tabular}{@{\extracolsep{2pt}}l*{6}{c}@{}}
\hline\hline

 & \% TR & \% TR & \% SP-U & \% SP-U & \% SF & \% SF \\
\hline
Trad-DA & $-$0.06 & $-$0.08 & 0.02 & 0.00 & 0.08 & 0.08 \\
 & (0.02) & (0.02) & (0.02) & (0.02) & (0.04) & (0.05) \\
Menu-DA & $-$0.19 & $-$0.18 & 0.04 & 0.05 & 0.02 & 0.03 \\
 & (0.03) & (0.02) & (0.02) & (0.03) & (0.04) & (0.05) \\
Menu-SP & $-$0.14 & $-$0.13 & 0.18 & 0.18 & 0.12 & 0.13 \\
 & (0.03) & (0.02) & (0.03) & (0.02) & (0.04) & (0.05) \\
Textbook-SP & $-$0.21 & $-$0.20 & 0.08 & 0.11 & 0.05 & 0.04 \\
 & (0.02) & (0.02) & (0.02) & (0.02) & (0.04) & (0.05) \\
Constant & 0.92 & 0.64 & 0.54 & 0.63 & 0.48 & 0.11 \\
 & (0.01) & (0.14) & (0.01) & (0.11) & (0.03) & (0.24) \\
Controls &  & X &  & X &  & X \\

\hline
R² & 0.16 & 0.60 & 0.10 & 0.51 & 0.02 & 0.31 \\
N & 542 & 542 & 542 & 542 & 542 & 542 \\
\hline\hline
\end{tabular}
}

    \caption{Treatment effects on outcome variables without controls and with controls}
    \label{tab:outcome-vars-with-controls}

    \begin{minipage}{\textwidth}  \footnotesize
    \textbf{Note:}
    Coefficients from OLS regressions of the 3 outcome variables on treatment indicators (with Null treatment the omitted category), both without and with the inclusion of a full set of controls in the regression, described in \autoref{app:demographics} in the Supplementary Materials. Robust standard errors in parentheses.
    \end{minipage}
\end{table}

Second, we find that adding controls does not change the relationship between \% SP-U and \% SF shown in \autoref{fig:sp-sf-detailed-relationship} by much. \autoref{tab:sf-sp-with-controls-within-treatment} shows mean \% SF when conditioning on SP-U $<$ 75\% vs.\ \% SP-U $\ge$ 75\%, within treatments.
Comparing regression results without and with inclusion of controls suggests that the relationship between \% SP-U and \% SF remains consistent across treatments when adding controls. %

\begin{table}[htbp]
    \centering
    \footnotesize
    Dependent variable: \% SF
    {
\def\sym#1{\ifmmode^{#1}\else\(^{#1}\)\fi}
\setlength\tabcolsep{2pt}
\begin{tabular}{@{\extracolsep{2pt}}l*{2}{c}@{}}
\hline\hline

 & (1) & (2) \\
\hline
Trad-DA $\times$ \% SP-U $<$ 75\% & 0.53 & 0.53 \\
 & (0.03) & (0.03) \\
Trad-DA $\times$ \% SP-U $\geq$ 75\% & 0.84 & 0.87 \\
 & (0.06) & (0.08) \\
Menu-DA $\times$ \% SP-U $<$ 75\% & 0.42 & 0.41 \\
 & (0.03) & (0.04) \\
Menu-DA $\times$ \% SP-U $\geq$ 75\% & 0.89 & 0.93 \\
 & (0.05) & (0.08) \\
Menu-SP $\times$ \% SP-U $<$ 75\% & 0.43 & 0.47 \\
 & (0.04) & (0.04) \\
Menu-SP $\times$ \% SP-U $\geq$ 75\% & 0.81 & 0.78 \\
 & (0.03) & (0.05) \\
Textbook-SP $\times$ \% SP-U $<$ 75\% & 0.45 & 0.43 \\
 & (0.04) & (0.04) \\
Textbook-SP $\times$ \% SP-U $\geq$ 75\% & 0.78 & 0.79 \\
 & (0.05) & (0.07) \\
Null $\times$ \% SP-U $<$ 75\% & 0.46 & 0.47 \\
 & (0.03) & (0.03) \\
Null $\times$ \% SP-U $\geq$ 75\% & 0.88 & 0.77 \\
 & (0.06) & (0.15) \\
Controls &  & X \\

\hline
R² & 0.20 & 0.43 \\
N & 542 & 542 \\
\hline\hline
\end{tabular}
}

    \caption{Within-treatment relation between \% SF and \% SP-U without controls and with controls}
    \label{tab:sf-sp-with-controls-within-treatment}

    \begin{minipage}{\textwidth}  \footnotesize
    \textbf{Note:}
    Coefficients from OLS regressions of \% SF on indicators of \% SP-U $<$ 75\% vs.\ \% SP-U $\ge$ 75\%, interacted with all five treatments, both without and with the inclusion of a full set of controls in the regression, described in \autoref{app:demographics} in the Supplementary Materials. Robust standard errors in parentheses.
    \end{minipage}
\end{table}

\clearpage

\setcounter{page}{1} \renewcommand{\thepage}{S\arabic{page}}
\numberwithin{figure}{section}
\numberwithin{table}{section}

\part*{\begin{center}Supplementary Materials\end{center}}

\section{Supplemental Information and Analysis}
\label{app:additional-analysis-secondary}

In this appendix, we present further details, results, and analyses of our data.

\subsection{Comparison of Analysis and Findings With Pre-Registration}
\label{app:prereg}

Our pre-registration can be found at
\url{https://aspredicted.org/7eq7e.pdf}.
Our pre-registration specified our treatments and main outcome variables (\% TR, \% SP-U, and \% SF), our sample size (see \autoref{app:sample-basics}), and our robustness check for outliers in completion time, (defined as bottom or top per-treatment 2.5\%; test conducted in \autoref{app:cognitive}).
Our main pre-registered hypothesis was that different descriptions would convey strategyproofness better or worse, and thus influence \% SP-U.

Across treatments, we mainly hypothesized regarding the different performance of Trad-DA and Menu-DA, and had weaker conjectures regarding the SP Property treatments.\footnote{
The treatment names we use in the paper are different than those we used in the pre-registration: Traditional DA Mechanics (Trad-DA) was named ``Traditional Mechanics'' (``Trad-Mech''), Menu DA Mechanics (Menu-DA) was ``Menu Mechanics'' (``Menu-Mech''), Menu SP Property (Menu-SP) was ``Menu Property'' (``Menu-Prop'') and Textbook SP Property (Textbook-SP) was ``Traditional Property'' (``Trad-Prop''). The Null treatment was named the same.
}
First, we hypothesized that Menu-DA would make it more complicated to understand DA mechanics than Trad-DA (which would be indicated by lower \% TR in Menu-DA on comparative questions). 
Second, we hypothesized that Menu-DA would outperform Trad-DA in \% SP-U, since in Menu-DA strategyproofness follows from a one-sentence proof, but the same is not true for Trad-DA.
Third, we hypothesized that the correlation between \% TR and \% SP-U in Menu-DA would be positive and stronger than that correlation for Trad-DA (and specifically, that conditional on a high percentile of \% TR, Menu-DA would outperform Trad-DA in \% SP-U).

While our results are organized differently than these hypotheses, they follow the analysis plan implied by our pre-registration by considering the mean levels of our main outcome variables. 
Our first three main results test and expand on the first two hypotheses, by comparing mean levels of \% TR and \% SP-U not only across Trad-DA and Menu-DA, but also across all other treatments.
In a secondary analysis, reported in \autoref{sec:results-samples-and-controls} and expanded in \autoref{app:tr-sp} in the Supplementary Materials, we test the third hypothesis.
Our fourth main result, on the correlation between \% SP-U and \% SF, is ex-post: We did not have clear hypotheses on how \% SF would compare across treatments and on whether \% SP-U and \% SF behavior would correlate, since this requires assumptions on participants' preferences.

In contrast to our pre-registered hypotheses, we find little difference in our main outcome measures between Trad-DA and Menu-DA, other than a difference in \% TR. 
However, we see significant differences between Menu-SP and Textbook-SP (and the other treatments); most significantly, Menu-SP leads to the highest rates of \% SP-U in our experiment.
We also find little difference in the correlations between \% TR and \% SP-U across Trad-DA and Menu-DA, and a strong, non-treatment-specific relation between \% SP-U and \% SF.
Our pre-registered robustness test verifies that completion-time outliers do not drive our main results.

\subsection{Demographics}
\label{app:demographics}

The last screen of the experiment elicits demographic characteristics of our participants.
We now describe the categories used and their mean levels across treatments.

The set of demographic characteristics and their categories are as follows: 
\begin{itemize}
    \item State of residence (a choice of a US state)
    \item Age ($[18,30)$, $[30,40)$, $[40,50)$, $[50,60)$, $[60,70)$, 70 or above; elicited as birth year and converted into age groups in the analysis).
    \item Number of people in household (1, 2, 3, 4, 5, 6 or above; elicited as any number and converted into one of these groups in the analysis).
    \item Number of people in household aged at least 18 (same categories).
    \item Gender (male, female, non-binary, other, prefer not to answer).
    \item Race (White and/or European-American, Black and/or African-American, Native American and/or First Nations, Hispanic and/or Latino, Asian and/or Pacific Islander, Middle-Eastern and/or North African, Multiracial and/or Mixed, other, prefer not to answer).
    \item Education (middle school or less, some high school, high school diploma, GED (HS equivalent), some college without finishing, two-year college degree / Associate degree / A.A. / A.S., four-year college degree / B.A. / B.S., some graduate school, Master’s degree (MA / MS / MBA / MFA / MDiv), advanced degree (PhD / MD / JD)).
    \item Primary education focus (humanities, social sciences, natural sciences or math, applied science or engineering, none).
    \item Knowledge of the DA mechanism (never heard, heard but don't remember, have vague knowledge, know some details, familiar with it, know it well).
    \item Participation in real-life DA application in the past or in the future (no participation, past participation, planned future participation).
    \item Marital status (married, widowed, separated, divorced, single, living with a significant other).
    \item Employment status (working (besides Prolific, if applicable), unemployed, retired, stay-at-home parent, student, other).
    \item Social views (very liberal, liberal, slightly liberal, moderate, slightly conservative, conservative, very conservative, other).
    \item Economic views (same categories).
    \item Identification with a political party (Republican, Democrat, Independent, other, none of the above).
    \item Vote in the 2020 presidential elections (Joe Biden, Donald Trump, other, did not vote).
    \item Combined household income ($[0,\$20\text{k})$, $[\$20\text{k},\$40\text{k})$, $[\$40\text{k},\$60\text{k})$, $[\$60\text{k},\$80\text{k})$, $[\$80\text{k},\$100\text{k})$, $[\$100\text{k},\$150\text{k})$, $[\$150\text{k},\$200\text{k})$, \$200k or above).
\end{itemize}

\begin{figure}[htbp]
\centering
\caption{Distribution of demographic characteristics in our sample (1/4)}
\label{fig:demographics1}
\includegraphics[width=\figscale\textwidth{}]{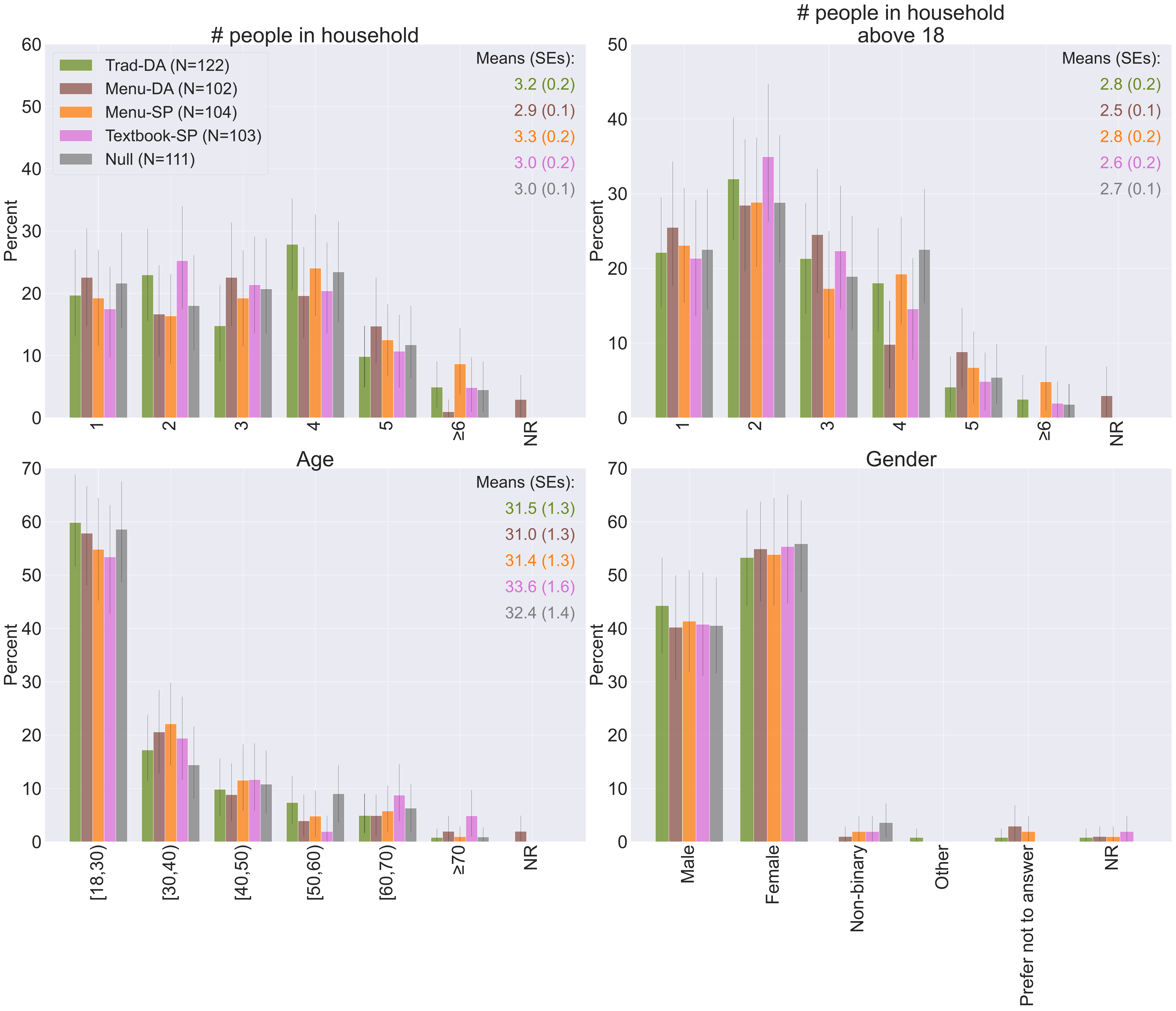}
\begin{minipage}{\textwidth}  \footnotesize
        \textbf{Notes:}
        ``NR'': No response. Error bars: 95\% confidence intervals.
    \end{minipage}
\end{figure}

\begin{figure}[htbp]
\centering
\caption{Distribution of demographic characteristics in our sample (2/4)}
\label{fig:demographics2}
\includegraphics[width=\figscale\textwidth{}]{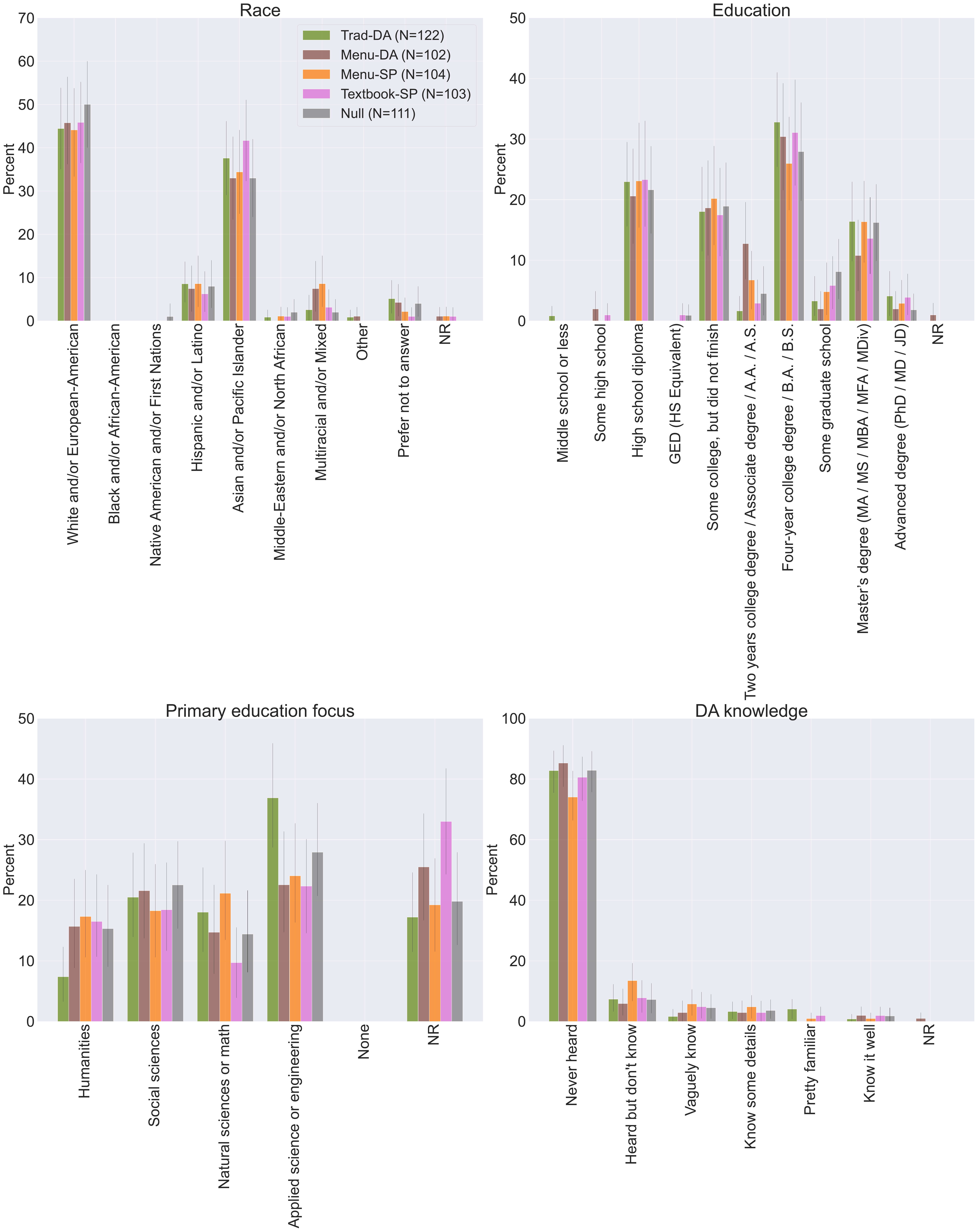}
\begin{minipage}{\textwidth}  \footnotesize
        \textbf{Notes:}
        See under \autoref{fig:demographics1}.
    \end{minipage}
\end{figure}

\begin{figure}[htbp]
\centering
\caption{Distribution of demographic characteristics in our sample (3/4)}
\label{fig:demographics3}
\includegraphics[width=\figscale\textwidth{}]{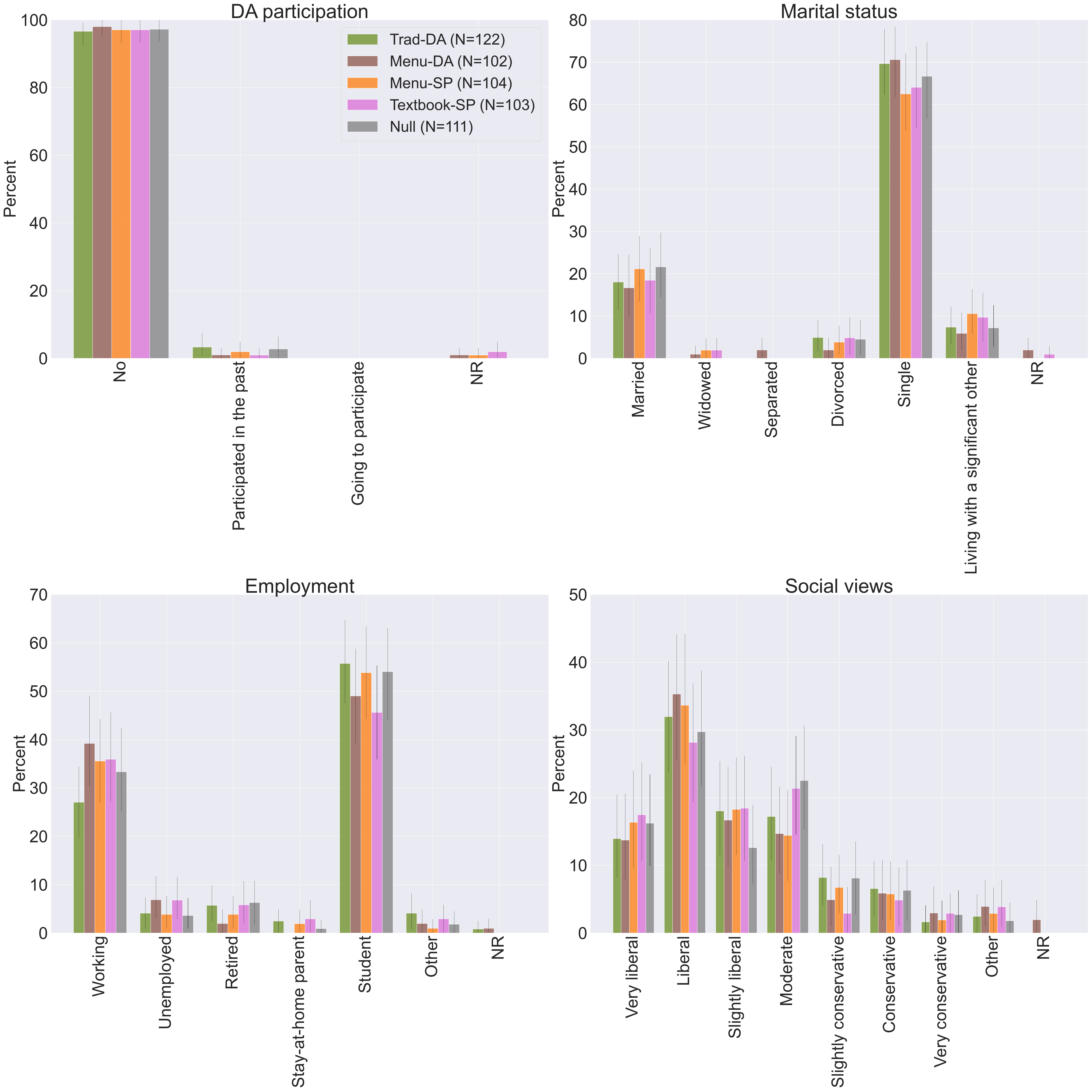}
\begin{minipage}{\textwidth}  \footnotesize
        \textbf{Notes:}
        See under \autoref{fig:demographics1}.
    \end{minipage}
\end{figure}

\begin{figure}[htbp]
\centering
\caption{Distribution of demographic characteristics in our sample (4/4)}
\label{fig:demographics4}
\includegraphics[width=\figscale\textwidth{}]{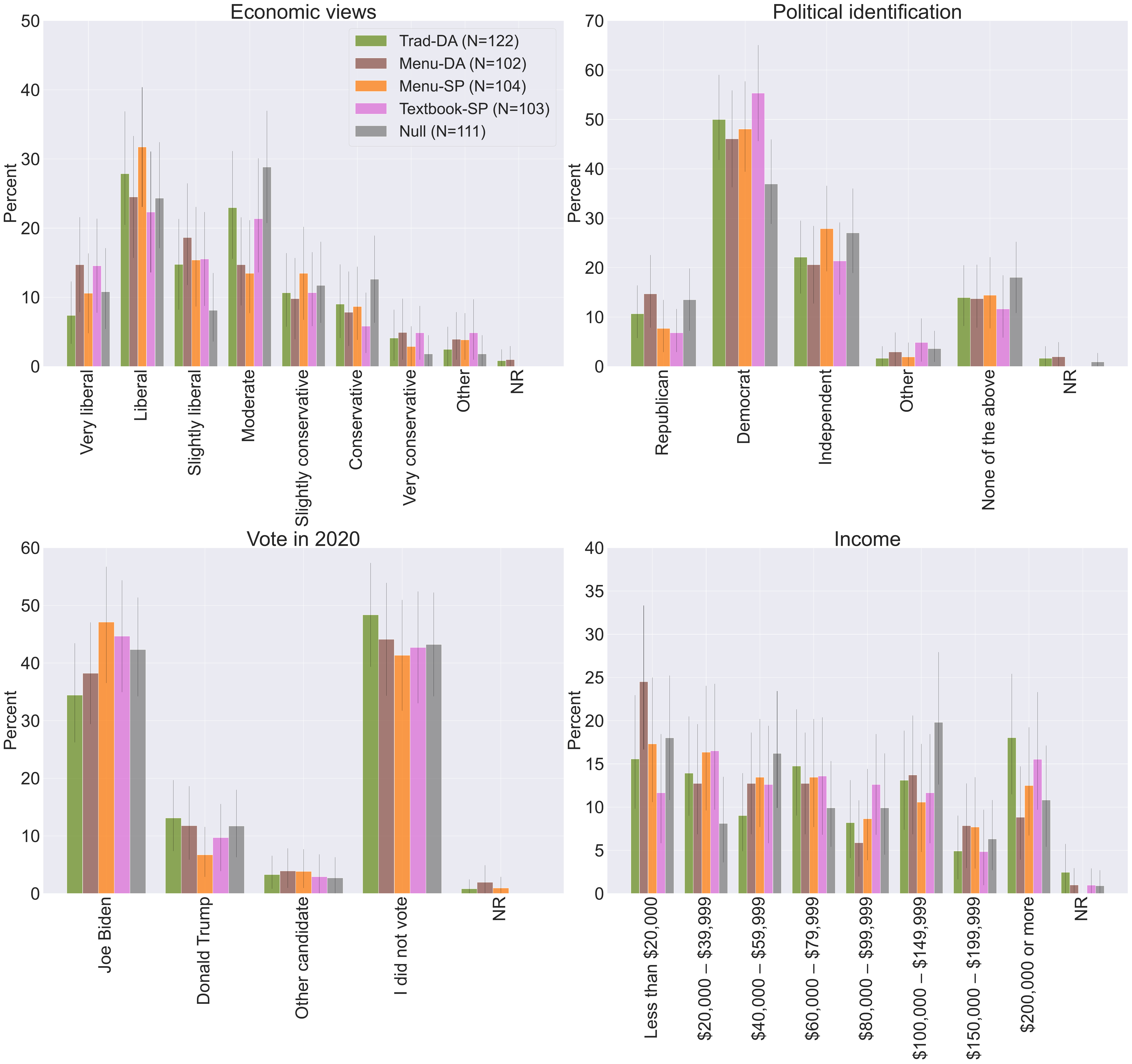}
\begin{minipage}{\textwidth}  \footnotesize
        \textbf{Notes:}
        See under \autoref{fig:demographics1}.
    \end{minipage}
\end{figure}

Figures 
\ref{fig:demographics1}, 
\ref{fig:demographics2}, 
\ref{fig:demographics3}, and 
\ref{fig:demographics4}
show the distribution of these characteristics in our sample across treatments (except for US state of residence). 

\clearpage

\subsection{Detailed Performance in Training Questions}
\label{app:tr-more-details}

We present the details of participants' performance in training questions. We first show the overall training-score distributions, and then show mean performance at the question level within treatments.

\subsubsection{Null Training Common to All Participants}

\begin{figure}[htbp]
    \centering
    \caption{Null training questions common to all participants}
    \includegraphics[width=\textwidth]{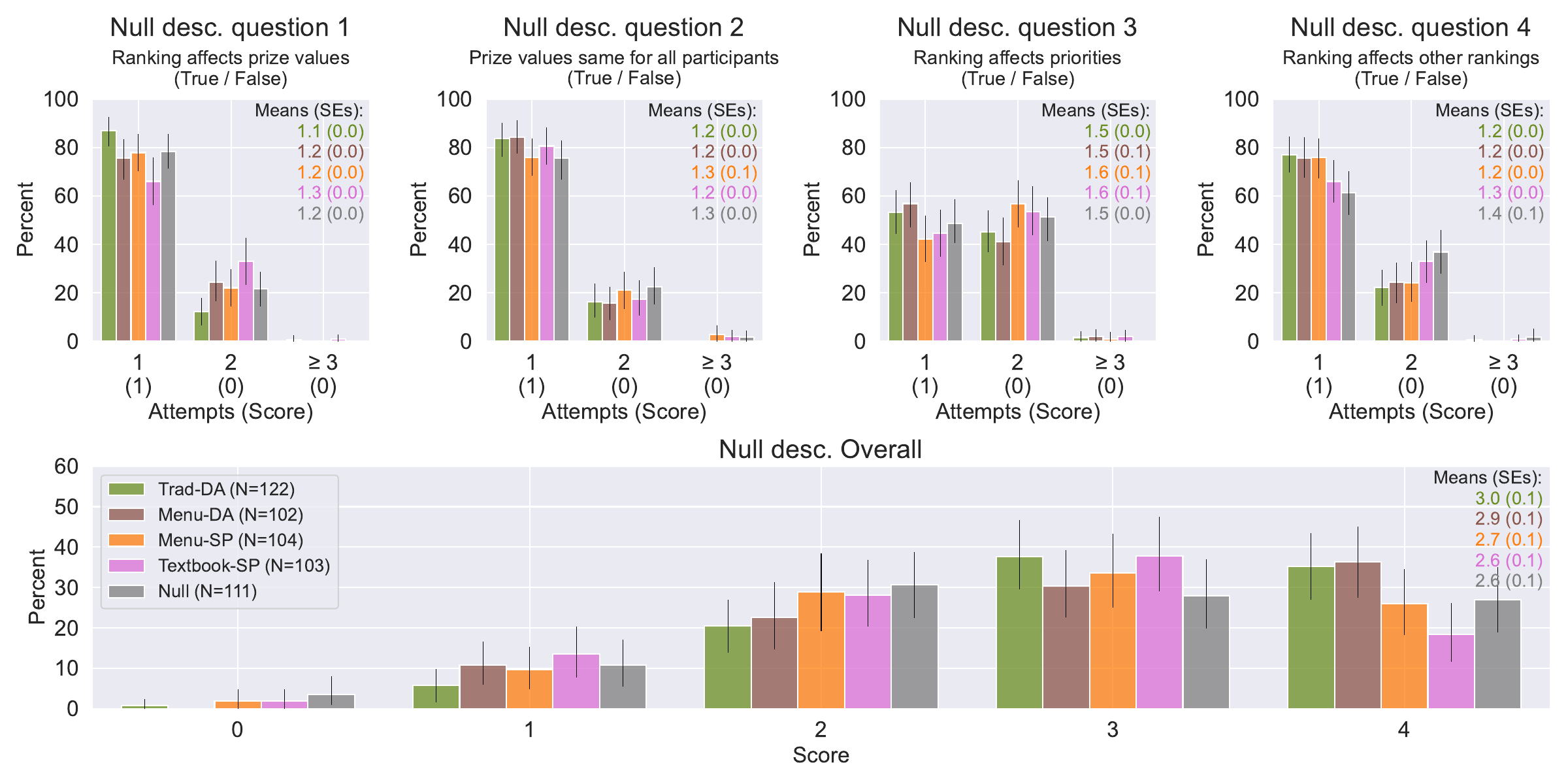}
    \label{fig:null-training-hists}
    \begin{minipage}{\textwidth}  \footnotesize
    \textbf{Notes:}
    \emph{Top row}: Performance at each of the four Null training round questions, by the number of attempts it took to answer the question correctly. Parentheses below amount of attempts show the score conditional on this amount.
    \emph{Bottom row}: Overall score (replicating \autoref{fig:all-training-hists}'s bottom row in terms of \% success).
    All panels include the mean values (SEs) in the upper right corner. 
    \emph{Both rows}: Error bars: 95\% confidence intervals.
    \end{minipage}
\end{figure}

\autoref{fig:null-training-hists} shows the distribution of participants' performance in the training round following the Null description, which is common to all participants and is not included in the treatment-specific training scores. %
There are four true/false questions about statements---shortly summarized above each top-row panel. 

In this figure and others like it in the appendix, instead of only showing the binary score for each training question, we show more details on the number of attempts it took to answer the question correctly. Recall that participants cannot advance beyond a question until getting it correctly and receiving a feedback of why it is correct. In all questions except for a few DA Mechanics questions, participants get a score of 1 (and a resultant monetary understanding bonus) only conditional on answering correctly at first attempt; we show the score awarded in parentheses below the number of attempts.
The overall score in the round is displayed at the bottom row of the figure (which is also shown in the bottom row of \autoref{fig:all-training-hists} in terms of \% success).

\subsubsection{DA Mechanics Training}

\begin{figure}[htbp]
    \centering
    \caption{DA Mechanics training questions: round 1}
    \includegraphics[width=\textwidth]{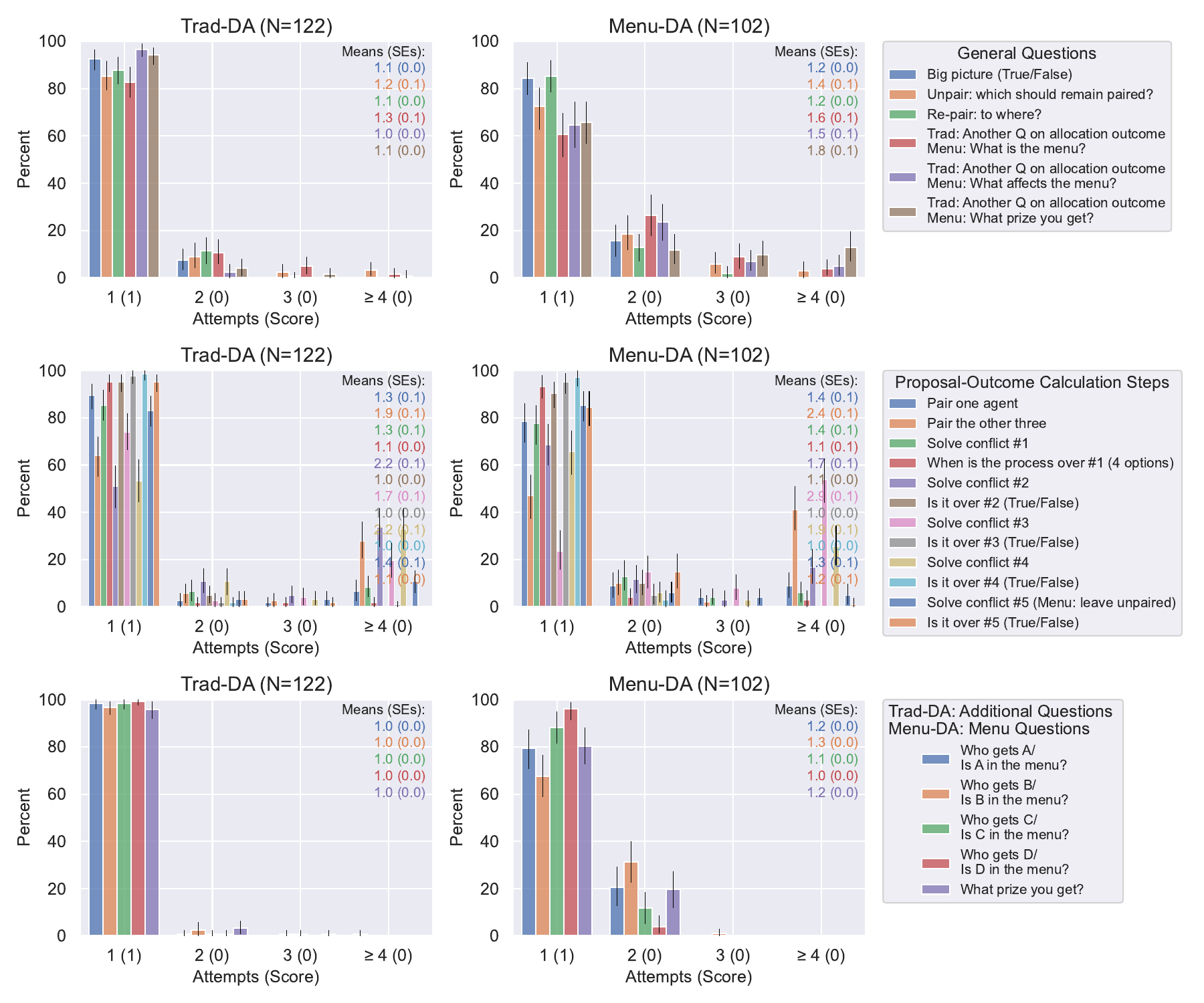}
    \label{fig:mech-training-hists1}
    \begin{minipage}{\textwidth}  \footnotesize
    \textbf{Notes:} \emph{Middle row}: $\geq 4$ attempts in ``Pair one agent,'' ``Pair the other three,'' and ``Solve~conflict~\#$i$'' (for $i=1,2,3,4,5$) mean that the participant did not successfully calculate these proposals-outcome steps within three attempts, and was shown the correct answer by the computer in order to proceed. 
    \emph{All rows}: Error bars: 95\% confidence intervals.
    \end{minipage}
\end{figure}

\begin{figure}[htbp]
    \centering
    \caption{DA Mechanics training questions: rounds 2--3}
    \includegraphics[width=\textwidth]{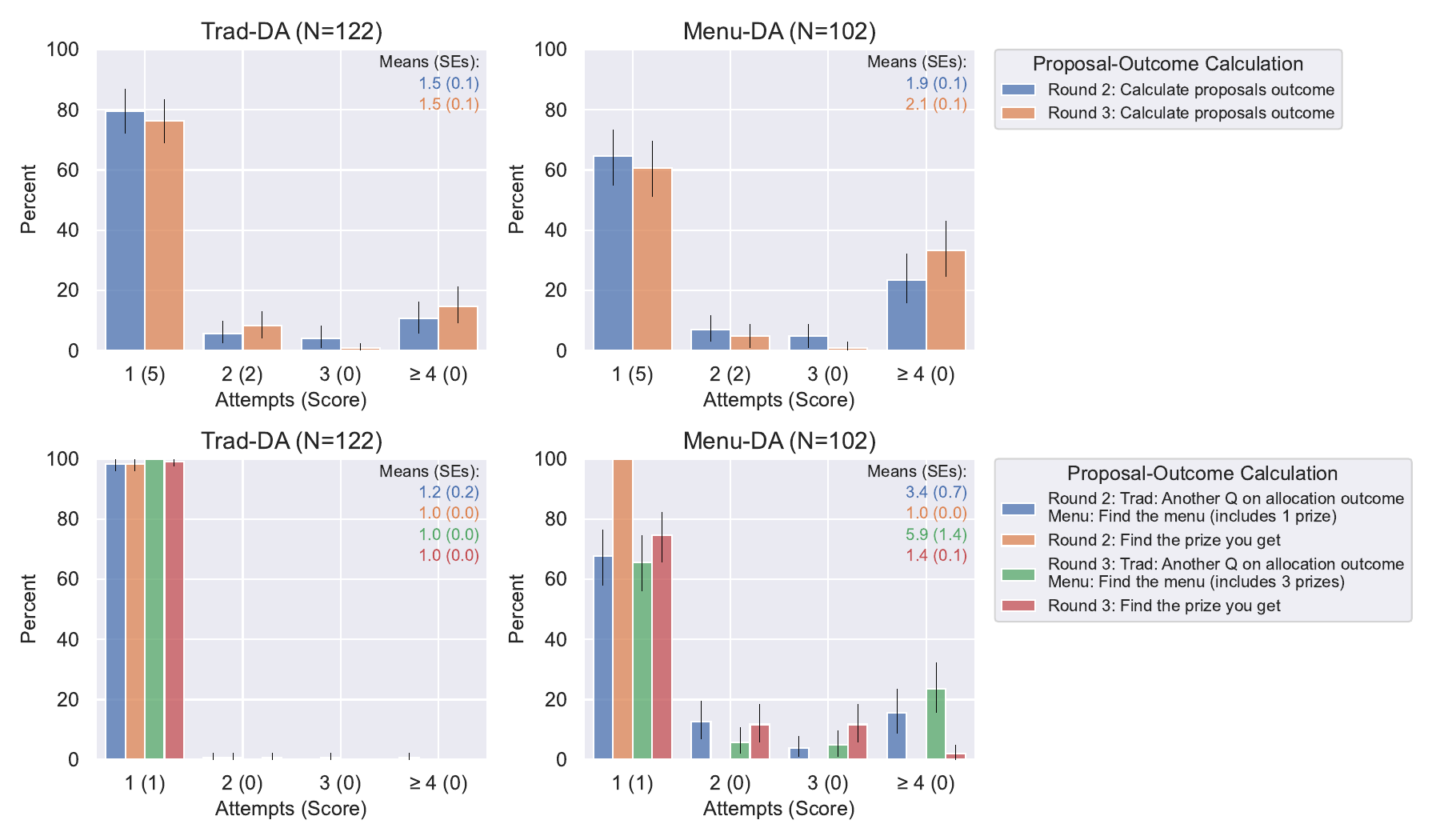}
    \label{fig:mech-training-hists2}
    \begin{minipage}{\textwidth}  \footnotesize
    \textbf{Notes:} \emph{Top row}: $\geq 4$ attempts mean that the participant did not successfully calculate the proposals outcomes within three attempts, and was shown the correct answer by the computer in order to proceed. 
    \emph{Both rows}: Error bars: 95\% confidence intervals.
    \end{minipage}
\end{figure}

\begin{figure}[htbp]
    \centering
    \caption{DA Mechanics walkthrough-video usage (round 2)}
    \includegraphics[width=\textwidth]{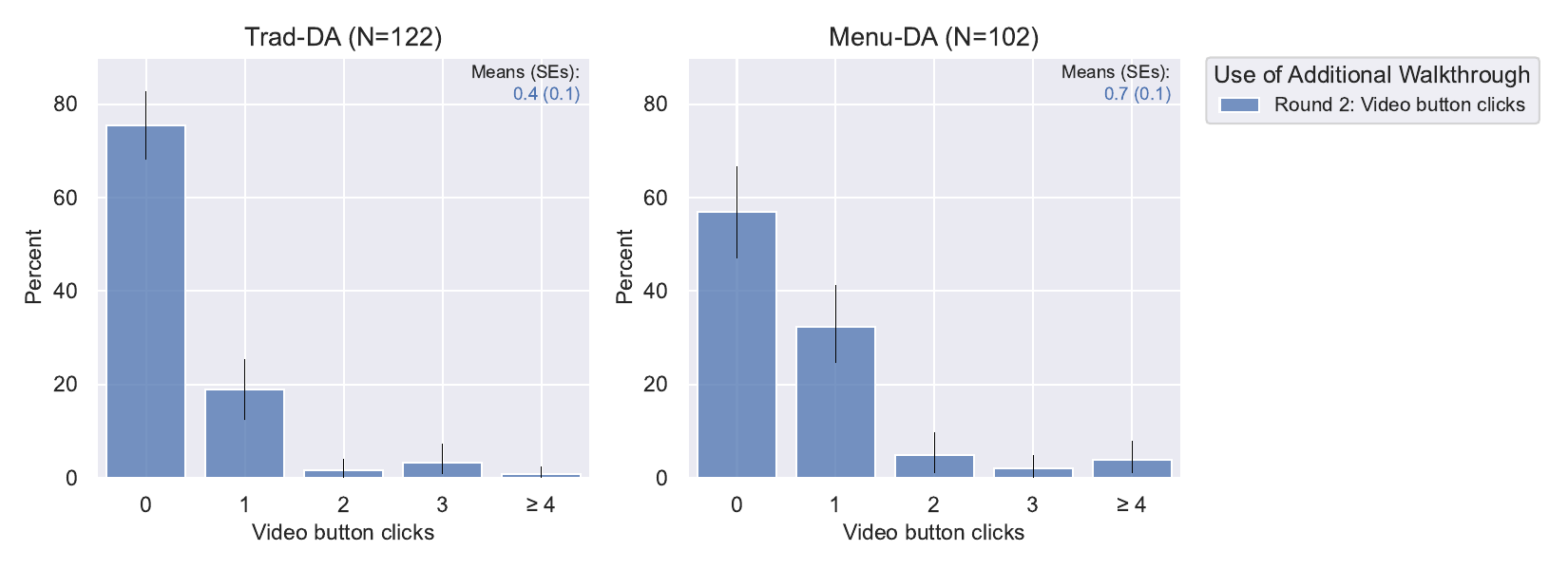}
    \label{fig:mech-training-hists3}
    \begin{minipage}{\textwidth}  \footnotesize
    \textbf{Notes:} Error bars: 95\% confidence intervals.
    \end{minipage}
\end{figure}

\begin{figure}[htbp]
    \centering
    \caption{DA Mechanics training scores}
    \includegraphics[width=\textwidth]{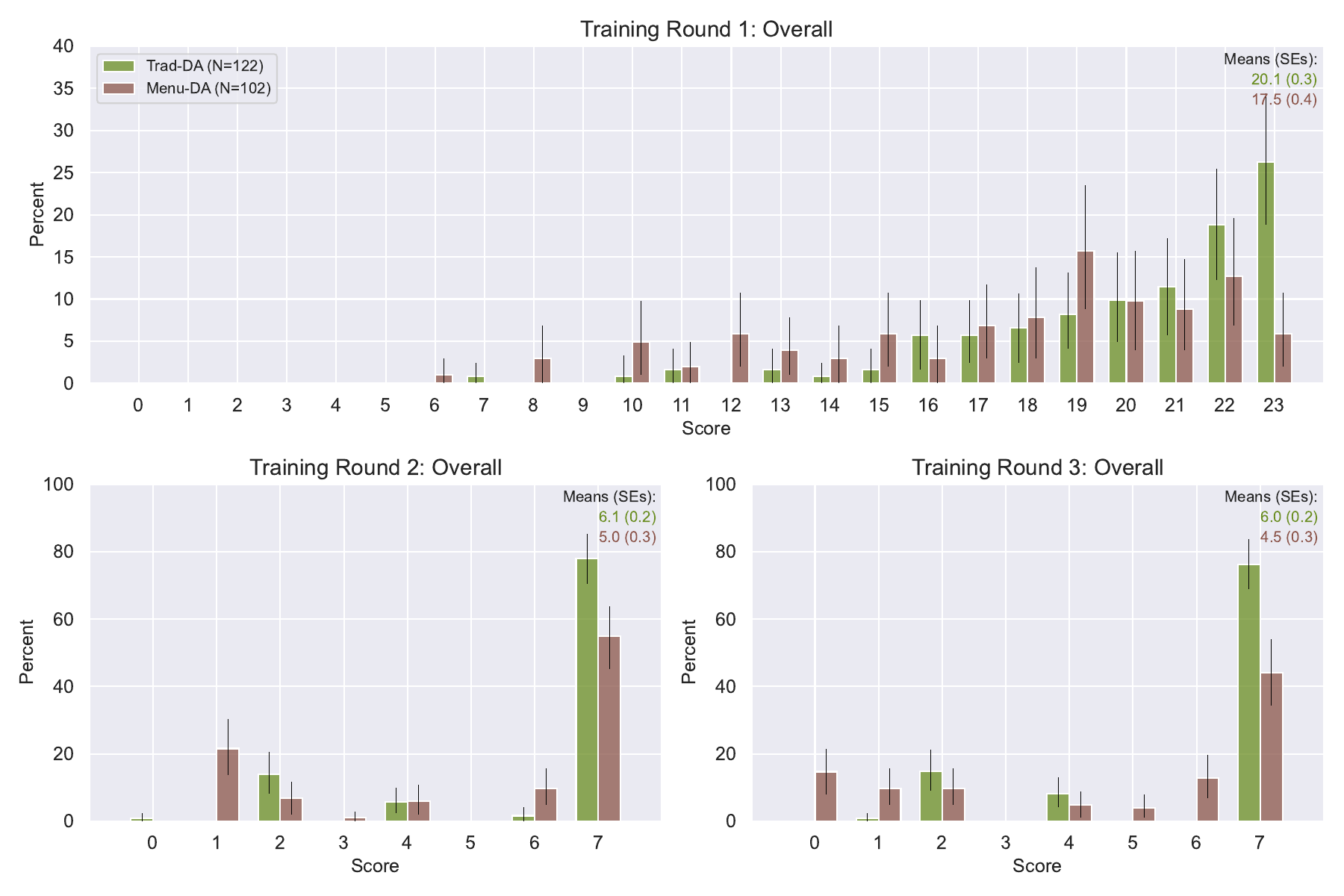}
    \label{fig:mech-training-hists4}
    \begin{minipage}{\textwidth}  \footnotesize
    \textbf{Notes:} Error bars: 95\% confidence intervals.
    \end{minipage}
\end{figure}

Figures \ref{fig:mech-training-hists1}, \ref{fig:mech-training-hists2}, \ref{fig:mech-training-hists3} and \ref{fig:mech-training-hists4} show participants' performance in the DA Mechanics training questions, including the DA-algorithm proposals-outcome calculations they needed to perform using the GUI, as well as their usage of an additional walkthrough video available in training round 2.

In DA Mechanics treatments, any error on an individual step of the DA-algorithm proposals-outcome calculation should, in theory, make the overall proposals outcome erroneous. 
However, despite a fairly high rate of errors on some calculation steps in training round 1, the overall success in rounds 2 and 3 is higher. 
This suggests that training round 1 was effective at teaching participants how to perform these calculations, perhaps through trial and error.

\clearpage 

\subsubsection{SP Property Training}

\begin{figure}[htbp]
    \centering
    \caption{SP Property training questions}
    \includegraphics[width=\textwidth]{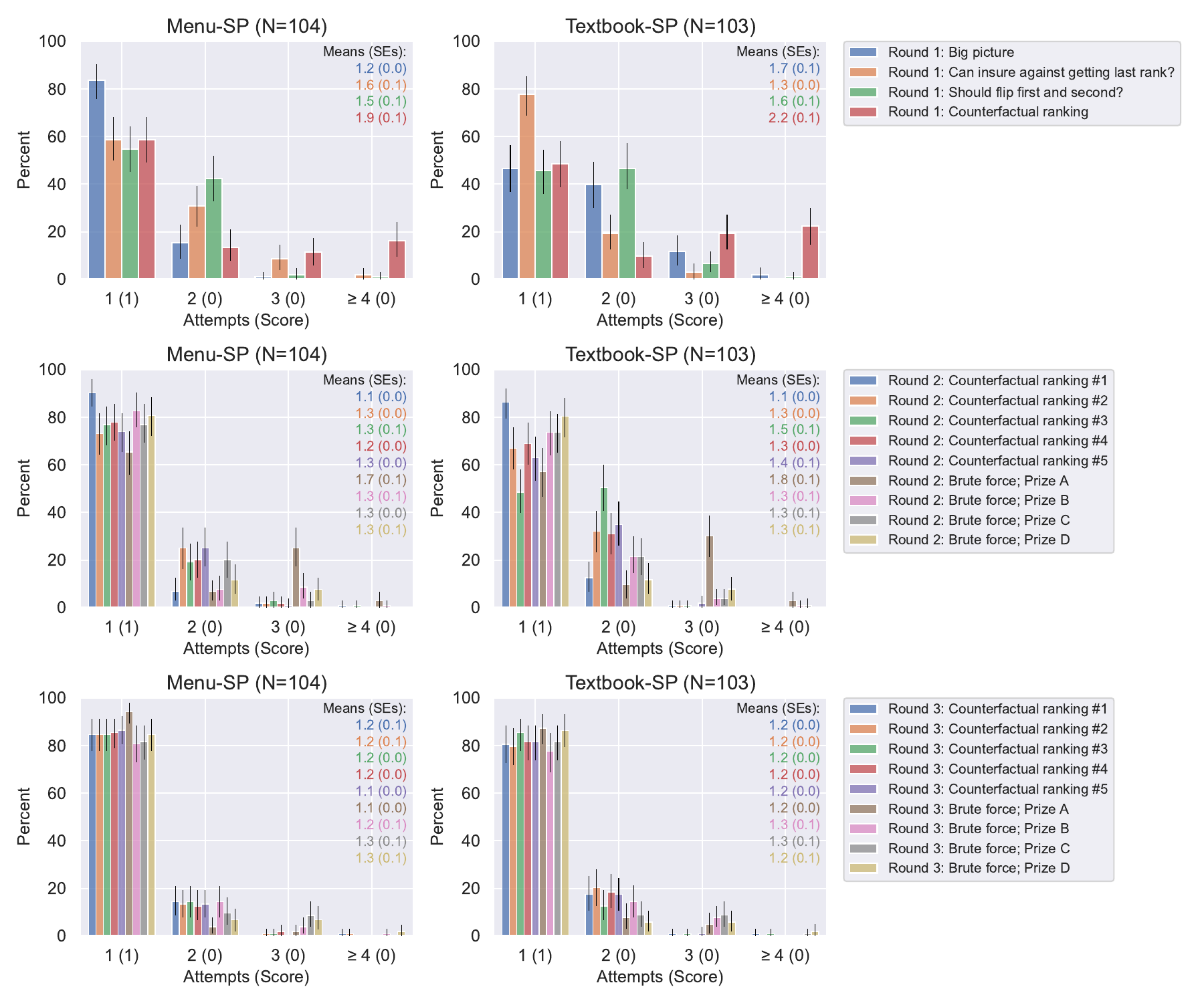}
    \label{fig:prop-training-hists1}
    \begin{minipage}{\textwidth}  \footnotesize
    \textbf{Notes:} In ``Counterfactual ranking'' and ``Brute force'' questions, participants know which prize they received after submitting a particular ranking, and are asked whether they could have received other prizes by submitting other rankings (given that the other participants' rankings and prize priorities do not change). Counterfactual-ranking questions ask which prizes could be received by submitting a specific alternative ranking. Brute-force question ask whether there exists any alternative ranking whose submission could lead to receiving a specific prize. Error bars: 95\% confidence intervals.
    \end{minipage}
\end{figure}

\begin{figure}[htbp]
    \centering
    \caption{SP Property training scores}
    \includegraphics[width=\textwidth]{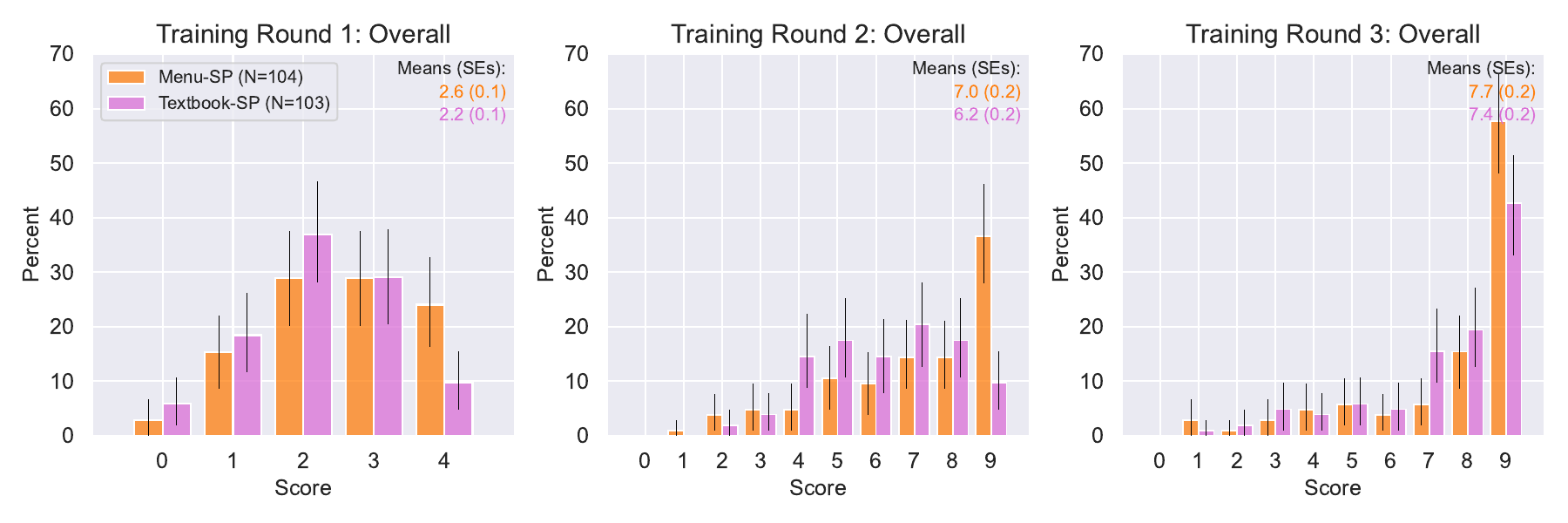}
    \label{fig:prop-training-hists2}
    \begin{minipage}{\textwidth}  \footnotesize
    \textbf{Notes:} Error bars: 95\% confidence intervals.
    \end{minipage}
\end{figure}

Figures \ref{fig:prop-training-hists1} and \ref{fig:prop-training-hists2} show participants' performance in the the SP Property training questions.

\subsubsection{Null Treatment Training}

\begin{figure}[htbp]
    \centering
    \caption{Null treatment training questions}
    \includegraphics[width=\textwidth]{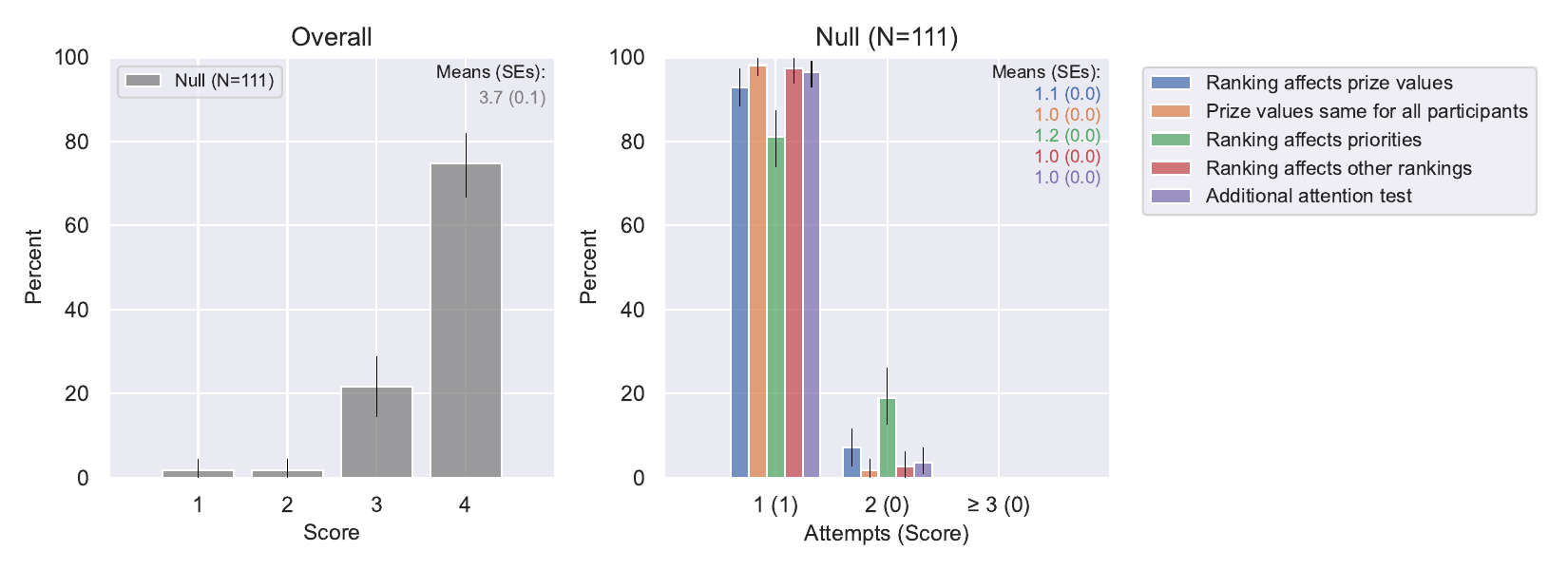}
    \label{fig:null-treatment-training-hists}
    \begin{minipage}{\textwidth}  \footnotesize
    \textbf{Notes:} Error bars: 95\% confidence intervals.
    \end{minipage}
\end{figure}

\autoref{fig:null-treatment-training-hists} shows participants' performance in the Null treatment's training questions. These questions exactly repeated the Null training questions common to all participants (though, like the other treatments, the Null treatment's training score measure ignores the initial Null training, and instead measures only the repeated Null training).

\subsection{Cognitive and Attention Scores}
\label{app:cognitive-attention}

We present the details of our cognitive and attention scores.

First, as discussed in \autoref{sec:design-cognitive}, the elicited cognitive score varies from 0 to 4. It uses the three questions of the Cognitive Reflection Task \citep{Frederick05}, and measures numeracy using one question from the Berlin Numeracy Test \citep{Cokely_Galesic_Schulz_Ghazal_Garcia-Retamero_2012}. The CRT questions are:
\begin{itemize}
    \item (Bat-and-ball question) A bat and a ball cost \$1.10 in total. The bat costs \$1.00 more than the ball. How much does the ball cost? Please enter an amount in cents. (Correct answer: 5).
    \item (Machines question) If it takes 5 machines 5 minutes to make 5 widgets, how long would it take 100 machines to make 100 widgets? Please enter a number of minutes. (Correct answer: 5).
    \item (Lily-pads question) In a lake, there is a patch of lily pads. Every day, the patch doubles in size. If it takes 48 days for the patch to cover the entire lake, how long would it take for the patch to cover half of the lake? Please enter a number of days. (Correct answer: 47).
\end{itemize}
The Berlin Numeracy Test question is:
\begin{itemize}
    \item (Choir question) Out of 1,000 people in a small town 500 are members of a choir. Out of these 500 members in the choir 100 are men. Out of the 500 inhabitants that are not in the choir 300 are men. What is the probability that a randomly drawn man is a member of the choir? Please enter a percent chance between 0 and 100. (Correct answer: 25).
\end{itemize}

\begin{figure}[htbp]
\centering
\caption{Distribution of cognitive score and its underlying questions across treatments
}
\label{fig:cognitive-distribution}
\includegraphics[width=\figscale\textwidth{}]{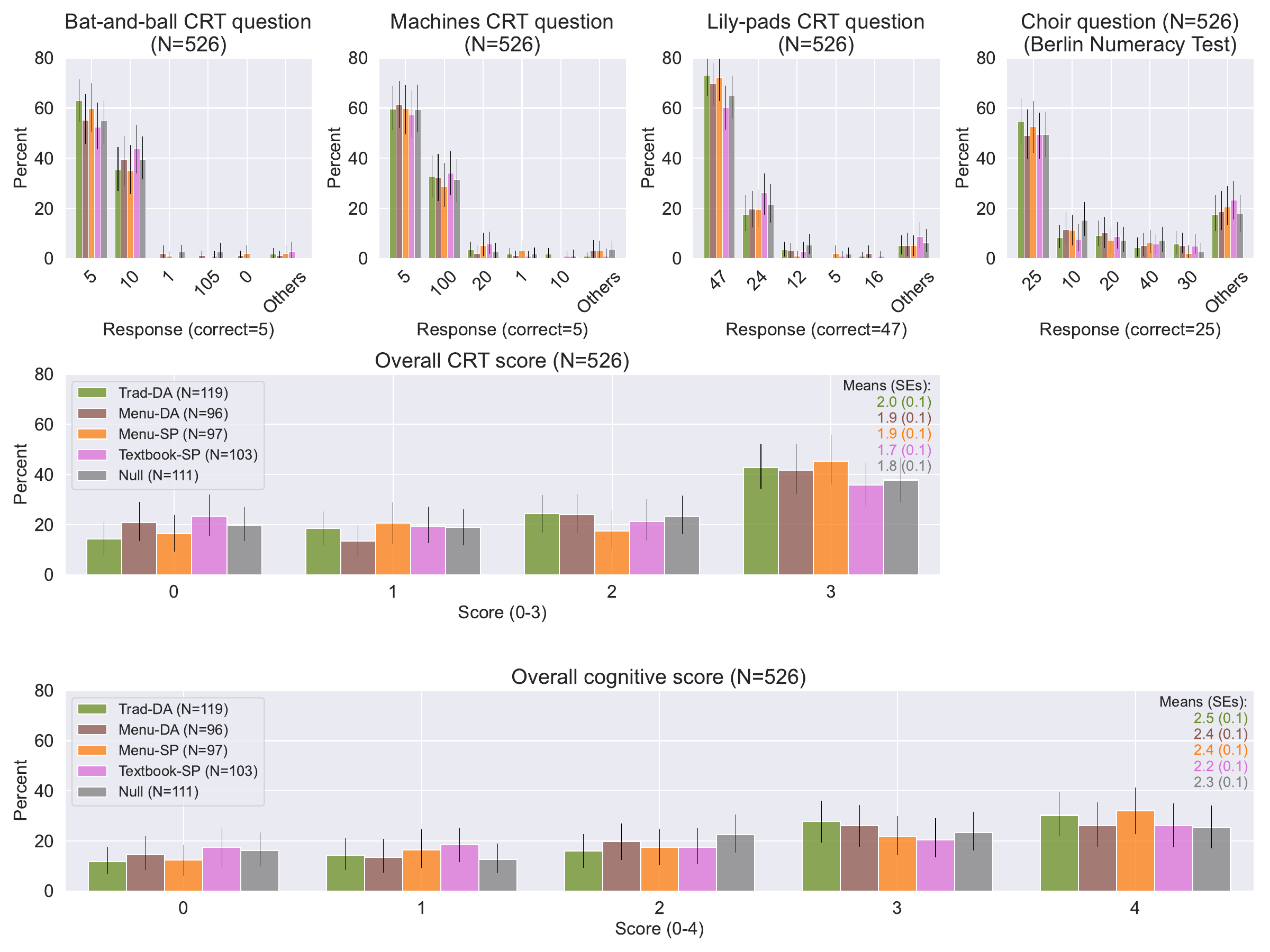}

\begin{minipage}{\textwidth}  \footnotesize
        \textbf{Notes:}
        The panels describe the distribution of responses to the four questions included in the total cognitive score, in order of response frequency, and the distribution of the overall scores. For screenshots of the questions, see \autoref{full-materials} in the Supplementary Materials. Error bars: 95\% confidence intervals.
    \end{minipage}
\end{figure}

\autoref{fig:cognitive-distribution} shows the distribution of cognitive score, as well the distribution of responses to the individual underlying questions, across treatments.
Due to a bug in the cognitive score elicitation interface in the first runs of the experiment, $16$ observations from the Prolific sample lack this data and are omitted from this analysis, and the relevant sample size for this analysis is thus $N=526$.

\begin{figure}[htbp]
\centering
\caption{Distribution of attention score and its underlying questions across treatments
}
\label{fig:attention-distribution}
\includegraphics[width=\figscale\textwidth{}]{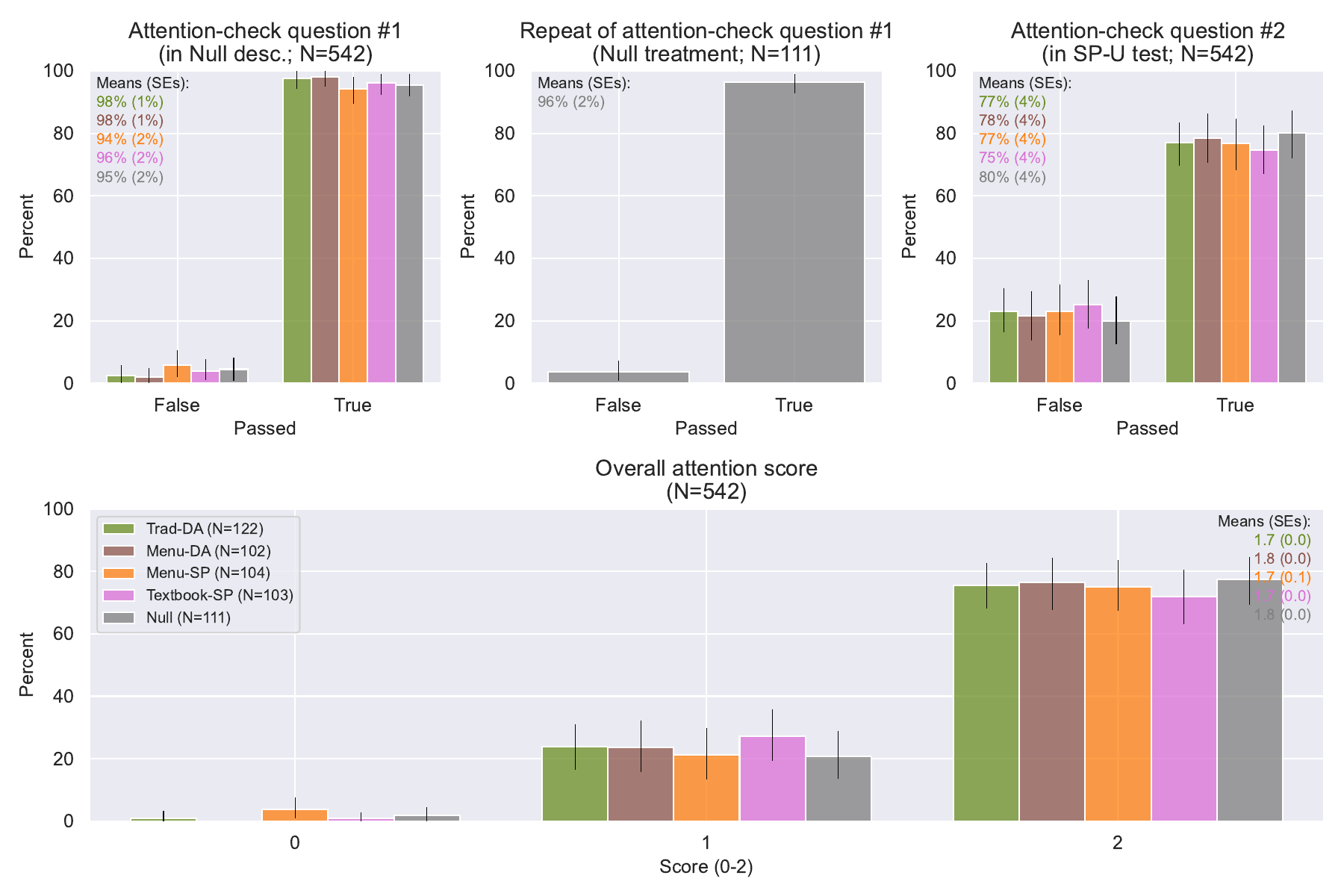}

\begin{minipage}{\textwidth}  \footnotesize
        \textbf{Notes:}
        The panels describe the distribution of performance in the two attention-check questions which appeared during the experiment, and the success rate in an additional attention-check question given to Null treatment participants only, which is not included in the overall attention score. Error bars: 95\% confidence intervals.
    \end{minipage}
\end{figure}

Second, the two attention-check questions were planted in the training questions following the Null description at the beginning of the experiment and in the middle of the SP understanding test, towards the end of the experiment. They include questions that explicitly instruct what response to submit, but are designed to look similarly to adjacent content, such that participants who skim quickly through the text are more likely to miss the explicit instructions and to answer incorrectly. \autoref{fig:attention-distribution} shows the performance in the attention-check questions and the total attention scores.
In the Null treatment, the first attention-check question is also repeated along the other questions from the Null training round common to all participants; this is shown in the figure but not included in the attention-score measure.

\clearpage 

\subsection{Additional Joint Distributions of Outcome Variables}
\label{app:new-joint-distrs}

\subsubsection{Univariate Regressions and Correlations Between Main Outcome Variables}
\label{app:univariate-regressions}

We investigate the relations between our main outcome variables, by reporting univariate regressions and correlations between them. 

\autoref{tab:basic-regressions} reports, separately for each treatment, three OLS regression coefficients using our main outcome variables (\% SP-U and \% SF regressed on \% TR, and \% SF regressed on \% SP-U), as well as the correlation coefficients between the three pairs. 
Focusing on the non-Null treatments, we make three observations which could serve as motivation for our other investigations.

\begin{table}[htbp]
    \begin{center}
    \caption{
    Main outcome variables: regression and correlation coefficients
    }
    \label{tab:basic-regressions}
    
\bgroup

\newcommand{\myTablePartGap}{0em}
\newcommand{\myTableCloseGroup}{-0.5em}
\newcommand{\myOneLineCell}[2]{#1 {\footnotesize #2}}

\small
\def\arraystretch{1.4}%
\begin{tabular}{ccccccccc}
  \toprule
  \makecell{\\ \\ Indep.\ } &
  \makecell{\\ \\ Dep.\ } &
  & \makecell{Trad \\DA \\ {\footnotesize $N = 122$} } 
  & \makecell{Menu \\DA \\ {\footnotesize $N = 102$} } 
  & \makecell{Menu \\ SP \\ {\footnotesize $N = 104$} } 
  & \makecell{Textbook \\ SP \\ {\footnotesize $N =103 $} } 
  & \makecell{Null \\ \\ {\footnotesize $N = 111$} }
  \\ \midrule
  \multirow{2}{*}{\% TR}
   &
  \multirow{2}{*}{\% SP-U} 
  & $\beta$
  & \myOneLineCell{0.34}{(0.08)}
  & \myOneLineCell{0.36}{(0.07)}
  & \myOneLineCell{0.70}{(0.10)}
  & \myOneLineCell{0.77}{(0.08)}
  & \myOneLineCell{0.09}{(0.07)}
  \\[\myTablePartGap]
  &  & $r$
  & \myOneLineCell{0.35}{(0.09)}
  & \myOneLineCell{0.38}{(0.08)}
  & \myOneLineCell{0.73}{(0.10)}
  & \myOneLineCell{0.80}{(0.08)}
  & \myOneLineCell{0.10}{(0.07)}
  \\ \midrule
  \multirow{2}{*}{\% SP-U}
  &
  \multirow{2}{*}{\% SF} 
  & $\beta$
  & \myOneLineCell{0.61}{(0.15)}	
  & \myOneLineCell{0.73}{(0.14)}	
  & \myOneLineCell{0.73}{(0.14)}	
  & \myOneLineCell{0.64}{(0.16)}	
  & \myOneLineCell{0.25}{(0.22)}
  \\[\myTablePartGap]
  &  & $r$
  & \myOneLineCell{0.24}{(0.06)}
  & \myOneLineCell{0.28}{(0.05)}	
  & \myOneLineCell{0.28}{(0.05)}	
  & \myOneLineCell{0.25}{(0.06)}	
  & \myOneLineCell{0.10}{(0.08)}
  \\ \midrule
  \multirow{2}{*}{\% TR}
  &
  \multirow{2}{*}{\% SF} 
  & $\beta$
  & \myOneLineCell{0.38}{(0.19)}
  & \myOneLineCell{0.38}{(0.15)}	
  & \myOneLineCell{0.58}{(0.17)}	
  & \myOneLineCell{0.50}{(0.18)}	
  & \myOneLineCell{0.07}{(0.18)}
  \\[\myTablePartGap]
  &  & $r$
  & \myOneLineCell{0.15}{(0.08)}
  & \myOneLineCell{0.15}{(0.06)}
  & \myOneLineCell{0.23}{(0.07)}
  & \myOneLineCell{0.20}{(0.07)}
  & \myOneLineCell{0.03}{(0.07)}
  \\ \bottomrule
\end{tabular}
\egroup

    \end{center}
    \begin{minipage}{\textwidth}  \footnotesize
    \textbf{Notes:}
    ``$\beta$'': Coefficients from OLS regressions of dependent (``Dep.'') on independent (``Indep.'') variables, within each treatment separately. ``$r$'': Estimated Pearson regression coefficients. Robust standard errors in parentheses. 
    \end{minipage}
\end{table}

First, higher training scores are associated with higher SP understanding.
This relationship is particularly strong for the SP Property treatments, which specifically aim to teach strategyproofness, but we also see a positive relation in the DA Mechanics treatments.
We further investigate this relation in \autoref{app:tr-sp} in the Supplementary Materials.
Second, the relation between \% SP-U and \% SF seems quite consistent across \emph{all} (non-Null) treatments, suggesting that understanding SP better corresponds to more SF behavior, regardless of the description. 
This gives additional motivation for our investigation into this relation in \autoref{sec:sp-sf}.
Third, the correlations between \% TR and \% SF seem overall weaker than the other relations in the table.
This finding is consistent with the idea that this relationship is essentially a noisy composition of those from first two rows of the table, i.e., that \% TR affects \% SF mostly via \% SP-U.
We further investigate this relation in \autoref{app:tr-to-sf} in the Supplementary Materials, and find additional support for this interpretation.

\subsubsection{Relationship Between \% SP-U and \% SF: More Details}
\label{app:sp-u-sf-more}

Following \autoref{tab:sf-sp123-joint-separate} showing OLS regressions of \% SF on Abstract and Practical, and total \% SP-U, \autoref{tab:sf-sp-joint-separate} adds a breakdown of Abstract into its three sub-measures AbstractHard, AbstractMedium, and AbstractEasy.

\begin{table}[hbtp]
    \centering
    \footnotesize
    Dependent variable: \% SF
    {
\def\sym#1{\ifmmode^{#1}\else\(^{#1}\)\fi}
\setlength\tabcolsep{2pt}
\begin{tabular}{@{\extracolsep{2pt}}l*{12}{c}@{}}
\hline\hline

 & (1) & (2) & (3) & (4) & (5) & (6) & (7) & (8) & (9) & (10) & (11) & (12) \\
\hline
\% AbstractHard & 0.15 & $-$0.04 &  &  &  &  &  &  & $-$0.01 & $-$0.06 &  &  \\
 & (0.07) & (0.08) &  &  &  &  &  &  & (0.07) & (0.08) &  &  \\
\% AbstractMedium &  &  & 0.15 & 0.03 &  &  &  &  & 0.06 & $-$0.02 &  &  \\
 &  &  & (0.04) & (0.06) &  &  &  &  & (0.05) & (0.06) &  &  \\
\% AbstractEasy &  &  &  &  & 0.17 & 0.11 &  &  & 0.05 & 0.07 &  &  \\
 &  &  &  &  & (0.04) & (0.06) &  &  & (0.05) & (0.06) &  &  \\
\% Practical &  &  &  &  &  &  & 0.43 & 0.41 & 0.41 & 0.40 &  &  \\
 &  &  &  &  &  &  & (0.03) & (0.04) & (0.03) & (0.04) &  &  \\
\% SP-U &  &  &  &  &  &  &  &  &  &  & 0.62 & 0.60 \\
 &  &  &  &  &  &  &  &  &  &  & (0.06) & (0.10) \\
Controls &  & X &  & X &  & X &  & X &  & X &  & X \\
Treatment &  & X &  & X &  & X &  & X &  & X &  & X \\

\hline
R² & 0.01 & 0.31 & 0.02 & 0.31 & 0.03 & 0.32 & 0.22 & 0.45 & 0.22 & 0.45 & 0.13 & 0.37 \\
N & 542 & 542 & 542 & 542 & 542 & 542 & 542 & 542 & 542 & 542 & 542 & 542 \\
\hline\hline
\end{tabular}
}

    \caption{Relation of all four \% SP-U sub-measures with SF behavior}
    \label{tab:sf-sp-joint-separate}

    \begin{minipage}{\textwidth}  \footnotesize
    \textbf{Notes:}
    See under \autoref{tab:sf-sp123-joint-separate}.
    \end{minipage}
\end{table}

\subsubsection{Relationship Between \% TR and \% SP-U}
\label{app:tr-sp}

We investigate the relationship between training score and SP understanding in more detail. 

\begin{figure}[htbp]
    \centering
    \caption{Joint distribution of \% TR and \% SP-U by treatment}
    \label{fig:tr-sp-u-detailed-relationship}
    
    \vspace{-0.1in}
    
    \includegraphics[width=\textwidth]{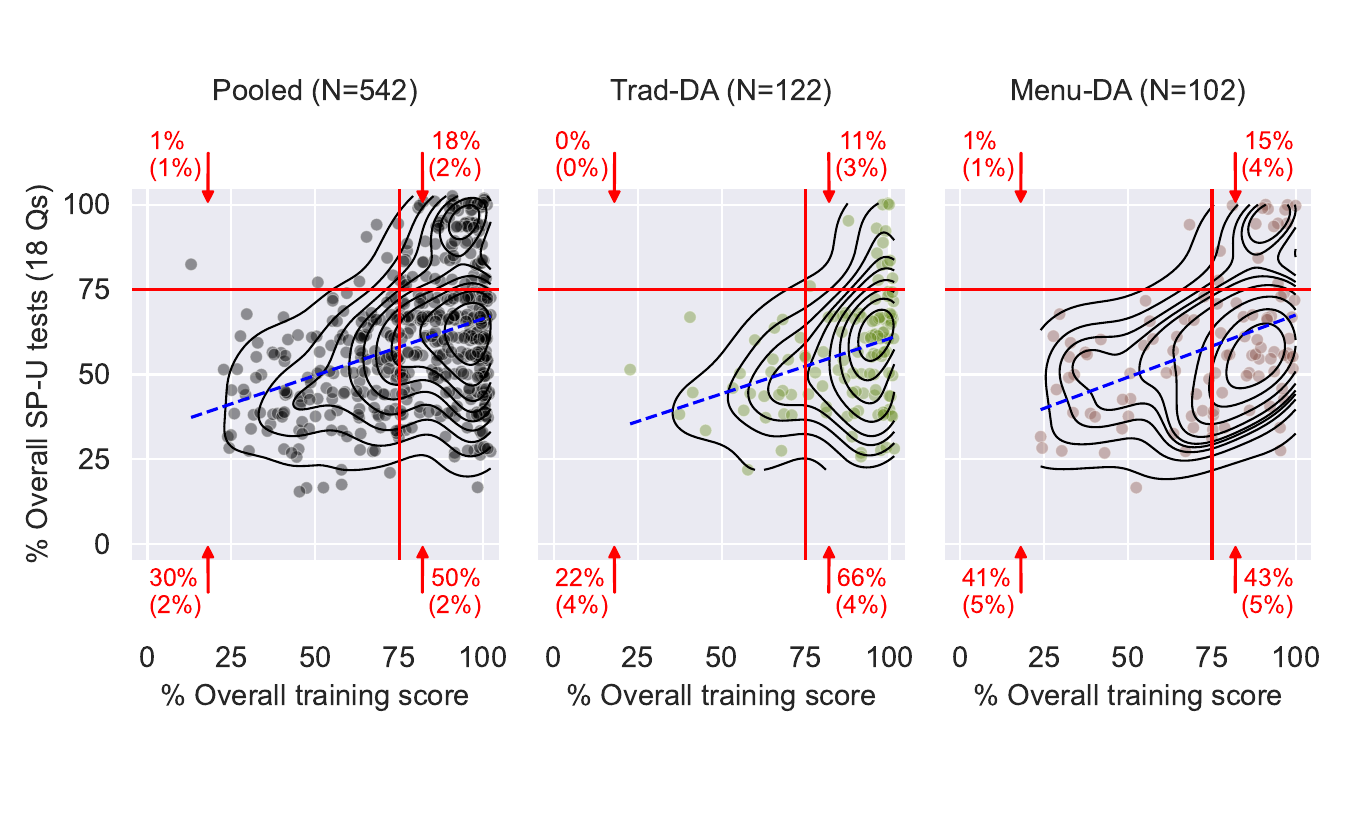}

    \vspace{-0.5in}

    \includegraphics[width=\textwidth]{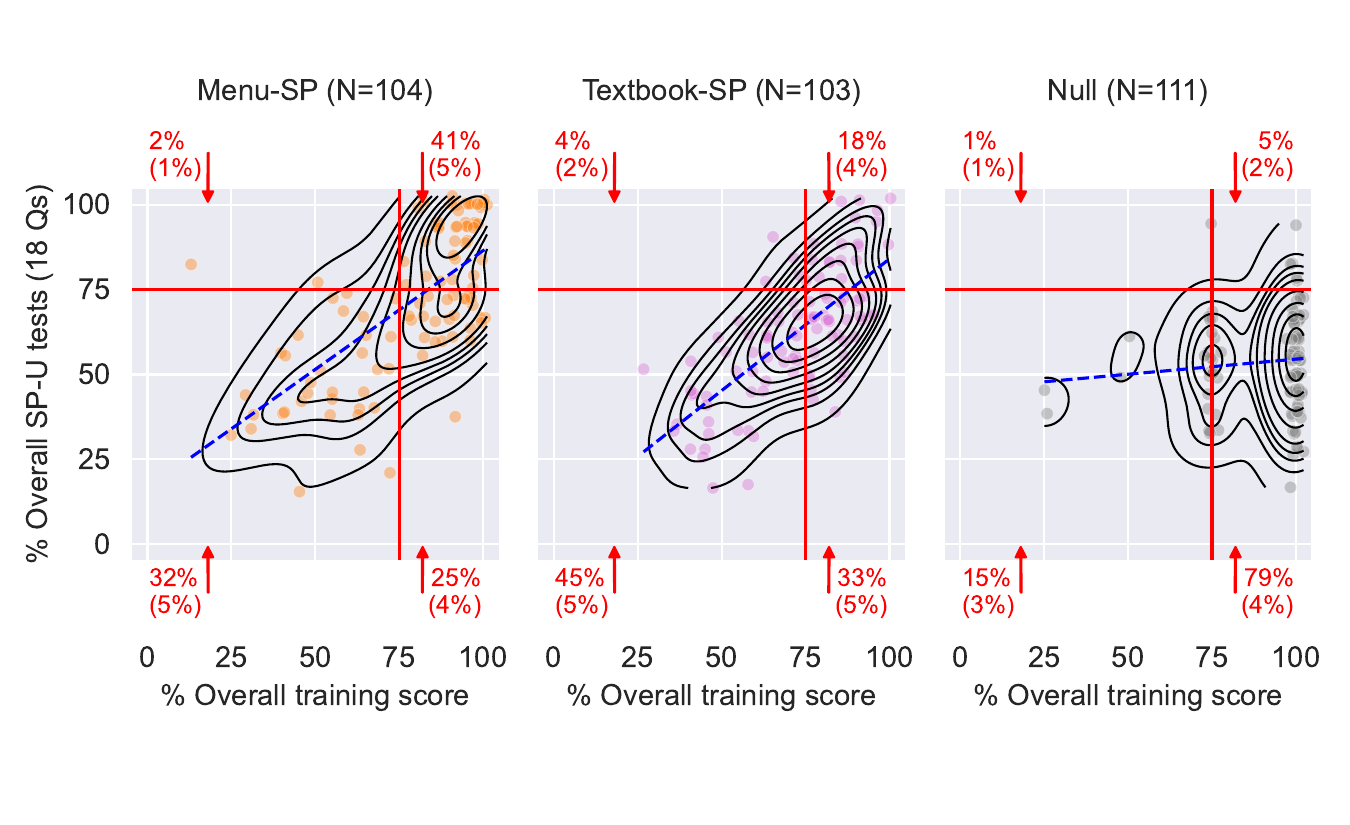}

    \vspace{-0.3in}
    \begin{minipage}{\textwidth}  \footnotesize
    \textbf{Note:}
    See under \autoref{fig:spu-sf-histogram-by-sp-u-groups}.
    \end{minipage}
\end{figure}

\autoref{fig:tr-sp-u-detailed-relationship} reports the joint distribution of \% TR and \% SP-U in all treatments. 
Note that, as discussed in \autoref{sec:results-training-score}, \% TR is not comparable across treatments (though it is similar within DA Mechanics treatments and within SP Property treatments), i.e., the horizontal axis represents a different measure for each mini-figure in \autoref{fig:tr-sp-u-detailed-relationship}. However, the distribution among the pooled data is also added for completeness.

We first focus on our two DA Mechanics treatments. 
Despite the positive regression coefficient between \% TR and \% SP-U in \autoref{tab:basic-regressions}, most participants with high \% TR do not reach particularly high levels of \% SP-U.
These findings support our second main result result, i.e., that understanding (the mechanics of) DA does not imply understanding strategyproofness. 
It provides within-treatment evidence (complementing our between-treatments evidence from \autoref{sec:results-strategyproofness-understanding}) that even a high level of understanding of DA Mechanics does not imply understanding of strategyproofness.

To further illustrate the above---and to provide an additional comparison between Trad-DA and Menu-DA which accounts for the fact that the training questions are more challenging in Menu-DA---we consider the fraction of participants who have \% SP-U above the threshold of 75\% identified in \autoref{sec:sp-u-sub-measures} among certain upper percentiles of the per-treatment distribution of \% TR.
We find:
\begin{itemize}
    \item Among participants in the top half of the \% TR distribution (\% TR $\ge$ 95\% in Trad-DA, and $\ge$ 78\% in Menu-DA; 52\% of participants in each treatment), \textbf{19\%} of participants in Trad-DA have \% SP-U $\ge$ 75\%; for Menu-DA, this fraction is \textbf{28\%}.
    \item  Among participants in the top quarter of the \% TR distribution (\% TR $\ge$ 97\% in Trad-DA, and $\ge$ 92\% in Menu-DA; 43\% and 29\% of participants, respectively), these fractions are \textbf{21\%} for Trad-DA and \textbf{30\%} for Menu-DA.
    \item  Among participants with perfect training scores in Trad-DA and Menu-DA (24\% and 4\% of participants, respectively), these fractions are \textbf{24\%} for Trad-DA and \textbf{25\%} for Menu-DA. 
\end{itemize}
The general similarity across Trad-DA and Menu-DA suggests that the two are fairly comparable in their ability to teach participants strategyproofness, even conditioned on (treatment-specific) high percentiles of understanding of the description.

Next, we focus on our SP Property treatments, which specifically aim to teach participants strategyproofness. We find that the relationship between \% TR and \% SP-U is much stronger in these treatments than in the DA Mechanics treatments.
This provides some empirical evidence that our SP Property treatments and our SP understanding test are aligned---they teach and test similar things.

\subsubsection{Relationship Between \% TR and \% SF}
\label{app:tr-to-sf}

We investigate the relationship between training score and SF behavior in more detail.

\begin{figure}[hbtp]
    \centering
    \caption{Joint distributions of \% TR and \% SF by treatment}
    \label{fig:tr-sp-or-sf-detailed-relationship}
    
    \vspace{-0.1in}
    
    \includegraphics[width=\textwidth]{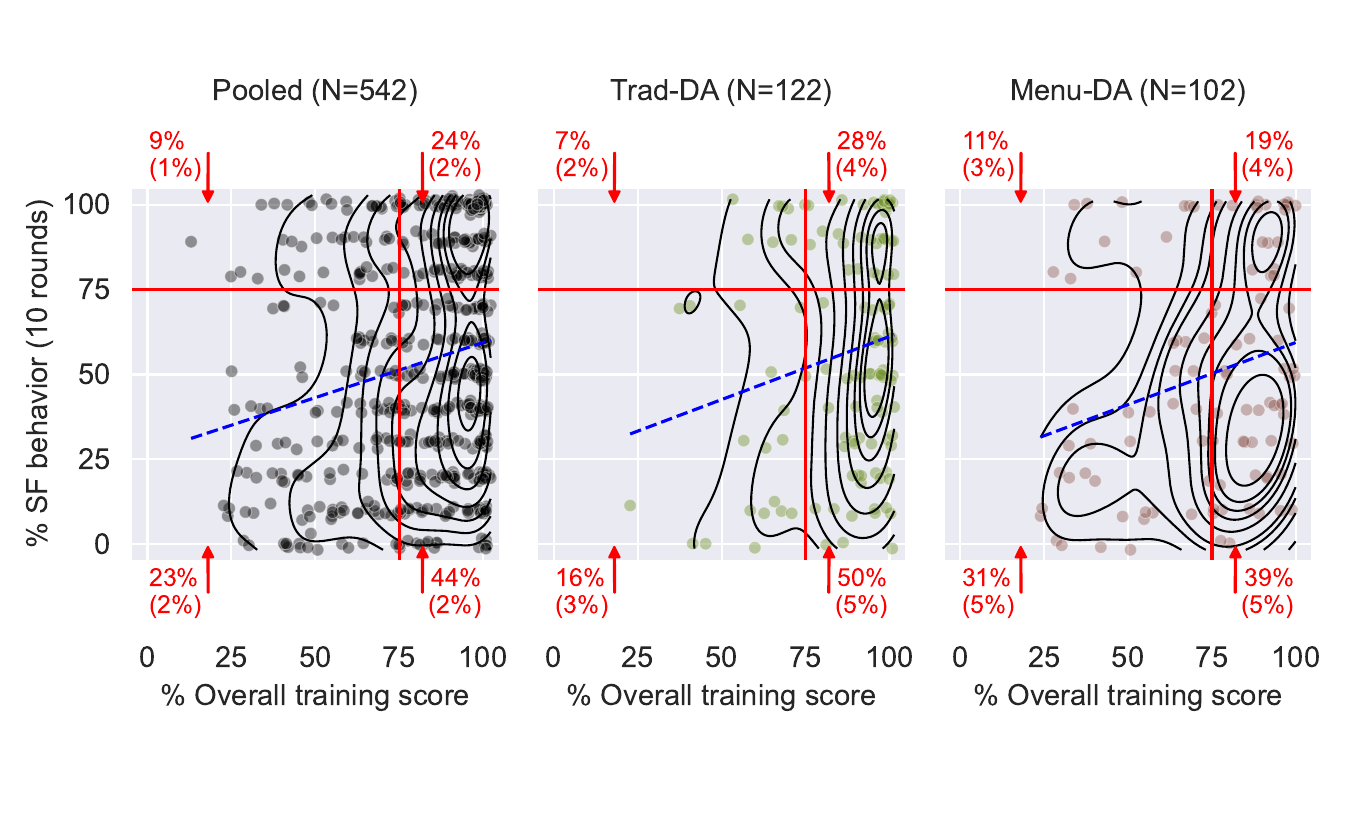}

    \vspace{-0.5in}

    \includegraphics[width=\textwidth]{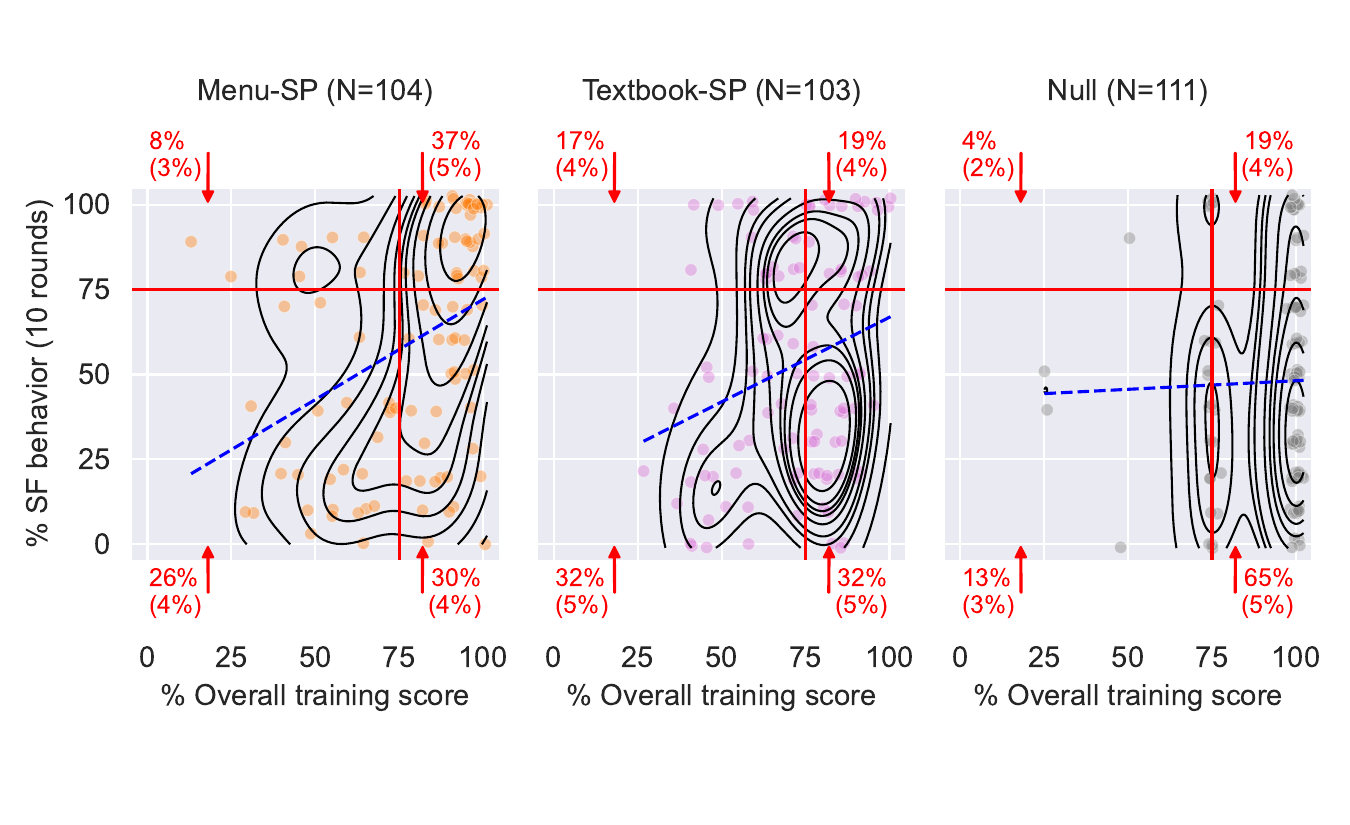}

    \vspace{-0.3in}
    \begin{minipage}{\textwidth}  \footnotesize
    \textbf{Note:}
    See under \autoref{fig:spu-sf-histogram-by-sp-u-groups}.
    \end{minipage}    
\end{figure}

\autoref{fig:tr-sp-or-sf-detailed-relationship} shows the full joint distribution of \% TR and \% SF. The distribution among the pooled data is also added for completeness (however, recall again that training scores are not
comparable outside the treatment groups of DA Mechanics, SP Property and Null).
The figure, along with the small correlation coefficients in \autoref{tab:basic-regressions}, suggests that the relation between \% TR and \% SF is overall weaker than the relations between \% TR and \% SP-U and between \% SP-U and \% TR, and may mostly reflect a (noisy) composition of the two. 

\section{Experiment Materials}
\label{app:select-screenshots-etc}

In this appendix, we provide experimental materials for easy reference, including screenshots of the main description texts, DA scenarios and parameter distributions used in incentivized DA rounds, and the full recruitment materials and procedures of running the experiment.

\subsection{Description Screenshots}
\label{app:select-screenshots}

We provide screenshots of the per-treatment descriptions screens of our experiment. 
For full screenshots, see \autoref{full-materials} in the Supplementary Materials.
Note that in all the description screens, successive parts of the text did not appear at once, and instead appeared one-by-one as the participant advanced through reading.

\begin{figure}
    \centering
    \includegraphics[width=0.7\textwidth]{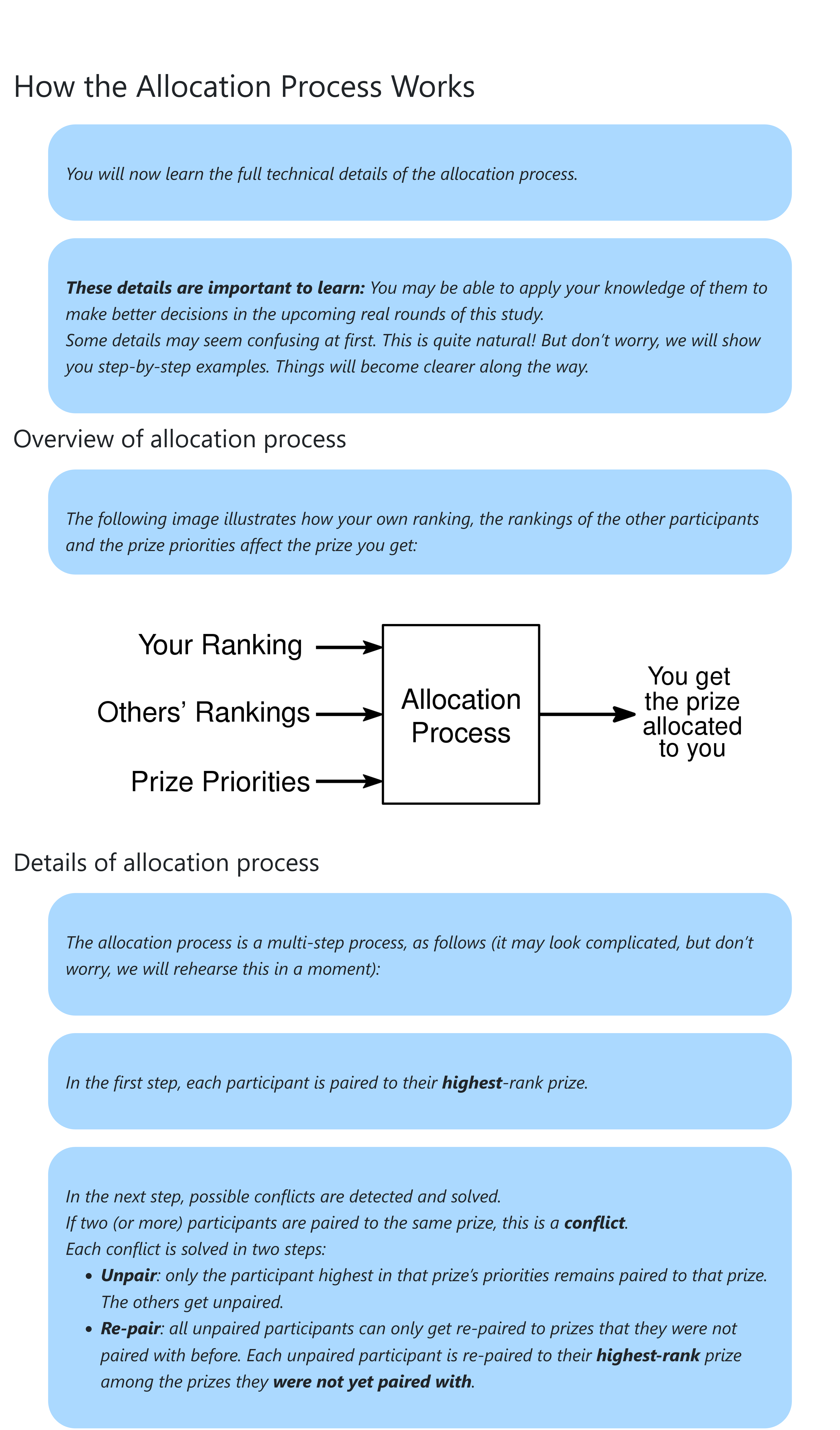}
    \caption{Traditional DA Mechanics description main text (1/2)}
    \label{fig:trad-mech-descr-1}
\end{figure}

\begin{figure}
    \centering
    \includegraphics[width=0.7\textwidth]{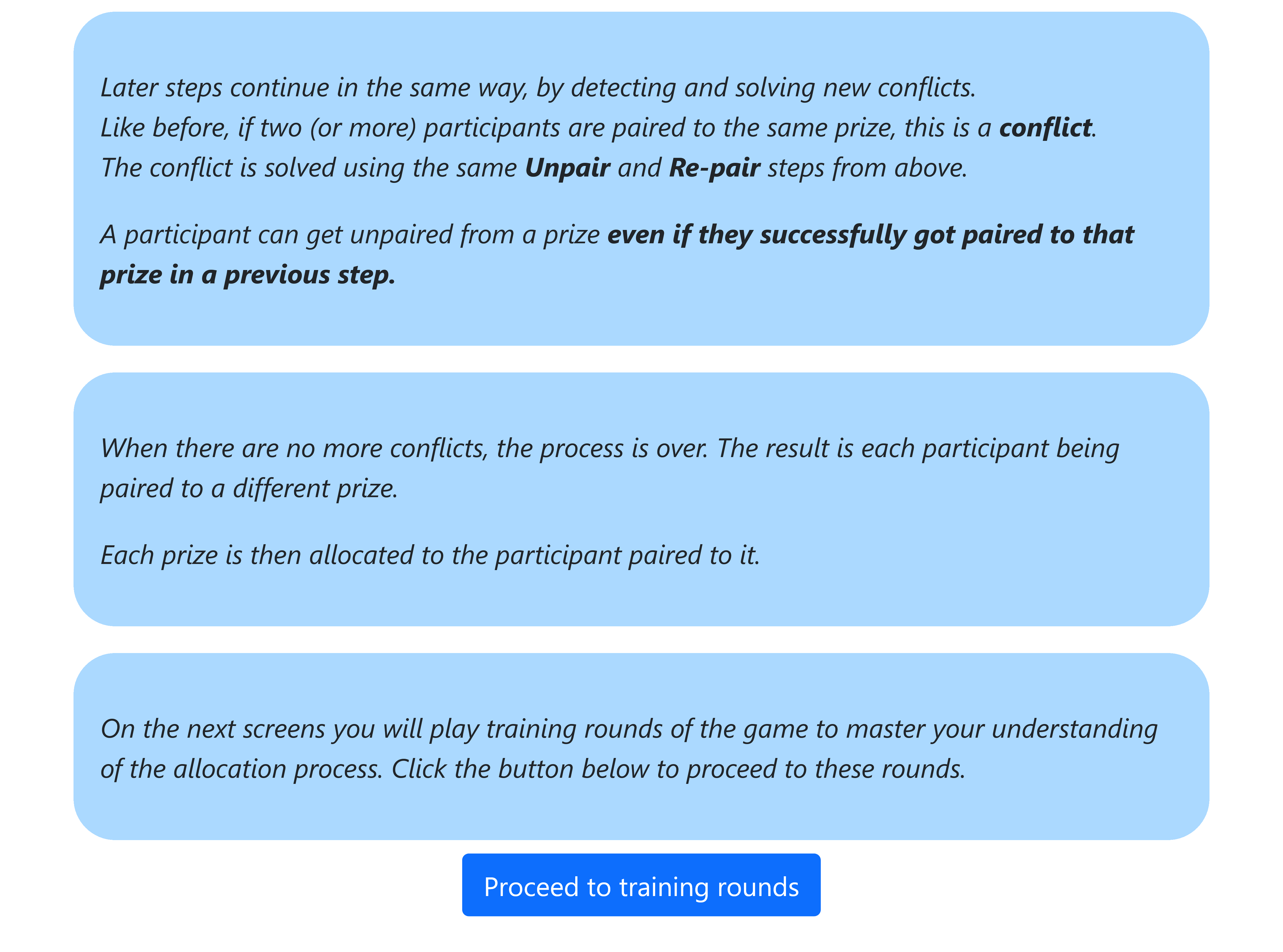}
    \caption{Traditional DA Mechanics description main text (2/2)}
    \label{fig:trad-mech-descr-2}
\end{figure}

\begin{figure}
    \centering
    \includegraphics[width=0.6\textwidth]{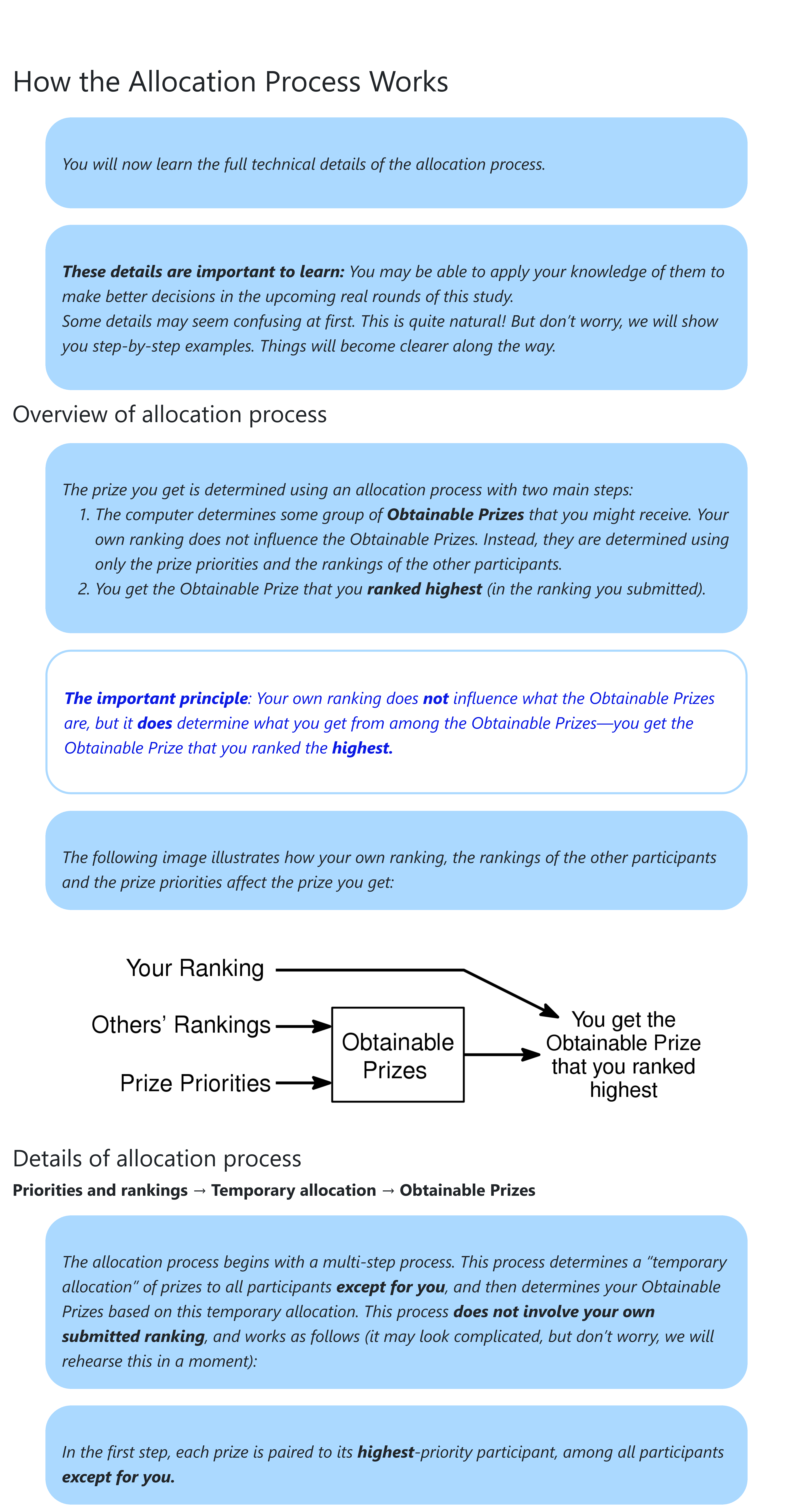}
    \caption{Menu DA Mechanics description main text (1/2)}
    \label{fig:menu-mech-descr-1}
\end{figure}

\begin{figure}
    \centering
    \includegraphics[width=0.55\textwidth]{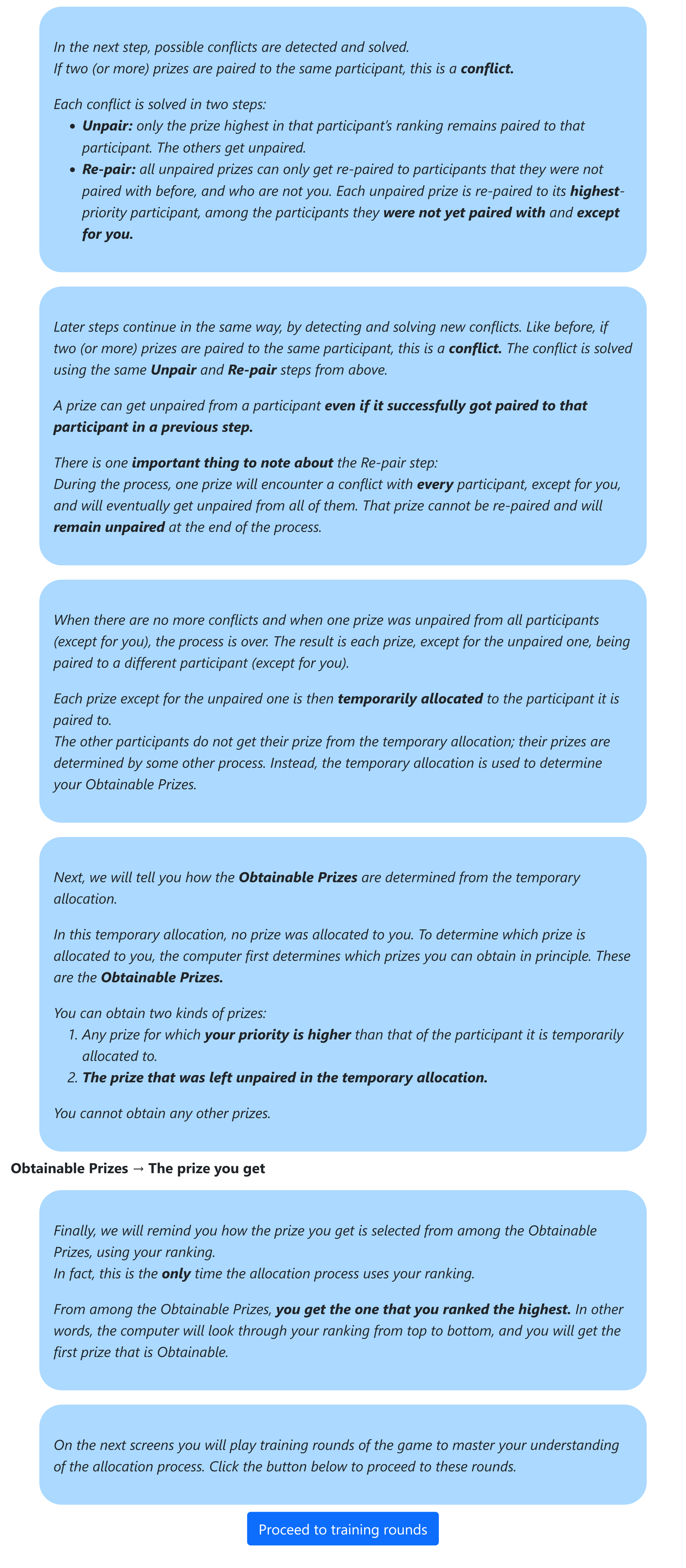}
    \caption{Menu DA Mechanics description main text (2/2)}
    \label{fig:menu-mech-descr-2}
\end{figure}

Figures \ref{fig:trad-mech-descr-1} and \ref{fig:trad-mech-descr-2} show the description provided in the Traditional DA Mechanics treatment.
Figures \ref{fig:menu-mech-descr-1}  and \ref{fig:menu-mech-descr-2} show Menu DA Mechanics.
We remark that participants are not necessarily expected to understand the details of this description immediately; as discussed in \autoref{sec:design-training}, they receive substantial additional coaching throughout the training rounds.



\begin{figure}
    \centering
    \includegraphics[width=0.7\textwidth]{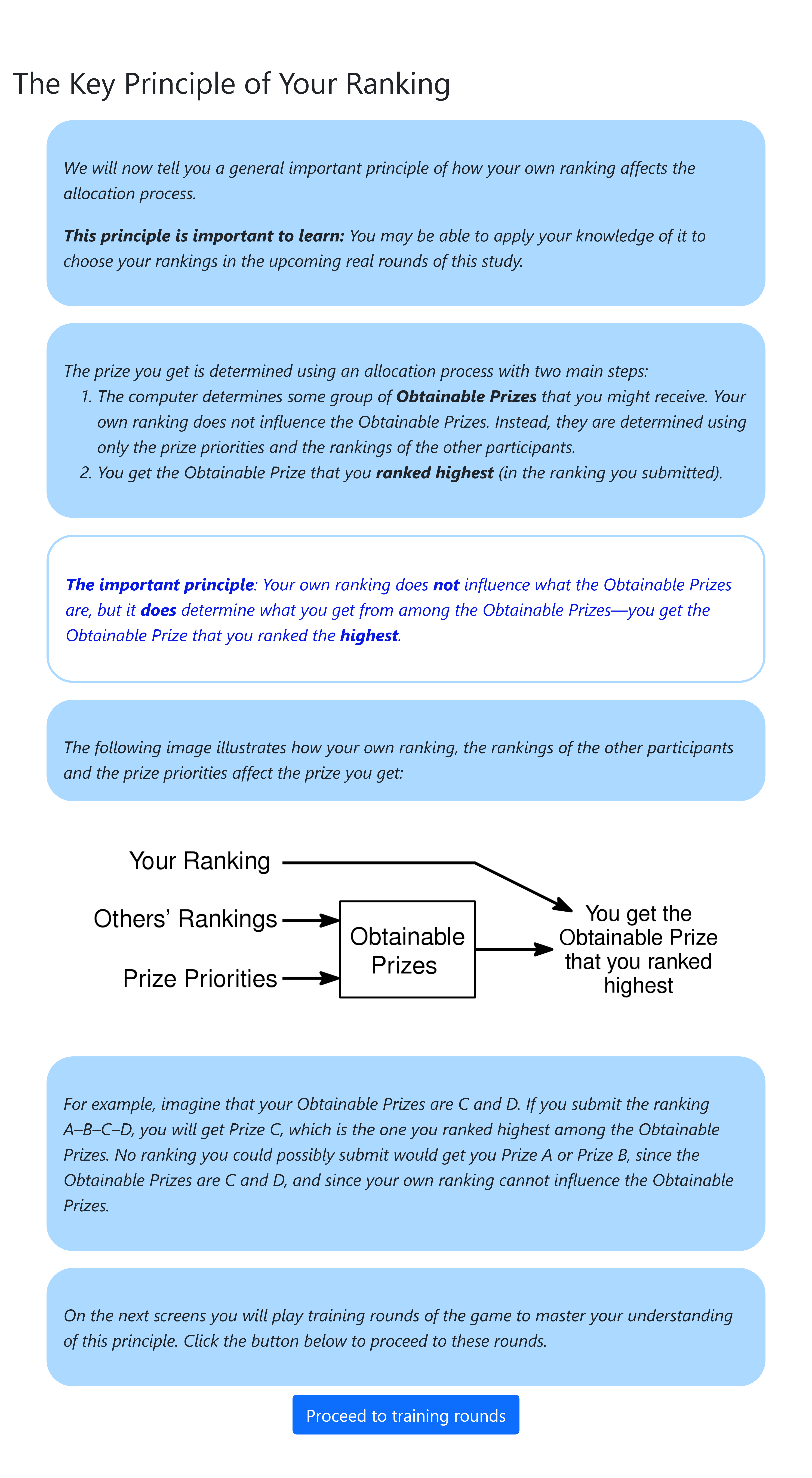}
    \caption{Menu SP Property description main text}
    \label{fig:menu-prop-descr-1}
\end{figure}

\begin{figure}
    \centering
    \includegraphics[width=0.6\textwidth]{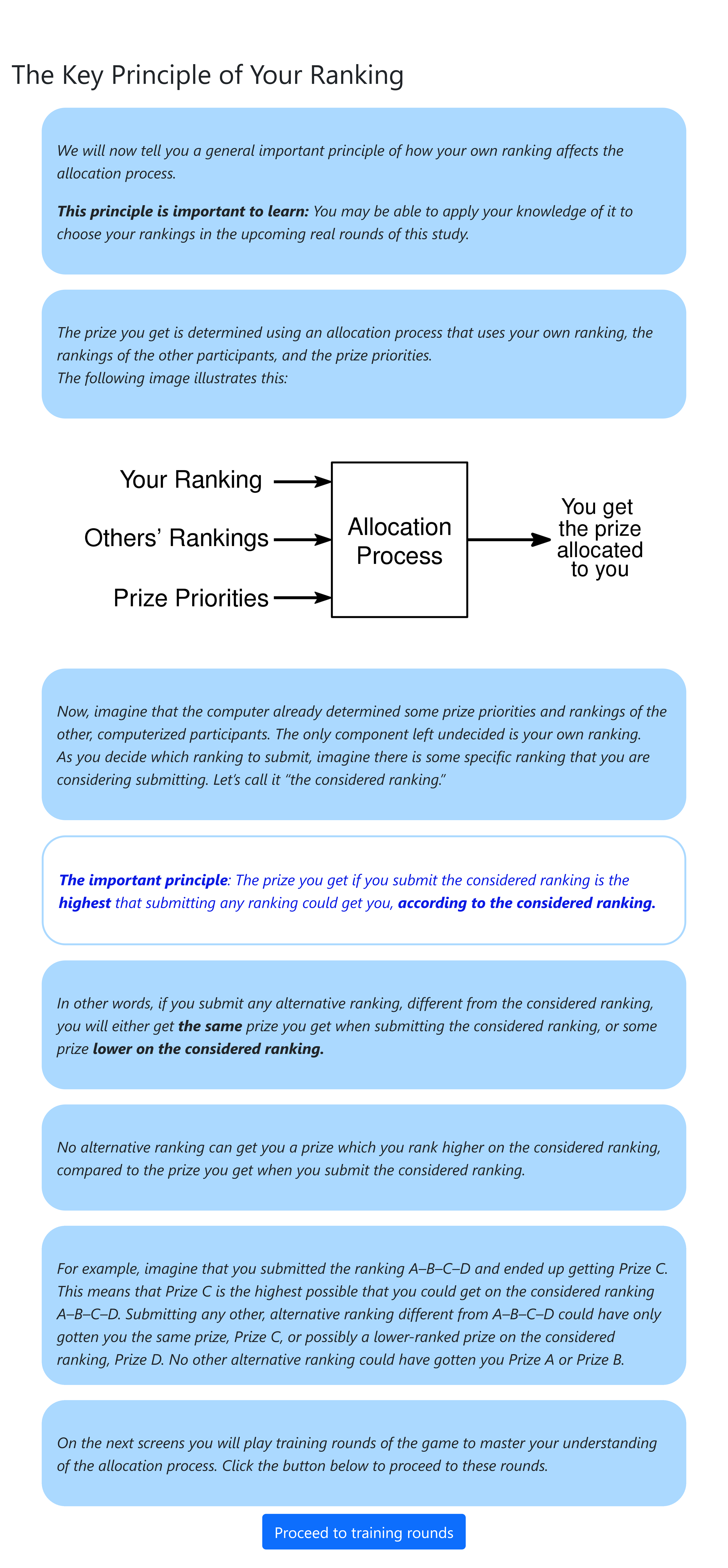}
    \caption{Textbook SP Property description main text}
    \label{fig:bench-prop-descr-1}
\end{figure}

Figures \ref{fig:menu-prop-descr-1} and \ref{fig:bench-prop-descr-1} show the main description texts of our two SP Property treatments.
Similarly to the DA Mechanics treatments, participants receive additional coaching on this description as they complete the training rounds, as discussed in \autoref{sec:design-training}.

\begin{figure}
    \centering
    \includegraphics[width=0.8\textwidth]{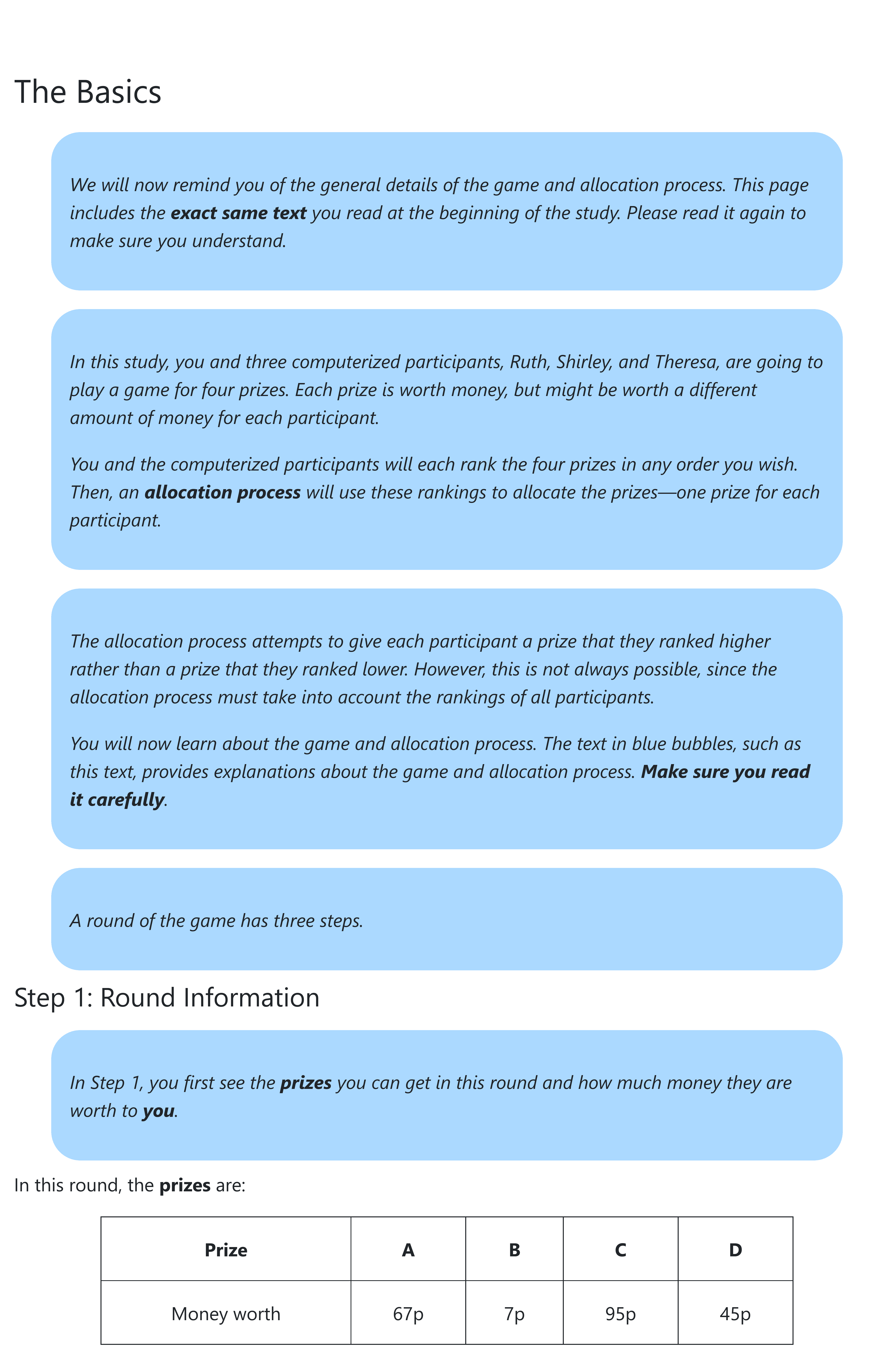}
    \caption{Null treatment description main text (1/3)}
    \label{fig:null-repeated-descr-1}
\end{figure}

\begin{figure}
    \centering
    \includegraphics[width=0.8\textwidth]{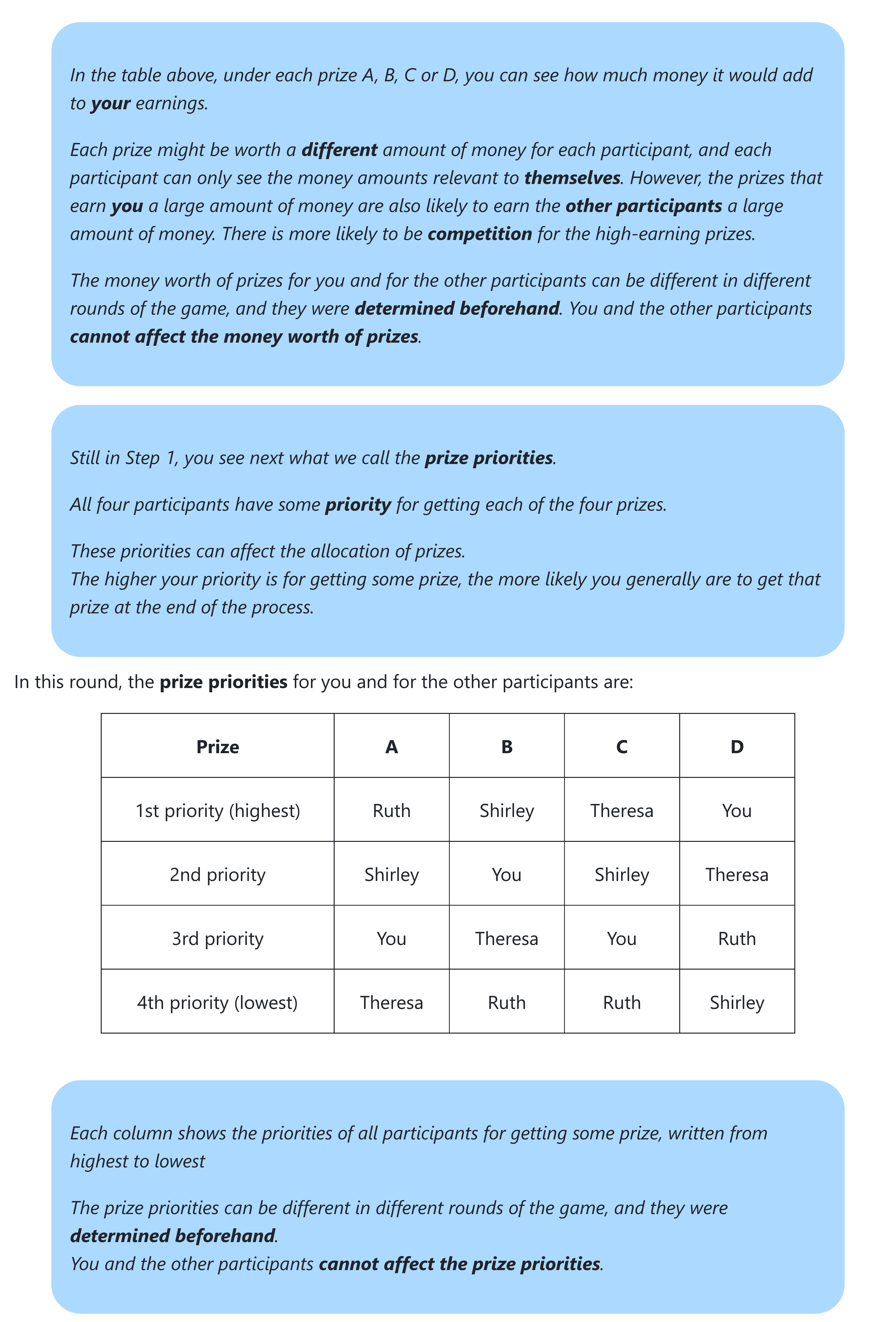}
    \caption{Null treatment description main text (2/3)}
    \label{fig:null-repeated-descr-2}
\end{figure}

\begin{figure}
    \centering
    \includegraphics[width=0.7\textwidth]{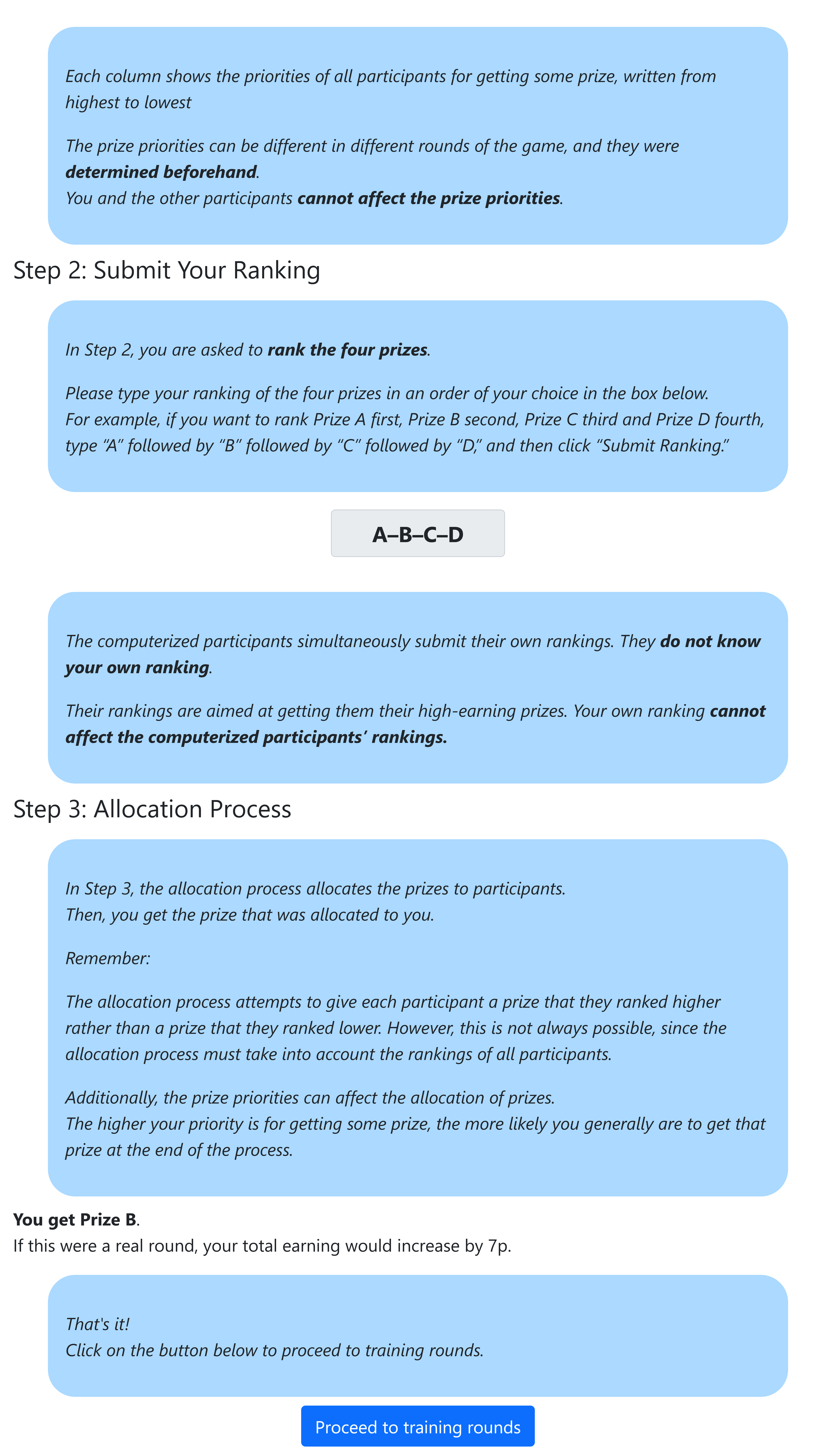}
    \caption{Null treatment description main text (3/3)}
    \label{fig:null-repeated-descr-3}
\end{figure}

Figures \ref{fig:null-repeated-descr-1} to \ref{fig:null-repeated-descr-3} show the full text of the Null treatment description.
Note that this text is identical to the Null description, which is given in the initial screens of all treatments (excluding the small bit of text at the beginning noting that information will be repeated).

\subsection{Scenarios in DA Mechanics Training Rounds}
\label{app:scenarios}

\begin{figure}[htbp]
    \centering
    \caption{DA Mechanics training scenarios}
    \label{fig:mech-tr-scenarios} 

    \begin{minipage}[t]{0.49\textwidth}
    \centering
        \begin{minipage}{0.8\textwidth}
        \subcaption{First training round}
        \end{minipage}
    \end{minipage}
    \begin{minipage}[t]{0.49\textwidth}
    \centering
        \begin{minipage}{0.8\textwidth}
        \subcaption{Third training round}
        \end{minipage}
    \end{minipage}

    \begin{minipage}[t]{0.49\textwidth}
        \vspace{0em}
        \frame{\includegraphics[width=\textwidth]{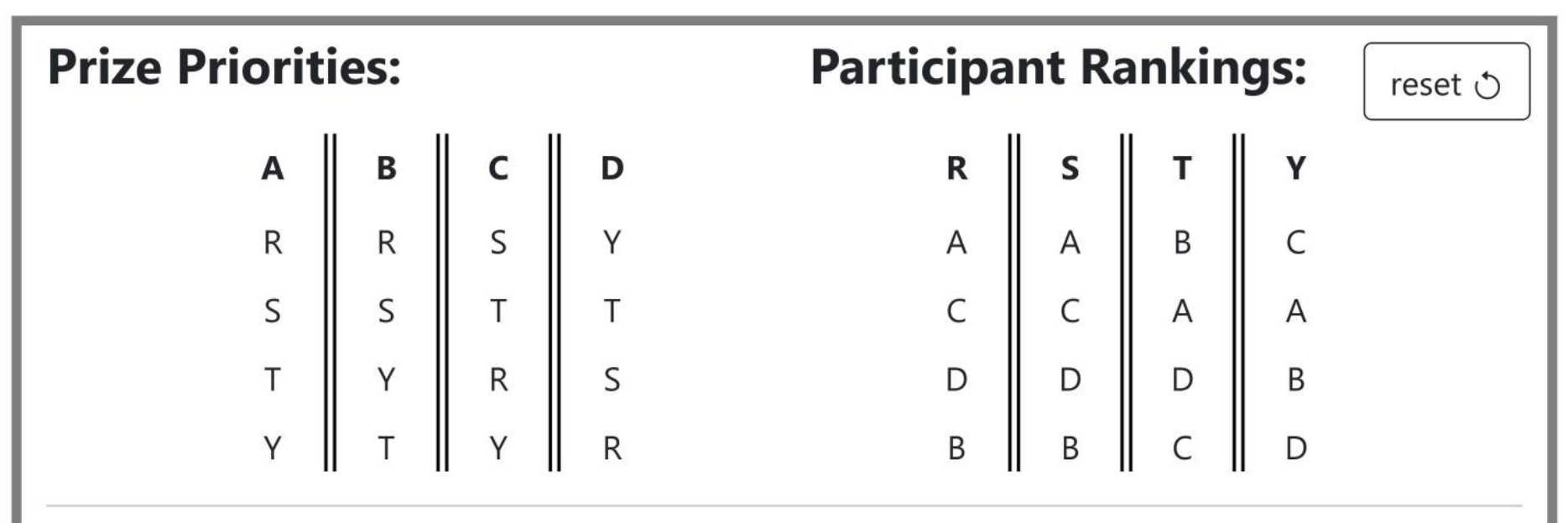}}
    \end{minipage}
    \begin{minipage}[t]{0.49\textwidth}
        \vspace{0em}
        \frame{\includegraphics[width=\textwidth]{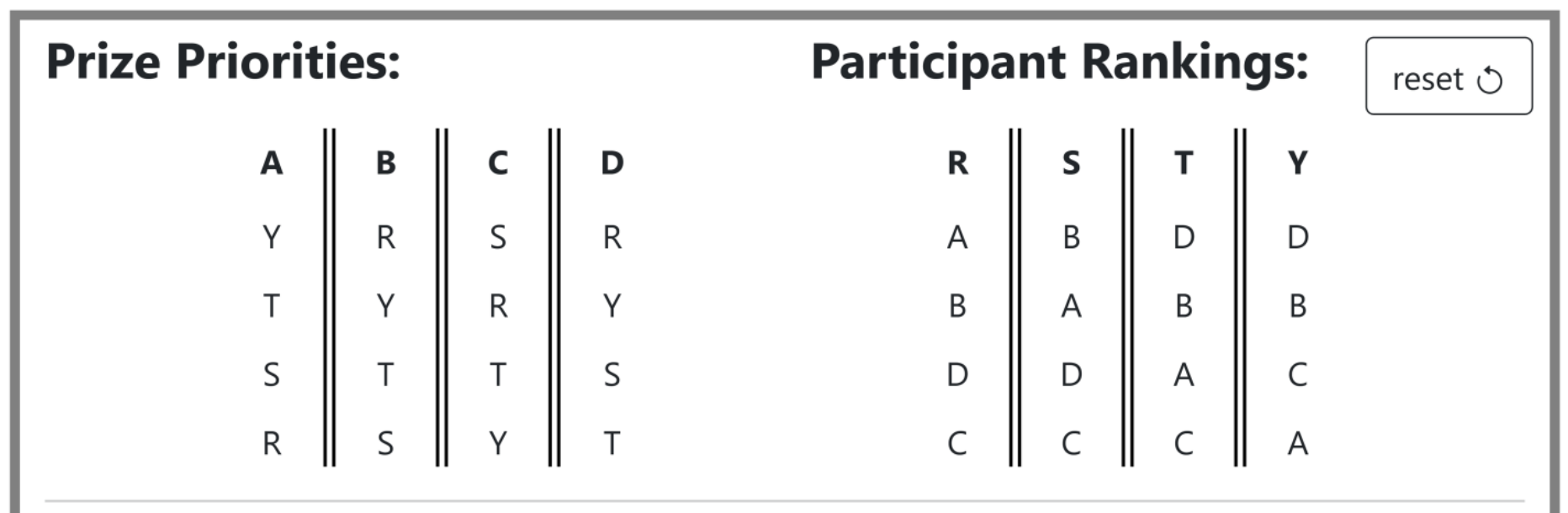}}
    \end{minipage}

    \vspace{0.5em} 
    \begin{minipage}{\textwidth} \footnotesize
        \textbf{Notes:}
        Screenshots are taken from the Trad-DA treatment.
        See \autoref{fig:mechanics-training-UI} for the scenarios used in the second training round.
    \end{minipage}
\end{figure}

    We describe the DA scenarios (i.e., the human and computerized participants' rankings and the prize priorities) used in the DA Mechanics training rounds.
    The same DA scenarios are presented in both Trad-DA and Menu-DA. 
    See \autoref{fig:mech-tr-scenarios} for the scenarios used in the first and third training rounds (\autoref{fig:mechanics-training-UI} in \autoref{sec:design-training} shows the second round).

    These scenarios are chosen so that running each of these algorithms (whether participant- or prize-proposing) would be of comparable complexity.
    In the first round, correctly completing the training GUI requires 9 total proposals in both Trad-DA and Menu-DA; the menu consists of Prizes B and D.
    In the second round, the GUI requires 8 total proposals in both Trad-DA and Menu-DA; the menu consists of Prize A.
    In the third round, the GUI requires 10 total proposals in Trad-DA and 8 in Menu-DA; the menu consists of Prizes A, C, and D.

\subsection{Randomization of Setting Components in Real DA Rounds}\label{app:randomization-setting}

We describe the underlying randomization procedures which determine the three setting components of real DA rounds that participants play. In each round, these components include (1) the monetary values of the four prizes A, B, C and D for the human participant, (2) each participant's priorities for getting the different prizes (the ``preferences of prizes'' over participants), and (3) the computerized participants' submitted rankings (unobserved by the human participant).

These three components are randomly drawn from a joint distribution each round, where each (human) participant $\times$ round draw is independent from others. Each round randomization is performed as follows:
\begin{enumerate}
    \item Prize values (for the human participant) are randomly and independently determined: 
    \begin{itemize}
        \item The highest value is drawn from a uniform distribution over $\{0.90,0.91,...0.99\}$.
        \item The second highest value is drawn from a uniform distribution over 
        $\{0.50,0.51,...0.89\}$.
        \item The third highest value is drawn from a uniform distribution over 
        $\{0.10,0.11,...0.49\}$.
        \item The lowest value is drawn from a uniform distribution over 
        $\{0,0.01,...0.09\}$.
        \item Each prize among Prize A, B, C and D gets one of the above values, with equal probabilities.
        \item All the above numeric values are determined the same in Prolific and Cornell, but the currency attached to them is British pounds (\pounds) in Prolific and US dollars (\$) at Cornell.
    \end{itemize} 
    
    \item For each prize, the four participants' priorities for getting it are sequentially determined from highest to lowest, by iteratively selecting a participant without replacement in four steps. In more detail:
    \begin{itemize}
        \item For all prizes except the highest-valued one from (1), each participant has an equal chance of having each of the four priorities, i.e., of being chosen at each step. 
        \item For the highest-valued prize, the human participant's chances are different from the computerized participants': at each step, each computerized participant has an $r_1$-times higher chance of being picked to that step's priority than human's chance.
        \item We use a value of $r_1=1.7>1$, such that the chances of having a high priority for getting the highest-valued prize are ``rigged'' against the human participant. (Note that a value of $r_1=1$ would generate symmetry across the human and computerized participants, and a value $r_1<1$ would rig the chances in favor of the human participant).
    \end{itemize}

    \item For each computerized participant, their submitted ranking is sequentially determined from top to bottom rank, by iteratively selecting a prize without replacement, where the chances to select among the different prizes at each step is a function of the order of prize values for the human participant from (1). In more detail:
    \begin{itemize}
        \item First, the participant's first (highest) rank is determined. The chance to select each prize at this rank is proportional to $r_2^{P-1}$, where $r_2$ is a fixed parameter (see below) and $P$ is the ordinal position of that prize when prizes are sorted from highest to lowest values for the human participant (i.e., $P=1$ for the highest-valued prize for the human participant and $P=4$ for the lowest-valued one for the human participant). 
        \item Then, that participant's second rank is similarly determined, after sorting the remaining three prizes from highest to lowest, and so on until the fourth rank.
        \item Due to the dependence of the ranking on the prize values for the human participant, the $r_2$ parameter effectively sets a type of correlation between the computerized participants' ranking and the human's participant straightforward (SF) ranking. We use a parameter value of $r_2=0.5$. (One can show that $r_2 \rightarrow 0$ would generate a perfect correlation of all rankings with the human participant's SF ranking, and $r_2=1$ would generate zero correlation and uniformly random rankings for the computrized participants.) 
    \end{itemize}
\end{enumerate}

The joint distribution of the three setting components following the above procedure is designed to achieve two main goals. First and most importantly, it was tested to induce sufficient variation in participants' submitted rankings---a necessary condition to identify of our treatments' effect on ranking behavior---and successfully replicates salient non-straightforward ranking patterns observed in previous studies (e.g., \citealt{Li17}, \citealt{DreyfussHR22}, \citealt{DreyfusGHR22}). 
Intuitively, it does so by making the setting feel ``hard'' and competitive, in order to encourage participants to think more carefully on their strategy. This is done by (1) giving participants a lower probability of having first priority for getting the highest-earning prize than to the computerized participants (recall that all these priorities are shown in a table each round), and (2) correlating the computerized participants' rankings with the participant's straightforward ranking.\footnote{
  Using a simpler, uniform distribution over the computerized participants' rankings instead results in an ``easy'' setting on average, with at least a 0.25 probability of getting the highest-earning prize whenever the participant ranks it highest, regardless of how the priorities for getting the prizes are distributed (since one can show that one uniformly random prize will go unmatched when considering DA run on only the computerized participants). }

Second, the distribution is designed to often induce large differences between prize values, such that conditional on NSF behavior being costly ex-post, i.e., earning-decreasing, the cost is significant. In most rounds of our experiment, the most common NSF ranking pattern of flipping the ranking of the highest- and second-highest-earning prizes is not costly ex-post. However, participants are neither informed on this nor can easily calculate this, as they do not know the computerized participants' rankings or their distribution. This common NSF ranking pattern is only costly when ranking the highest-earning prize first would win it, and making such conditions frequent would interfere with our goal of creating a competitive setting, where it is unlikely to win the highest-earning prize. Evidence from \citet{DreyfusGHR22}---that NSF behavior is most prevalent when the probability of winning the highest reward is low, i.e., when such behavior is unlikely to affect the participant's final outcome and hence is not costly---suggests that increasing the expected cost of NSF behavior would also interfere with maintaining overall sufficient rates of such behavior. In order to promote better identification of ranking patterns, we opt for inducing high rates of non-costly NSF behavior over inducing lower rates of costly NSF behavior.

\subsection{Experiment Procedures}
\label{app:experiment-procedures}

We describe the full protocol used to run the experiment on Prolific and at the Cornell Business Simulation Lab (BSL), and provide more details on the sessions in each of the platforms.

\subsubsection{Prolific}

\begin{figure}[htbp]
\centering
\caption{Prolific recruitment text}
\label{fig:prolific-recruitment}
\frame{\includegraphics[width=\figscale\textwidth{}]{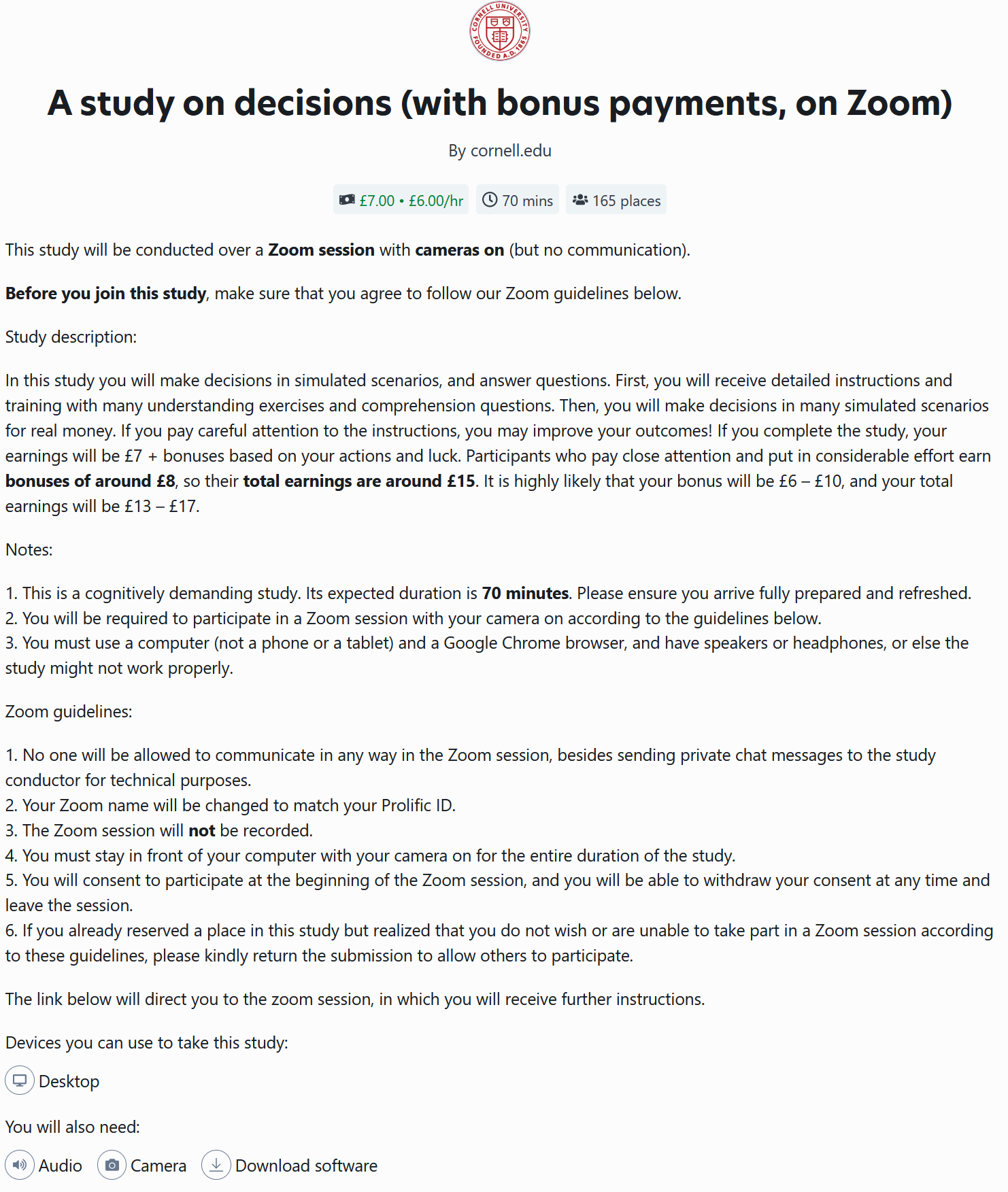}}

\begin{minipage}{\textwidth}  \footnotesize
        \textbf{Notes:}
        A screenshot of the page used to recruit Prolific participants to our experiment. The number of slots mentioned at the top (``165 places'') is session-specific.
    \end{minipage}
\end{figure}

\paragraph{Protocol.}
Participants were recruited using the page shown in \autoref{fig:prolific-recruitment}. Only Prolific participants who passed the pre-screening criteria of living in the US and having at least 99\% past approval rate and at least 50 past completed tasks were able to see the page. Participants who wished to participate could click a button (not shown in the screenshot) that redirected them to the video-conference session where the experiment took place (see below).

In the recruitment text, participants were explicitly informed of the length of the study and on it being cognitively demanding, and were asked to prepare accordingly. They were informed on the general nature of the experiment and on their expected payment (based on pilot runs). 

Participants had to run the experiment using a computer, rather than a tablet or a phone, use a specific browser (due to technical reasons of software stability) and verify they have functional camera and audio output in their computer. 

Importantly, they were required to join a video-conference session (using Zoom) with the experiment conductor in order to participate, and were asked to follow specific guidelines in the Zoom session: (1) avoid any kind of communication with each other, (2) use their Prolific ID instead of their name, and (3) stay visible in front of their camera but stay muted and not communicate through messages (as mentioned next, in addition to asking participants to avoid such behavior, Zoom settings were changed to technically prevent as much of it as possible).

The experiment Zoom sessions were run according to the following protocol. Early preparations prior to the session included creating a Zoom room for the session, and changing Zoom settings to disable the chat function of participants with one another, enable a waiting room, disable any kind of screen sharing, emojis, ability to unmute onself, captioning and recording. The Zoom room password (and link given to participants) was changed between sessions to prevent past participants from re-joining the room using their link.

In addition, early preparations of the oTree environment and server running the experiment were done before each session, and included resetting all data and runtime variables.

The Zoom session was opened shortly prior to posting the recruitment message on Prolific. The session conductor was named ``CONDUCTOR,'' and was required to join the session using another device named ``BACKUP,'' in case the main device disconnects.

Participants typically joined the session gradually over a few hours, but the session's protocol was designed to create an experience as similar as possible for all participants despite their different timing of joining the session. As participants entered the Zoom waiting room, they were first renamed as ``Pending approval \#'' (where \# was 1, 2, 3 and so on, in the participants' order of connection).  After renaming, a participant was let in the meeting and was sent the following message: ``Dear participant, please turn on your camera. Please enter the following link to read the general guidelines for this study: \url{https://jescstudies.my.canva.site/}'' (in case the conductor noted that the participant's camera is already on, the message read ``Dear participant, please enter...'').

\begin{figure}[htbp]
\centering
\caption{Study guidelines read by Prolific participants once entering the Zoom session}
\label{fig:prolific-guidelines}

\frame{\includegraphics[width=\figscale\textwidth{}]{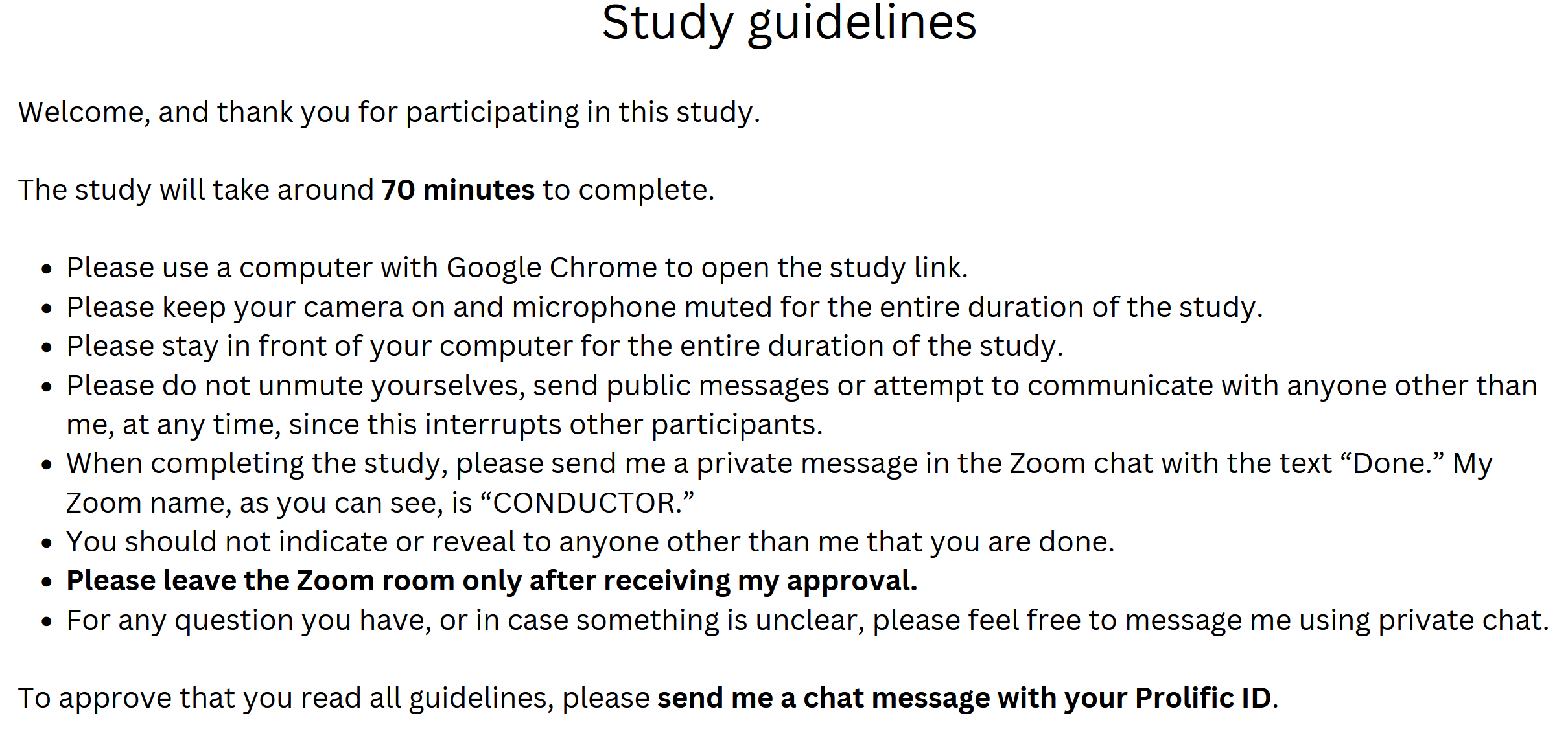}}

\begin{minipage}{\textwidth}  \footnotesize
        \textbf{Notes:}
        A screenshot of the page to which participants were redirected shortly after joining the Zoom session and before starting the experiment.
    \end{minipage}
\end{figure}

The participant then saw the slide shown in \autoref{fig:prolific-guidelines}. The slide contained a repetition of the guidelines for the zoom session, including instructions on how and when to communicate with the session conductor to signal that they completed the experiment or to ask for assistance. Participants were asked to leave the session only upon getting approval from the conductor. Following the guidelines slide, each participant sent the conductor a message with their Prolific ID and then received the link to the experiment using the following message: ``Thank you. Remember to stay in front of your computer with your camera on for the  entire duration of the study, and send me “Done” when you  complete it. The study link is: [LINK].''

As participants started the experiment, they were renamed to their Prolific ID to help keep track of their identities inside the session, and were documented in a spreadsheet with their ID and time of starting the experiment. Typically, there was no communication with participants from this point until the point when they completed the experiment. After informing on completion to the conductor, they received the following message: ``Thank you for completing the study. You can leave the meeting,'' and their end time was documented in the spreadsheet.

Participants who asked questions about the experiment itself were answered ``Unfortunately, I can't help you with the study itself. Please try to answer to the best of your understanding.'' In case of a technical difficulty, the conductor provided assistance in trying to solve the problem.

After the session has ended, participants' completion was approved in the Prolific platform and they were paid according to their responses in the experiment, typically within a day.

\paragraph{Sessions.}

291 Prolific participants participated in a total of 3 Zoom sessions on August 3, 7 and 8, 2023. 

The August 3 session was a final pilot including 43 participants. We opted (and pre-registered) to include it in our data, since at that point the experiment no longer changed except for two small bugs which were fixed toward the final two sessions.

The session on August 7 included 72 participants, and the session on August 8 included 176 participants.

\subsubsection{Cornell Business Simulation Lab (BSL)}

\paragraph{Protocol.}

\begin{figure}[htbp]
\centering
\caption{Cornell BSL recruitment text}
\label{fig:cornell-recruitment}

\frame{\includegraphics[width=\figscale\textwidth{}]{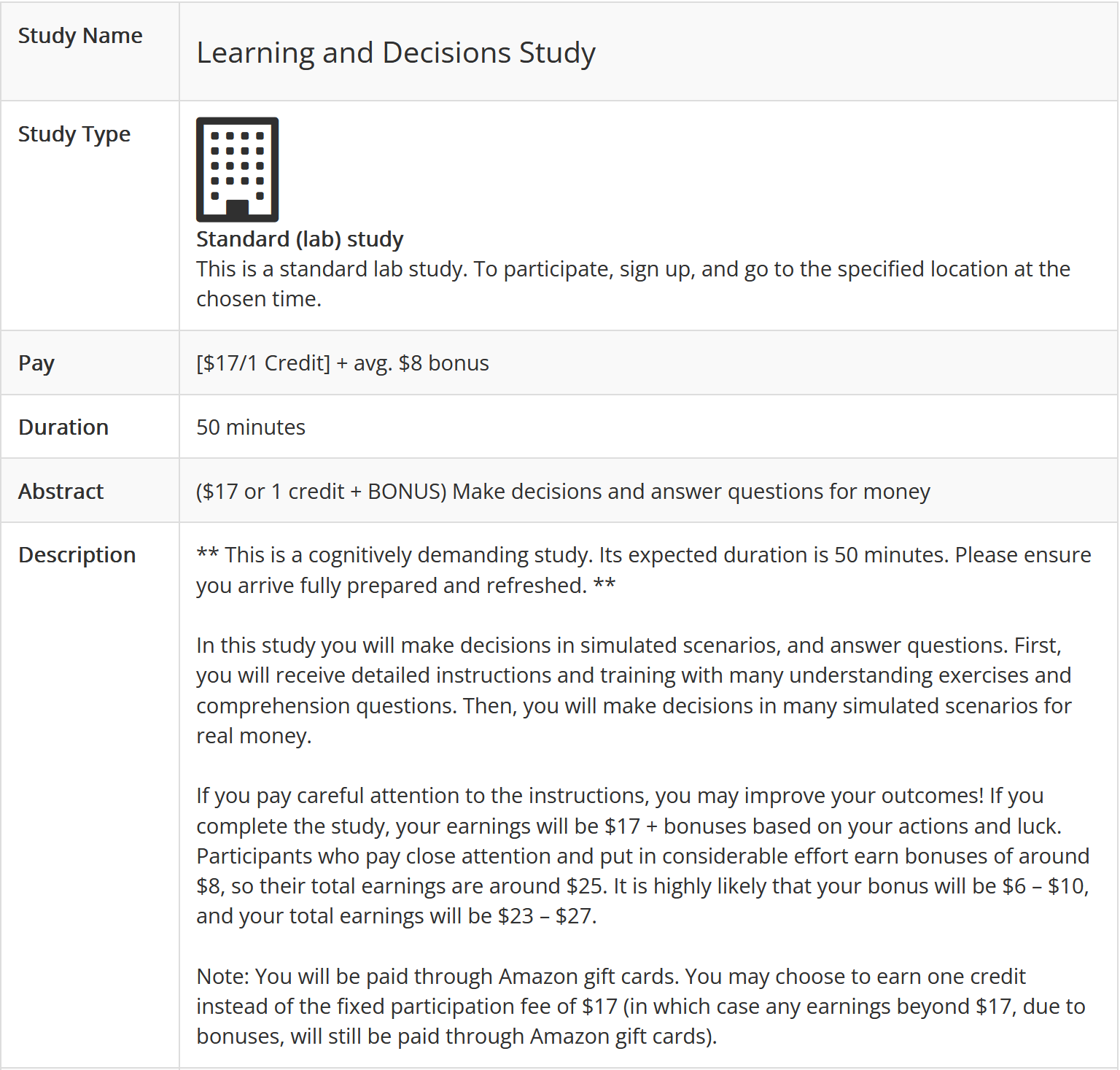}}

\begin{minipage}{\textwidth}  \footnotesize
        \textbf{Notes:}
        A screenshot of the page used by the Cornell Business Simulation Lab (BSL) to recruit Cornell students to our experiment. 
    \end{minipage}
\end{figure}

Participants were recruited using the page shown in \autoref{fig:cornell-recruitment}. Only Cornell students were able to register to the experiment, and could choose a future physical lab session out of a few possibilities, typically at standard work hours. 

In the recruitment text, participants were explicitly informed of the length of the study and on it being cognitively demanding, and were asked to prepare accordingly. They were informed on the general nature of the experiment and on their expected payment (based on pilot runs). 

In contrast to Prolific sessions to which participants could join at any time during some time window, BSL sessions required all participants to start at the same time. Sessions at the same day were scheduled at least 1.5 hours apart to enable enough time for completing the experiment (in the rare case where a participant did not complete the experiment before another session began, they could stay at the lab and complete it while the next session took place). Sessions were planned to include at most 20 participants each, but in practice registration was sparse and most sessions included less than 10 participants.

Participants used lab computers to access the experiment, but the experiment was run on a remote server. Prior lab preparations included opening the experiment link in the browser at all lab stations.

Each session was run using the following protocol. The session conductor read participants the following text aloud: 
\begin{quote}
    Welcome and thank you for participating in this study. 

    There is no deception in this study---everything you will read on the screen during this study is true. 
    
    Please wear the headphones in your station during the entire study and follow the instructions on your screen. 
    
    If you are done in less than 50 minutes you can do whatever you would like while seated in your station, but you still have to stay in your station until the 50 minutes are over. You should not indicate or reveal to anyone that you are done.
    
    If you are done and 50 minutes have already passed, please raise your hand quietly and we will come to your station to dismiss you. 
    
    In case you have any question or problem, please also raise your hand quietly and we will reach out to help.
\end{quote}
As the text mentions, participants were required to stay at the lab for at least 50 minutes, to avoid potential effects of participants randomized into shorter treatments leaving before participants who were randomized into longer treatments. If a participant joined the session late but by no more than 10 minutes, they were allowed to join and the above text was read to them again privately.

Participants were typically paid by the lab within a few days.

\paragraph{Sessions.}

295 Prolific participants participated over 32 days, each including between 1 and 3 sessions. This large number of small sessions was required to reach our pre-registered sample size since sign-up rates were lower than expected. 

On October 11, 2023, one session including 16 participants was accidentally set incorrectly, such that instead of randomizing participants into treatments they were all allocated into the Trad-DA treatment. We do not count this session towards the 50 participants-per-treatment threshold defined by our pre-registration, and instead add these participants on top of at least 50 Trad-DA participants which were collected in the other sessions. In our robustness analysis we control for session-date fixed effects and do not find them to change our results.

\section{Full Screenshots}\label{full-materials}

For Appendix D, which contains screenshots of every screen of every treatment, 
see the supplementary material on the authors' websites and in the ancillary files on arXiv. 
For screenshots of the per-treatment description screens, see \autoref{app:select-screenshots} in the Supplementary Materials.

\end{document}